\documentclass[onecolumn,prc,floatfix,showpacs,preprintnumbers,nofootinbib,%
superscriptaddress]{revtex4}
\usepackage{mathrsfs}
\usepackage{amssymb}	
\usepackage{amsmath}
\usepackage{graphicx}
\usepackage{subdepth}
\usepackage{color}
\usepackage{xcolor}
\definecolor{lcolor}{rgb}{0.5,0,0}
\definecolor{citcolor}{rgb}{0,0.3,0.0}

\usepackage[latin1]{inputenc}

\usepackage[breaklinks,colorlinks,urlcolor=blue,citecolor=citcolor,linkcolor=lcolor]{hyperref}

\usepackage{tikz}

\usepackage{mciteplus}

\newcommand{\ktp}{{\mathbf{k}'}}
\newcommand{\ptp}{{\mathbf{p}'}}
\newcommand{\ptpp}{{\mathbf{p}''}}

\newcommand{\rt}{{\mathbf{r}}}
\newcommand{\xt}{{\mathbf{x}}}

\newcommand{\yt}{{\mathbf{y}}}

\newcommand{\pt}{{\mathbf{p}}}
\newcommand{\qt}{{\mathbf{q}}}
\newcommand{\kt}{{\mathbf{k}}}
\newcommand{\nt}{{\mathbf{n}}}
\newcommand{\mt}{{\mathbf{m}}}

\newcommand{\hht}{\mathbf{h}}
\newcommand{\lt}{\mathbf{l}}

\newcommand{\epst}{\boldsymbol{\varepsilon}}
\newcommand{\gammat}{\boldsymbol{\gamma}}

\newcommand{\ovec}{{\vec{0}}}
\newcommand{\kvec}{{\vec{k}}}
\newcommand{\pvec}{{\vec{p}}}
\newcommand{\ppvec}{{{\vec{p}}{\, '}}}
\newcommand{\pppvec}{{{\vec{p}}{\, ''}}}
\newcommand{\kpvec}{{{\vec{k}}{\, '}}}
\newcommand{\kppvec}{{{\vec{k}}{\, ''}}}
\newcommand{\qvec}{{\vec{q}}}
\newcommand{\qpvec}{{{\vec{q}}{\, '}}}
\newcommand{\xvec}{{\vec{x}}}

\newcommand{\lo}{{\textnormal{LO}}}

\newcommand{\epsl}{{\epsilon\hspace{-0.9ex}/}}

\newcommand{\dk}{{\widetilde{\mathrm{d} k}}}
\newcommand{\dkp}{{\widetilde{\mathrm{d} k}{'}}}
\newcommand{\dkpp}{{\widetilde{\mathrm{d} k}{''}}}

\newcommand{\dpp}{{\widetilde{\mathrm{d} p}{'}}}
\newcommand{\dppp}{{\widetilde{\mathrm{d} p}{''}}}
\newcommand{\dqp}{{\widetilde{\mathrm{d} q}{'}}}
\newcommand{\dq}{{\widetilde{\mathrm{d} q}}}

\newcommand{\ud}{\, \mathrm{d}}

\newcommand{\tr}{\, \mathrm{Tr} \, }

\newcommand{\nf}{{N_\mathrm{F}}}
\newcommand{\tf}{{T_\mathrm{F}}}
\newcommand{\half}{\frac{1}{2}}

\newcommand{\cf}{C_\mathrm{F}}
\newcommand{\ca}{C_\mathrm{A}}

\newcommand{\nr}[1]{(\ref{#1})}

\newcommand{\fig}{Fig.~}
\newcommand{\figs}{Figs.~}
\newcommand{\eq}{Eq.~}

\newcommand{\eqs}{Eqs.~}

\newcounter{diag}
\newcommand{\namediag}[1]{\refstepcounter{diag} \thediag \label{#1}}
\renewcommand{\thediag}{(\alph{diag})}

\newcommand{\epsmsbar}{\varepsilon_{\overline{\textnormal{MS}}}}

\begin{document}

\author{T. Lappi}
\affiliation{
Department of Physics, %
 P.O. Box 35, 40014 University of Jyv\"askyl\"a, Finland}
\affiliation{
Helsinki Institute of Physics, P.O. Box 64, 00014 University of Helsinki,
Finland}

\author{R. Paatelainen}
\affiliation{
Department of Physics, %
 P.O. Box 35, 40014 University of Jyv\"askyl\"a, Finland}
\affiliation{
Instituto Galego de Fisica de Altas Enerxias (IGFAE), 
Universidade de Santiago de Compostela,
E-15782 Santiago de Compostela, Galicia, Spain}

\title{
The one loop gluon emission light cone wave function
}


\begin{abstract}
Light cone perturbation theory has become an essential tool to calculate cross sections for various small-$x$ dilute-dense processes such as deep inelastic scattering and forward proton-proton and proton-nucleus collisions. Here we set out to do one loop calculations in an explicit helicity basis in the four dimensional helicity scheme. As a first process we calculate light cone wave function for one gluon emission to one-loop order in Hamiltonian perturbation theory on the light front. We regulate ultraviolet divergences with transverse dimensional regularization and soft divergences with using a cut-off on longitudinal momentum. We show that when all the renormalization constants are combined, the ultraviolet divergences can be absorbed into the standard QCD running coupling constant, and give an explicit expression for the remaining finite part.
\end{abstract}

\maketitle

\section{Introduction}

Hamiltonian perturbation theory in the light cone (or light front) 
form~\cite{Kogut:1969xa,Bjorken:1970ah,Lepage:1980fj,Brodsky:1997de}  has become a standard tool in understanding hadronic scattering processes within a first-principles QCD approach. The calculational inconveniences of light cone perturbation theory (LCPT)  compared to standard covariant perturbation theory are balanced by several features that make its physical interpretation more transparent. Light cone gauge LCPT only involves physical degrees of freedom, i.e. spin- or helicity states of partons, enabling an interpretation in terms of a constituent picture of hadrons. The factorization between long distance hadronic physics both in the incoming and final state hadrons on one hand, and hard perturbative QCD scattering on the other hand, is naturally implemented in LCPT. This allows for a simultaneous  description of inclusive and exclusive processes in a consistent framework. 

More recently LCPT has found a new area of application in understanding the nonlinear QCD physics of gluon saturation.  Gluon saturation is most naturally understood in the ``color glass condensate'' (CGC) effective theory~\cite{Iancu:2003xm,Weigert:2005us,Gelis:2010nm,Albacete:2014fwa}, which describes the soft small-$x$ degrees of freedom in the high energy hadron as a classical field, radiated by color sources representing the large momentum partons. This classical color field can then be probed in various scattering processes by different dilute probes, whose interactions with the target are described in the high energy limit by an eikonal Wilson line on the light cone. The most natural way to describe the structure of the dilute probe (a real or virtual photon, or a quark or gluon from a probe hadron) in terms of a Fock state decomposition of partonic states is the light cone wave function of LCPT. This provides a formalism to factorize cross sections into light-cone wave functions describing the structure of the probe developing on a long timescale before the interaction, and Wilson line operators that describe the instantaneous scattering of the probe Fock state on the target.

Early loop calculations in  LCPT~\cite{Thorn:1979gv,Mustaki:1990im,Perry:1992sw,Zhang:1993is,Zhang:1993dd,Harindranath:1993de} explored the stucture of divergences in the longitudinal and transverse momentum integrals and recovered the one-loop renormalization constants known from covariant theory. Due to the more complicated mathematical structure resulting from the breaking of explicit rotational symmetry, LCPT has never been the formulation of choice for high order loop calculations. More recently, however, calculations of dilute-dense scattering processes in the CGC picture have increasingly started advancing to the NLO level, for example for the small-$x$ evolution equations~\cite{Balitsky:2008zza,Balitsky:2013fea,Kovner:2013ona,Balitsky:2014mca,Beuf:2014uia,Lappi:2015fma,Iancu:2015vea,Iancu:2015joa,Lappi:2016fmu,Lublinsky:2016meo}, inclusive DIS cross sections~\cite{Beuf:2011xd,Balitsky:2012bs,Beuf:2016wdz}    and single~\cite{Altinoluk:2011qy,JalilianMarian:2011dt,Chirilli:2012jd,Stasto:2013cha,Kang:2014lha,Altinoluk:2014eka} and double~\cite{Ayala:2016lhd,Boussarie:2014lxa,Boussarie:2016ogo} inclusive particle production in the hybrid formalism.  Many of these calculations have been, or could be, performed very naturally in LCPT. In particular, the concept of ``light-cone wave function'' appears frequently in calculations that are factorized into the partonic structure of the probe, and its eikonal interaction with the target.

The primary purpose of this paper is to develop techniques for systematically performing loop calculations in light cone perturbation theory. As a first step in this program we will calculate to one loop order the quark-to-quark-gluon splitting light cone wave function, i.e. the probability amplitude for finding a quark and a gluon state component in the quantum state of an interacting quark. Although the expressions for the diagrammatic rules for LCPT calculations can be found in many references (see in particular \cite{Pauli:2000gw,Kovchegov:2012mbw}), we find that they can be given in a particularly simple form using an explicit spin/helicity basis for  both quarks and gluons. This is natural to combine with the four dimensional helicity (FDH) scheme for dimensional regularization, as will be discussed in more detail in Sec.~\ref{sec:rules}.

We start this paper by a brief exposition 
of the LCPT rules and the concept of the light cone wave function in Sec.~\ref{sec:lcpt}. We then calculate the one-loop contributions, both divergent and finite parts, in Sec.~\ref{sec:diags}, assembling the results in Sec.~\ref{sec:results}. We finally end with a discussion of applications and future extensions of this calculation in Sec.~\ref{sec:disc}.

\section{LC conventions and LCPT rules}\label{sec:lcpt}

\subsection{LC coordinates}

In light-cone coordinates a four-vector $x^{\mu}$ is given by the components
\begin{equation}
x^{\mu} =  (x^+,x^-,\mathbf{x}) \quad \text{with}\quad  \mathbf{x} = (x^1,x^2).
\end{equation}
The component $x^+$ is the light cone time along which the states are evolved,  $x^-$ is the longitudinal coordinate and $\mathbf{x}$  the transverse position. We denote spatial three-vectors as $\xvec = (x^+,\xt)$. In this work, we use the Kogut-Soper (KS) conventions \cite{Kogut:1969xa} in which $x^+$ and $x^-$ are related to the usual Minkowski coordinates by
\begin{equation}
 x^\pm= \frac{1}{\sqrt{2}}(x^0\pm x^3).
\end{equation}
The metric tensor is
\begin{equation}
g^{\mu\nu} = g_{\mu\nu} =   \left( \begin{array}{cccc}
0 &1&0&0 \\
1 &0&0&0 \\
0 &0&-1&0 \\
0 &0&0&-1 
  \end{array} 
\right),
\end{equation}
and thus the inner product of two four-vectors is
\begin{equation}
x\cdot y  = x^+y^- + x^-y^+ - \xt \cdot \yt,
\end{equation}
where $x^+ = x_{-}$ and $x^- = x_{+}$. The canonical conjugate of the longitudinal coordinate $x^-$ is the longitudinal momentum $p^+$, and the evolution in light cone time $x^+$ is generated by the light cone energy $p^-$. In LCPT all particles are on mass shell, with
\begin{equation}
p^- = \frac{\mathbf{p}^2 + m^2}{2p^+}.
\end{equation}

\subsection{Normalization of Fock states}

In the LC gauge the QCD Hamiltonian (see e.g. \cite{Kovchegov:2012mbw, Brodsky:1997de}) can be expressed entirely in terms of physical degrees of freedom, i.e. each interaction vertices correspond to a real dynamical process. To quantize the Hamiltonian, we expand the dynamical free quark field $\tilde{\Psi}$ and transverse gluon field $\tilde{A}^{\mu}_a$ at $x^+ =0$ in terms of the creation and annihilation operators 
\begin{equation}
\tilde{\Psi}^i(x^-,\mathbf{x}) = \int \dk \sum_{h=\pm}\biggl [b^i_h(\vec k)u_h(\vec k)e^{-ik\cdot x} + d^{i\dagger}_h(\vec k)v_h(\vec k)e^{+ik\cdot x} \biggr ]
\end{equation}
and 
\begin{equation}
\tilde{A}^{\mu}_a(x^-,\mathbf{x}) = \int \dk \sum_{\lambda=\pm}\biggl [a^a_\lambda(\vec k)\epsilon^{\mu}_\lambda(\vec k)e^{-ik\cdot x} + a^{a\dagger}_\lambda(\vec k)\epsilon^{\ast \mu}_\lambda(\vec k)e^{+ik\cdot x} \biggr ],
\end{equation}
where we denote the two fermion spin states $\pm 1/2$ by $h=\pm$ for notational simplicity.
These field operators satisfy the (anti)commutation relations 
\begin{equation}
\begin{split}
\biggl \{b^i_h(\kvec ),b^{j\dagger}_\sigma(\pvec ) \biggr \}  = \biggr \{d^i_h(\kvec ),d^{j\dagger}_\sigma(\pvec ) \biggl \} = 2k^+(2\pi)^3\delta^{(3)}(\kvec - \pvec )\delta_{h,\sigma}
\delta^{ij}
\end{split}
\end{equation}
and
\begin{equation}
\begin{split}
\biggl [a^a_{\lambda}(\kvec), a^{b\dagger}_{\lambda'}(\pvec)\biggr ]  = 2k^+(2\pi)^3\delta^{(3)}(\kvec - \pvec)\delta_{\lambda,\lambda'} \delta^{ab}.
\end{split}
\end{equation}
Here  $h$ and $\lambda$ are the quark and gluon helicities and $i,j,a,b$  SU(N) color indices. The momentum space integral measure is as in \cite{Kovchegov:2012mbw}
\begin{equation}
\label{eq:phasespace}
\int \dk
= \int \frac{\ud^4 k}{(2\pi)^4} (2\pi)\delta(k^2 -m^2) 
=  \int \frac{\ud k^+ \ud^2 \kt}{2 k^+ (2\pi)^3},
\end{equation}
where $k^+ > 0$. We always work in  LC gauge $\epsilon^+=0$ with transverse physical
polarizations $k\cdot \epsilon_{\lambda}(k) =0$, thus the gluon polarization vector simplifies to
\begin{equation}
\epsilon^{\mu}_{\lambda}(k) = (0, \epsilon^-_\lambda(k), \epst_\lambda)  \quad\text{with}\quad \epsilon^-_\lambda(k) = \frac{\kt \cdot \epst_\lambda}{k^+},
\end{equation}
and  (note the notation $\epsilon$ for a 4-dimensional and $\epst$ for a 2-dimensional vector)
\begin{equation}
 \epsilon_\lambda(k) \cdot  \epsilon_{\lambda'}(k')
= - \epst_\lambda \cdot \epst_{\lambda'}.
\end{equation}
The 2-dimensional physical gluon polarization vectors are
\begin{equation}
 \epst_{\lambda=\pm} = \frac{1}{\sqrt{2}}\left( \begin{array}{c}1\\ \pm i\end{array} \right).
\end{equation}
Note that these do not depend on the momentum of the gluon.
The polarization vectors satisfy 
\begin{equation}
\begin{split}
 \epst_{\lambda}^*\cdot \epst_{\lambda'} & =  \delta_{\lambda,\lambda'}\\
 \epst_{\lambda}^* & =  \epst_{-\lambda}\\
 \epst_{\lambda} \cdot \epst_{\lambda'} & = \epst_{\lambda}^*\cdot \epst_{\lambda'}^* 
=  \delta_{\lambda,-\lambda'}.
\end{split}
\end{equation}

\subsection{Elementary vertices}
\label{sec:elementvert}

Using the light-cone QCD Hamiltonian (e.g.~\cite{Kovchegov:2012mbw}) one can construct the necessary elementary quark and gluon vertices. In this paper  we only deal with massless quarks, and therefore helicity is conserved at the gluon emission vertex. The vertices can be very compactly expressed in the helicity  basis. This enables a very efficient computation of loop diagrams (compared to e.g. the calculations in Ref.~\cite{Beuf:2016wdz}), with the help of computer algebra tools. The well known downside of working with explicit polarizations is that all our particles have exactly two helicity states even in $4-2\varepsilon$ spatial dimensions, i.e. we are working in the FDH regularization scheme~\cite{Bern:1991aq,Bern:2002zk}. This would become problematic  at higher orders in perturbation theory.

\begin{figure*}[t]
\centerline{
\includegraphics[width=6.28cm]{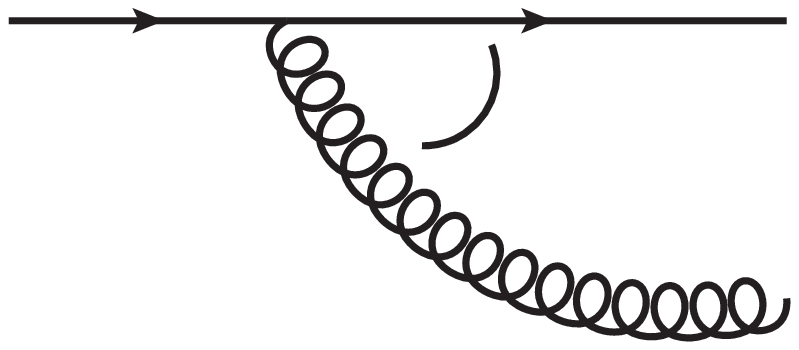}
\begin{tikzpicture}[overlay]
\node[anchor=south east] at (-5cm,2.6cm) {$\pvec,h,i$};
\node[anchor=south west] at (-2cm,2.6cm) {$\ppvec \equiv \pvec-\kvec,h,j$};
\node[anchor=south west] at (-2cm,0.7cm) {$\kvec,\lambda,a; \quad k^+ = z p^+$};
\node[anchor=north west] at (-2.5cm,2cm) {$\qt \equiv \kt - z\pt$};
\end{tikzpicture}
\rule{8em}{0pt}
\reflectbox{\includegraphics[width=6.28cm]{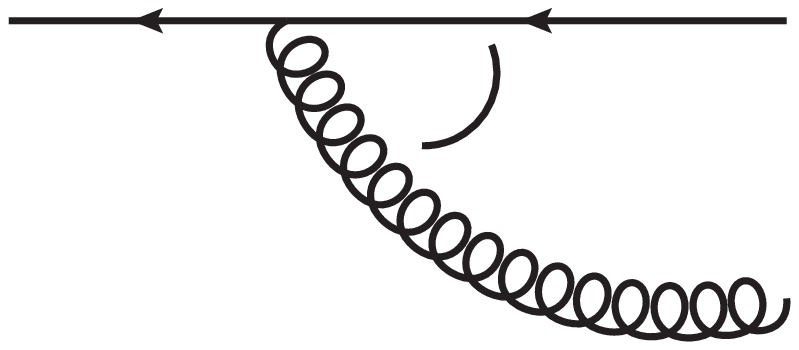}}
\begin{tikzpicture}[overlay]
\node[anchor=south west] at (-1.5cm,2.6cm) {$\pvec,h,i$};
\node[anchor=south west] at (-5cm,2.6cm) {$\ppvec \equiv \pvec-\kvec,h,j$};
\node[anchor=south west] at (-7cm,0.7cm) {$\kvec,\lambda,a; \quad k^+ = z p^+$};
\node[anchor=north east] at (-4.1cm,2cm) {$\qt \equiv \kt - z\pt$};
\end{tikzpicture}
}
\caption{Left: Gluon emission vertex from a quark,
$V_{\lambda,h}^{i;j,a}(\qt,z)$,
where $i,j$ are quark colors, $h$ the quark helicity, $a$ the gluon color and
$\lambda$ the gluon helicity,
\eq\nr{eq:vertexqtoqg}.
Right: Gluon absorption vertex into quark,
$V_{\lambda,h}^{j,a;i}(\qt,z)$ \eq\nr{eq:vertexqgtoq}.}
\label{fig:vertexqtoqg}
\end{figure*}

The simplest vertex is that for the emission of a gluon of momentum $\kvec$ and helicity $\lambda$ from a quark of momentum $\pvec$ and helicity $h$; it is most naturally expressed in terms of the longitudinal momentum fraction $z = k^+/p^+$ (note $0\leq z \leq 1$) and the center-of-mass transverse momentum $\qt = \kt-z\pt$. We denote this vertex, shown in \fig\ref{fig:vertexqtoqg} (left), as
\begin{equation}\label{eq:vertexqtoqg}
V_{\lambda,h}^{i;j,a}(\qt,z)
= 
-gt^{a}_{ji}
\biggl [ \bar{u}_h(p') \epsl^*_\lambda(k) u_h(p)\biggr ] = \frac{-2gt^{a}_{ji}}{z\sqrt{1-z}} 
\left( \delta_{\lambda,h} + (1-z) \delta_{\lambda,-h}\right) \qt \cdot \epst^*_\lambda.
\end{equation}
The quark absorption vertex,  \fig\ref{fig:vertexqtoqg} (right), is just the complex conjugate:
\begin{equation}\label{eq:vertexqgtoq}
V_{\lambda,h}^{j,a;i}(\qt,z)=
-gt_{ij}^a \biggl [ \bar{u}_h(p) \epsl_\lambda(k) u_h(p') \biggr ] = 
 \frac{-2gt_{ij}^a}{z\sqrt{1-z}} 
\left( \delta_{\lambda,h} + (1-z) \delta_{\lambda,-h}\right) \qt \cdot \epst_\lambda .
\end{equation}
Note that in \eqs\nr{eq:vertexqtoqg} and \nr{eq:vertexqgtoq} $\ppvec$ is the quark momentum with the smaller plus-component, 
final state in the emission and initial state in the absorption. 

\begin{figure*}[t]
\centerline{\includegraphics[width=6.28cm]{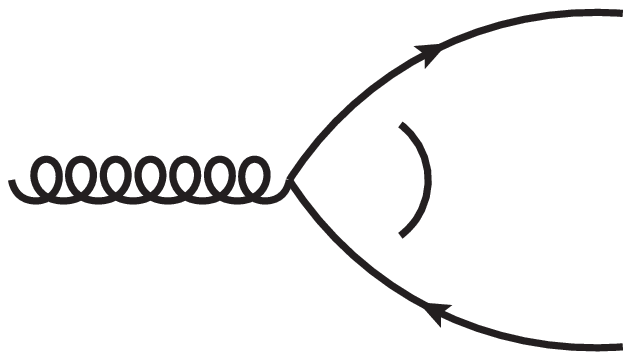}
\begin{tikzpicture}[overlay]
\node[anchor=south west] at (-5cm,2.1cm) {$\pvec,\lambda,a$};
\node[anchor=south west] at (-2.2cm,2.4cm) {$\kvec,h,i; \quad k^+ = z p^+$};
\node[anchor=south west] at (-2cm,0.7cm) {$\pvec-\kvec,-h,j$};
\node[anchor=north east] at (-0cm,2cm) {$\qt \equiv \kt - z\pt$};
\end{tikzpicture}
\rule{8em}{0pt}
\reflectbox{\includegraphics[width=6.28cm]{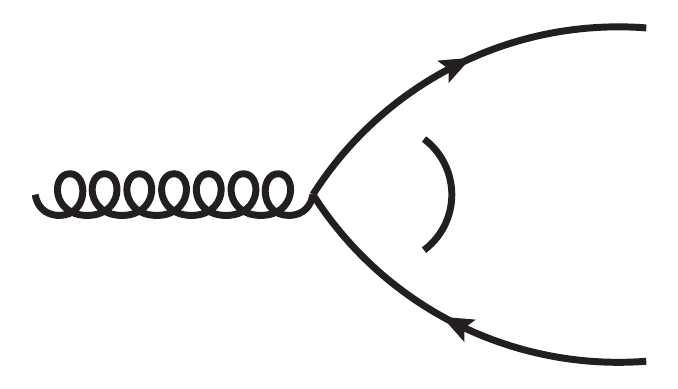}}
\begin{tikzpicture}[overlay]
\node[anchor=south west] at (-2cm,2.1cm) {$\pvec,\lambda,a$};
\node[anchor=south east] at (-4cm,0.7cm) {$\kvec,h,i; \quad k^+ = z p^+$};
\node[anchor=south east] at (-4cm,2.4cm) {$\pvec-\kvec,-h,j$};
\node[anchor=north east] at (-4.5cm,2cm) {$\qt \equiv \kt - z\pt$};
\end{tikzpicture}
}
\caption{Left: Gluon splitting vertex into quark-antiquark pair,
$A_{\lambda,h}^{a;j,i}(\qt,z)$ \eq\nr{eq:vertexgtoqqb}.
Right: Quark-antiquark annihilation vertex into gluon
$A_{\lambda,h}^{j,i;a}(\qt,z)$ \eq\nr{eq:vertexqqbtog}.
}\label{fig:vertexgtoqqb}
\end{figure*}

The vertex for a gluon (momentum $\pvec$) splitting into a quark (momentum $\kvec$ and helicity $h$) and antiquark ($\ppvec$, for massless particles the antiquark has helicity $-h$) is (see \fig\ref{fig:vertexgtoqqb} left):
\begin{equation}\label{eq:vertexgtoqqb}
A_{\lambda,h}^{a;j,i}(\qt,z)=
-gt^a_{ij}\biggl [
\bar{u}_h(k) \epsl_\lambda(p) v_{-h}(p') \biggr ]= 
\frac{-2gt^a_{ij}}{\sqrt{z(1-z)}} 
\left( z \delta_{\lambda,h} - (1-z) \delta_{\lambda,-h} \right) \qt \cdot \epst_\lambda.
\end{equation}
The natural momentum of the splitting (corresponding to that 
of the quark) is $\qt = \kt-z\pt$. 
The quark-antiquark annihilation vertex into a gluon (now with a minus sign for an incoming antiquark) is minus the complex conjugate of \eq\nr{eq:vertexgtoqqb},
\begin{equation}\label{eq:vertexqqbtog}
A_{\lambda,h}^{j,i;a}(\qt,z)=
-gt^a_{ji}
\biggl [-\bar{v}_{-h}(p') \epsl^*_\lambda(p) u_{h}(k)\biggr ] = 
\frac{2gt^a_{ji}}{\sqrt{z(1-z)}} 
\left( z \delta_{\lambda,h} -(1- z) \delta_{\lambda,-h}\right) \qt \cdot \epst^*_\lambda,
\end{equation}
see \fig\ref{fig:vertexgtoqqb} right.
\begin{figure*}[t]
\centerline{
\includegraphics[width=6.28cm]{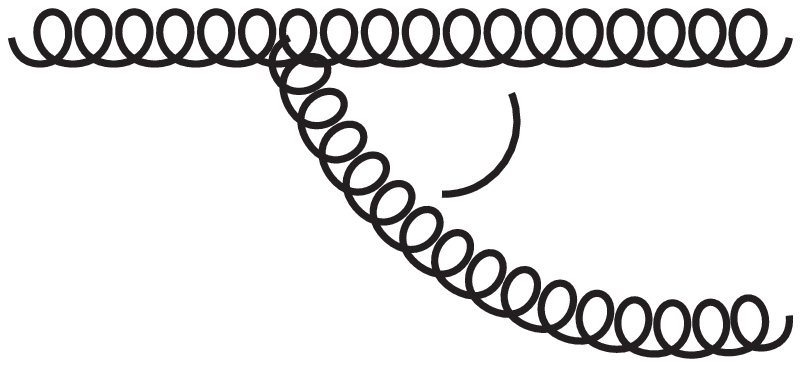}
\begin{tikzpicture}[overlay]
\node[anchor=south west] at (-6cm,2.7cm) {$\pvec,\lambda_1,a$};
\node[anchor=south west] at (-2cm,2.7cm) {$\ppvec \equiv\pvec-\kvec,\lambda_2,b$};
\node[anchor=south west] at (-2cm,0.7cm) {$\kvec,\lambda_3,c; \quad k^+ = z p^+$};
\node[anchor=north west] at (-2.3cm,2cm) {$\qt \equiv \kt - z\pt$};
\end{tikzpicture}
\rule{8em}{0pt}
\reflectbox{\includegraphics[width=6.28cm]{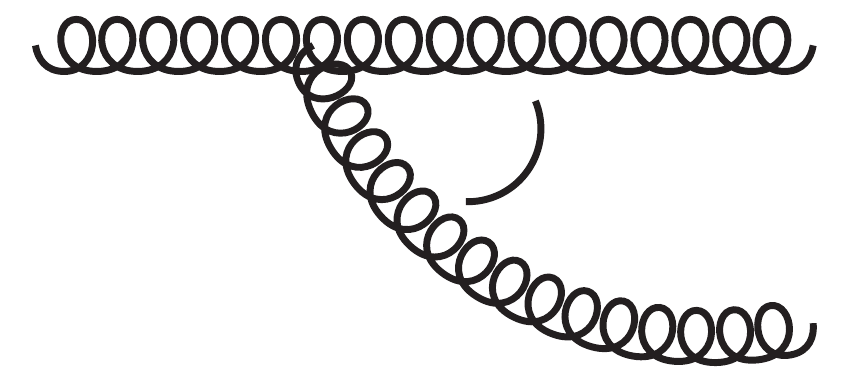}}
\begin{tikzpicture}[overlay]
\node[anchor=south west] at (-1.5cm,2.7cm) {$\pvec,\lambda_1,a$};
\node[anchor=south west] at (-6cm,2.7cm) {$\ppvec \equiv \pvec-\kvec,\lambda_2,b$};
\node[anchor=south west] at (-7cm,0.7cm) {$\kvec,\lambda_3,c; \quad k^+ = z p^+$};
\node[anchor=north east] at (-4.3cm,2cm) {$\qt \equiv \kt - z\pt$};
\end{tikzpicture}
}
\caption{Left: Gluon splitting vertex 
$\Gamma^{a; bc}_{\lambda_1; \lambda_2,\lambda_3}(\qt,z)$ \eq\nr{eq:vertexgtogg}, where $a, b, c$ are the gluon colors and $\lambda_1, \lambda_2, \lambda_3$  gluon helicities. 
Right: Gluon merging vertex 
$\Gamma^{ bc;a}_{ \lambda_2,\lambda_3;\lambda_1}(\qt,z)$ \eq\nr{eq:vertexggtog}.
}\label{fig:vertexgtogg}
\end{figure*}
The elementary vertex for $1\rightarrow 2$ gluon splitting (\fig\ref{fig:vertexgtogg} left) is given by\footnote{Changing signs from \cite{Kovchegov:2012mbw} so that  momenta flow from left to right.}
\begin{equation}
\Gamma^{a; b,c}_{\lambda_1; \lambda_2,\lambda_3}
(\qt,z)=
 ig f^{abc} \biggl [ (p+k)\cdot \epsilon_{\lambda_2}^*(p')
      \epsilon_{\lambda_1}(p)\cdot \epsilon_{\lambda_3}^*(k)
+ (-p'-p)\cdot \epsilon_{\lambda_3}^*(k) 
      \epsilon_{\lambda_1}(p)\cdot \epsilon_{\lambda_2}^*(p')
+ (-k+p')\cdot \epsilon_{\lambda_1}(p) 
      \epsilon_{\lambda_3}^*(k) \cdot \epsilon_{\lambda_2}^*(p')
\biggr ].
\end{equation}
In the LC gauge this can be simplified to
\begin{equation}\label{eq:vertexgtogg}
\Gamma^{a; b,c}_{\lambda_1; \lambda_2,\lambda_3}
(\qt,z)
=
 -2ig f^{abc} \biggl [  \frac{\qt \cdot \epst_{\lambda_2}^*}{1-z} 
      \delta_{\lambda_1,\lambda_3}
 + \frac{\qt \cdot \epst_{\lambda_3}^*}{z} 
      \delta_{\lambda_1,\lambda_2}
- \qt \cdot \epst_{\lambda_1} 
      \delta_{\lambda_3,-\lambda_2}
\biggr ].
\end{equation}
Similarly, the $2\rightarrow 1$ gluon merging vertex shown in \fig\ref{fig:vertexgtogg} (right) is the same, except incoming lines change into outgoing ones, 
which changes the overall sign and $\epst$ into $\epst^*$, i.e. changing \nr{eq:vertexgtogg} to its complex conjugate:
\begin{equation}\label{eq:vertexggtog}
\Gamma^{ b,c;a}_{ \lambda_2,\lambda_3;\lambda_1}
(\qt,z) =
 +2ig f^{abc} \biggl [  \frac{\qt \cdot \epst_{\lambda_2}}{1-z} 
      \delta_{\lambda_1,\lambda_3}
 + \frac{\qt \cdot \epst_{\lambda_3}}{z}
      \delta_{\lambda_1,\lambda_2}
- \qt \cdot \epst_{\lambda_1}^*
      \delta_{\lambda_3,-\lambda_2}
\biggr ].
\end{equation}
Incidentally, note that the three interference terms in the squared vertex
\nr{eq:vertexgtogg}$\times$\nr{eq:vertexggtog} give
$\sim 1/(z(1-z)) -1/z -1/(1-z) =0$, so that the three different polarization terms of the vertex  do not interfere in a gluon propagator correction diagram.

\subsection{Instantaneous vertices}
\label{sec:instvert}

\begin{figure}[t]
\centerline{\includegraphics[width=6.4cm]{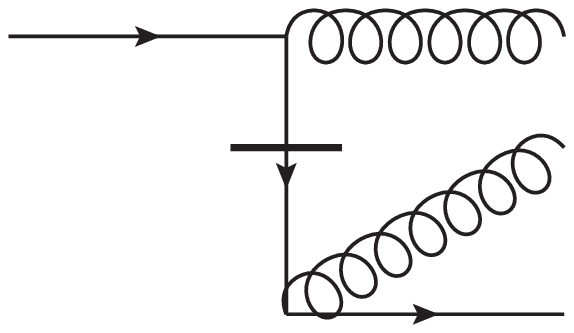}
\begin{tikzpicture}[overlay]
\node[anchor=south west] at (-5cm,3.2cm) {$\pvec,h,i$};
\node[anchor=south west] at (-1cm,0.2cm) {$\ppvec \equiv \pvec-\kvec-\kpvec,h,j\quad 
p'^+ = (1-z)p^+$};
\node[anchor=south west] at (-1cm,3.3cm) {$\kvec,\lambda,a; \quad k^+ = z'z p^+$};
\node[anchor=south west] at (-1cm,2cm) {$\kpvec,\lambda',b; \quad k'^+ = (1-z')z p^+$};
\end{tikzpicture}
}
\caption{Instantaneous quark interaction, \eq\nr{eq:qtoqgginstq}.
}\label{fig:qtoqgginstq}
\end{figure}

\begin{figure}[t]
\centerline{\includegraphics[width=4cm]{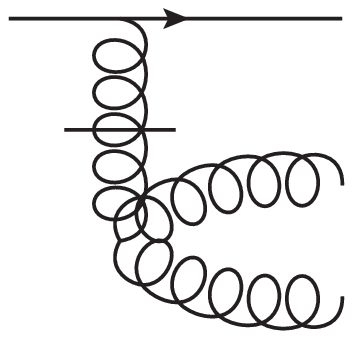}
\begin{tikzpicture}[overlay]
\node[anchor=south west] at (-4cm,3.3cm) {$\pvec,h,i$};
\node[anchor=south west] at (-1cm,3.3cm) {$\ppvec \equiv \pvec-\kvec-\kpvec,h,j\quad 
p'^+ = (1-z)p^+$};
\node[anchor=south west] at (-1cm,2cm) {$\kvec,\lambda,a; \quad k^+ = z'z p^+$};
\node[anchor=south west] at (-1cm,0.6cm) {$\kpvec,\lambda',b; \quad k'^+ = (1-z')z p^+$};
\end{tikzpicture}
}
\caption{Instantaneous gluon interaction, \eq\nr{eq:qtoqgginst}
}\label{fig:qtoqgginst}
\end{figure}

\begin{figure}[t]
\centerline{\includegraphics[width=4.2cm]{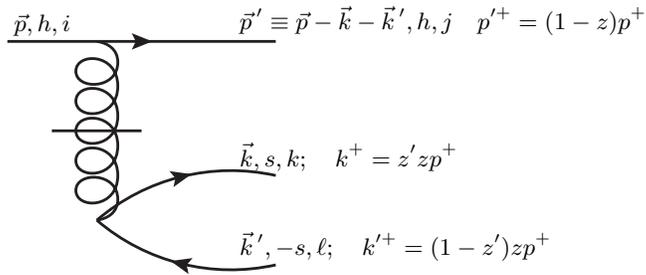}
\begin{tikzpicture}[overlay]
\node[anchor=south west] at (-4cm,3.3cm) {$\pvec,h,i$};
\node[anchor=south west] at (-1cm,3.3cm) {$\ppvec \equiv \pvec-\kvec-\kpvec,h,j\quad 
p'^+ = (1-z)p^+$};
\node[anchor=south west] at (-1cm,1.5cm) {$\kvec,s,k; \quad k^+ = z'z p^+$};
\node[anchor=south west] at (-1cm,0.3cm) {$\kpvec,-s,\ell; \quad k'^+ = (1-z')z p^+$};
\end{tikzpicture}
}
\caption{Instantaneous gluon interaction with quark-antiquark pair creation, \eq\nr{eq:qtoqqbinst}.
}\label{fig:qtoqqbarqinst}
\end{figure}

As we will discuss in more detail below, the instantaneous interaction terms do not contribute to the one-loop gluon emission wave function. They are, however, needed for the 3-particle final states discussed in Sec.~\ref{sec:3partfinal}. We will not present here the full set of instantaneous vertices (see~\cite{Pauli:2000gw}) but merely the ones needed for our calculation. There are three of these.

First, the instantaneous quark diagram \fig\ref{fig:qtoqgginstq} is given by the following matrix element
\begin{equation}
\label{eq:qtoqgginstqv1}
g^2t^{b}_{j l}t^{a}_{l i}\biggl [\frac{\bar{u}_h(p') \epsl^*_{\lambda' }(k')\gamma^{+}\epsl^*_\lambda (k)\ u_h(p)}{2(p^+-k^+)}\biggr ].
\end{equation}
The numerator in \eq\nr{eq:qtoqgginstqv1} simplifies to
\begin{equation}
 \epsl^*_{\lambda' }(k)\gamma^{+}\epsl^*_\lambda (q) = (\gammat \cdot \epst^{\ast}_{\lambda'})\gamma^{+}(\gamma \cdot \epst^{\ast}_{\lambda})
\end{equation}
and thus with the parametrization as shown in \fig\ref{fig:qtoqgginstq} , we get 
\begin{equation}
\label{eq:qtoqgginstq}
g^2t^{b}_{j l}t^{a}_{l i}\biggl [\bar{u}_h(p')  \frac{(\gamma \cdot \epst^{\ast}_{\lambda'})\gamma^{+}(\gamma \cdot \epst^{\ast}_{\lambda})}{2(p^+-k^+)} \ u_h(p)\biggr ] = 
2g^2t^{b}_{j l}t^{a}_{l i}\frac{\sqrt{1-z}}{1-zz'}\delta_{\lambda, -\lambda'}\delta_{\lambda, h}.
\end{equation}
Secondly, the instantaneous gluon diagram \fig\ref{fig:qtoqgginst} is given by the following matrix element 
\begin{equation}
\label{eq:qtoqgginst}
-g^2f^{abc}t^{c}_{ji} \frac{k'^+-k^+}{(k^+ + k'^+)^2}\biggl [\bar{u}_h(p') \gamma^{+} u_h(p)\biggr ] \epst^{\ast}_{\lambda} \cdot \epst^{\ast}_{\lambda'}
=
-2ig^2 f^{abc}t^{c}_{ji} \frac{\sqrt{1-z}}{z} (1-2z')\delta_{\lambda,-\lambda'}.
\end{equation}
Similarly, the instantaneous gluon with quark-antiquark pair creation diagram \fig\ref{fig:qtoqqbarqinst}
simplifies to the following matrix element
\begin{equation}
\label{eq:qtoqqbinst}
g^2\frac{t^{a}_{ji}t^{a}_{k\ell}}{(p^+-p'^+)^2}\biggl [\bar{u}_h(p') \gamma^{+} u_h(p)\biggr ]\biggl [\bar{u}_{s}(k) \gamma^{+} v_{-s}(k')\biggr ]
= 4 g^2 t^{a}_{ji}t^{a}_{k\ell} \frac{\sqrt{1-z}}{z}\sqrt{z'(1-z')}.
\end{equation}


\subsection{LCPT rules}
\label{sec:rules}

The diagrammatic LCPT rules for calculating initial-state light-cone wave functions are following: First, draw all topologically distinct $x^+$-ordered diagrams for a given physical process at the desired order in the coupling $g$. Second, calculate the perturbative contribution from each diagram according to the following rules:
\begin{enumerate}
\item Assign an on-shell four-momentum $p^{\mu}$ to each line such that the momentum is flowing from left to right ($x^+$-direction)
\item For each elementary vertex (or instantaneous diagram) include the relevant expression from sections \ref{sec:elementvert} and \ref{sec:instvert}, and a factor of $(2\pi)^3\delta^{(3)}(\vec p_{\rm final} - \vec p_{\rm initial})$, such that in the vertex the total $\vec p =(p^+,\mathbf{p})$ is conserved.
\item For each intermediate state include a LC energy denominator factor 
\begin{equation}
\label{eq:LCenergydenom}
\frac{1}{\sum_m p^-_{im} - \sum_n p^-_{fn} + i0_+}  \equiv \frac{1}{\Delta_{if}^- + i0_+}, 
\end{equation}
where the sum $\sum_m $ runs over all incoming particles present in the initial state $i$ and the sum $\sum_n$ over all the particles in the corresponding intermediate state $f$.  
\item For each internal line, sum over helicities and integrate using $\int \dk$ measure for quarks and gluons.
\item Include, if necessary, a standard symmetry factor $1/S$ which takes care of the possible permutations of fields, a factor $(-1)$ for quark loops and for quark lines beginning and ending at the initial state.
\end{enumerate}

Three different kinds of divergences can appear in perturbative calculations: ultraviolet, collinear and soft. In order to keep a maximally transparent physical interpretation of the nature of these divergences, we will regularize all three of them separately. Ultraviolet (UV) divergences appear from loop (or final state phase space) integrals over transverse momenta. These will be regulated by performing transverse momentum integrals in $2-2\varepsilon$ transverse dimensions, see Appendix~\ref{sec:integrals}. Since QCD is a renormalizable theory, all UV divergences can be removed by a renormalization of the parameters of the theory. For massless quarks there is only one such parameter: the coupling constant $g$.  All UV divergences must therefore disappear with coupling constant renormalization. Collinear divergences appear in the limit of small transverse momentum. Depending on the physical process, these will either cancel between real and virtual terms, or be absorbed into DGLAP-evolution of parton distributions or fragmentation functions (see e.g.~\cite{Chirilli:2011km}). To keep these separate from the UV divergences we will here regulate them by inserting a mass regulator $\lambda_g$ when needed. The third kind of divergence is the soft one, appearing in the limit of zero longitudinal momentum. In the physical context of high energy scattering these need to be absorbed into small-$x$ renormalization group evolution of scattering amplitudes, at one loop in terms of the BK equation~\cite{Balitsky:1995ub,Kovchegov:1999yj}. These will be regulated with a cutoff in longitudinal momentum, where all longitudinal momenta are assumed to be greater than a cutoff parameter $\alpha$ times $p^+$; the longitudinal momentum of the incoming particle. A technical disadvantage of LCPT compared to covariant perturbation theory with explicit rotational symmetry is that these divergences appear separately, and can mix in unhysical ways in intermediate stages of the calculation. The benefit gained from the associated extra work, however, is that the different physics (running coupling, DGLAP evolution and BK evolution) is explicit in the calculation.

Although transverse integrals are performed in $2-2\varepsilon$ dimensions, we will work in an explicit helicity basis, where quarks and gluons both have exactly 2 helicity states. Thus regularization scheme corresponds to the four dimensional helicity (FDH) scheme~\cite{Bern:1991aq,Bern:2002zk}. The elementary vertices depend on scalar products between 2-dimensional polarization vectors and $2-2\varepsilon$-dimensional internal momenta, whose consistent treatment requires care. The precise implementation of the FDH scheme requires that one first performs the momentum integrations, leaving a result that only involves scalar products between polarization vectors. These can then be evaluated in 2 dimensions together with the polarization sum. At higher loop orders than considered here, the FDH scheme would break the unitarity of the theory (see~\cite{Kilgore:2011ta,Boughezal:2011br}). However, the FDH scheme is much simpler for calculations, in particular for LCPT where one of the central features of the theory is that one is working with physical degrees of freedom, i.e. on-shell particles in explicit helicity eigenstates. For an example of a recent LCPT loop calculation in conventional dimensional regularization see~\cite{Beuf:2016wdz}.

\subsection{Fock state decomposition and light cone wave function}\label{sec:fock}

In this paper, our object of interest is not directly a scattering amplitude, but the light cone wave function. This concept is particularly useful in the context of scattering off an external classical potential in the high energy limit~\cite{Bjorken:1970ah}; this is precisely the situation in dilute-dense scattering processes in the CGC framework.

We want to express the full physical incoming particle state (a quark, in the case of this paper) as a simultaneous perturbative and Fock state decomposition in terms of the ``bare'' eigenstates of the noninteracting Hamiltonian. This expansion has the usual form of the ``old-fashioned'' textbook quantum mechanical perturbation theory, which we write as
\begin{equation}
|\Psi\rangle_{\textnormal{int}}
= |\Psi\rangle + 
\sum_{n_1}\frac{|n_1\rangle \langle n_1 | H |\Psi\rangle}{\Delta^-_{1 \Psi}}
+
\sum_{n_1,n_2}\frac{|n_2\rangle \langle n_2 | H |n_1\rangle \langle n_1 | H |\Psi\rangle}{\Delta^-_{2 \Psi}\Delta^-_{1 \Psi}}
+\dots 
\end{equation}
It is convenient to separate~\cite{Bjorken:1970ah} from this sum the terms where an intermediate state 
$|n_i\rangle$ is proportional to the state $|\Psi\rangle$ and absorb them into a (re)normalization of the LO term in the expansion
\begin{equation}\label{eq:expprime}
|\Psi\rangle_{\textnormal{int}}
= \sqrt{Z_\Psi}\left[ |\Psi\rangle + 
\left.\sum_{n_1}\right.' \frac{|n_1\rangle \langle n_1 | H |\Psi\rangle}{\Delta^-_{1 \Psi}}
+
\left.\sum_{n_1,n_2}\right.' \frac{|n_2\rangle \langle n_2 | H |n_1\rangle \langle n_1 | H |\Psi\rangle}{\Delta^-_{2 \Psi}\Delta^-_{1 \Psi}}
+\dots
\right]
\end{equation}
where $\sum'$ means that intermediate states proportional to $|\Psi\rangle$ are excluded from the sum.
The incoming state renormalization $Z_\Psi$ can be calculated either 
directly  by calculating the incoming particle propagator correction diagrams or from the normalization requirement
\begin{equation}
{}_{\textnormal{int}} \langle\Psi
|\Psi\rangle_{\textnormal{int}}
= 
\langle\Psi |\Psi\rangle,
\end{equation}
which leads to
\begin{equation}\label{eq:defZ}
Z_\Psi^{-1} = 
1 + \frac{1}{\langle\Psi |\Psi\rangle}
\left.\sum_{n}\right.' \frac{| \langle n | H |\Psi\rangle|^2}
{(\Delta^-_{n \Psi})^2}
+ \dots
\end{equation}
Here note that the one particle state is normalized to 
$2p^+(2\pi)^3\delta^3(0)$, whereas the matrix element
$\langle n | H |\Psi\rangle$ has a factor $(2\pi)^3\delta^3(0)$,
so there is effectively a factor $1/(2p^+)$ in the normalization 
compared to the calculation of the propagator correction diagrams when using this formula to calculate $Z_\Psi$. Note also that in LCPT the wave function renormalization constant can depend on the longitudinal momentum of the incoming particle~\cite{Thorn:1979gv,Mustaki:1990im,Perry:1992sw,Zhang:1993is,Zhang:1993dd,Harindranath:1993de}.

Specifically for the case of one incoming quark, the decomposition is
\begin{multline}\label{eq:psibare}
|q(\pvec)\rangle_{\textnormal{int}}
= \sqrt{Z_q(p^+)}
\bigg[ |q(\pvec)\rangle + 
\int \dpp \dq (2\pi)^3 \delta^3(\pvec-\ppvec-\qvec)
\psi^{q\to qg}(\pvec,\qvec) \left|q(\ppvec) g(\qvec) \right\rangle
\\
+
\frac{1}{\sqrt{2}}
\int \dpp \dq \dk (2\pi)^3 \delta^3(\pvec-\ppvec-\qvec-\kvec)
\left[
\psi^{q\to qgg}(\pvec,\qvec) 
\left|q(\ppvec) g(\qvec) g(\kvec)\right\rangle
+
\psi^{q\to qq\bar{q}}(\pvec,\qvec) 
\left|q(\ppvec) q(\qvec) \bar{q}(\kvec)\right\rangle
\right]
+\dots
\bigg].
\end{multline}
The expression \nr{eq:psibare} \emph{defines} the light cone wave functions $\psi^{q\to qg}$, 
$\psi^{q\to qgg},$ $\psi^{q\to qq\bar{q}}$ etc, including the symmetry factor $1/\sqrt{2}$ for Fock states containing  two identical particles. At leading order, their power counting in the QCD coupling is $\psi^{q\to qg} \sim g$; $\psi^{q\to qgg} \sim \psi^{q\to qq\bar{q}}  \sim g^2$. We will in this paper compute the one-loop (i.e. $\sim g^3$) contribution to one-gluon emission wave function  $\psi^{q\to qg}$. For explicitness and reference, we will also write down the tree level $\sim g^2$ light cone wave functions for the wave functions with three particle final states $\psi^{q\to qgg},\ \psi^{q\to qq\bar{q}}$.

The expansion \nr{eq:psibare} is the one that one would actually use in the calculation of scattering off an external potential. The wave function $\psi^{q\to qg}$ is, however, not straightforward to compare to a one-loop $q\to qg$ vertex in covariant perturbation theory, e.g. to see that all the UV divergences can be absorbed into a renormalization of the coupling constant. In particular, $\psi^{q\to qg}$ does not include incoming propagator correction diagrams (they are absorbed into the nornalization $\sqrt{Z_q(p^+)}$, but does include the final state propagator correction diagrams (diagrams \ref{diag:qwavefeafter}, \ref{diag:gtoggloop} and \ref{diag:gtoqqbloop} in \figs\ref{fig:qwavefeafter}, \ref{fig:gtoggloop} and \ref{fig:gtoqqbloop}). To make contact with covariant perturbation theory, one should define  renormalized free particle states $|q(\pvec)\rangle_{R} = \sqrt{Z_q(p^+)}|q(\pvec)\rangle$, 
 $|g(\kvec)\rangle_{R} = \sqrt{Z_g(k^+)}|g(\kvec)\rangle$. In terms of these we have
\begin{equation}
\sqrt{Z_q(p^+)}\psi^{q\to qg}(\pvec,\qvec) \left|q(\ppvec) g(\qvec) \right\rangle
=
\frac{\sqrt{Z_q(p^+)}}{\sqrt{Z_q(p'^+)Z_g(q^+)}} 
 \psi^{q\to qg}(\pvec,\qvec)
\left|q(\ppvec) g(\qvec) \right\rangle_{R},
\end{equation}
where the denominator $\sqrt{Z_q(p'^+)Z_g(q^+)}$ cancels half of the outgoing particle propagator correction diagrams included in $\psi^{q\to qg}$, and the numerator $\sqrt{Z_q(p^+)}$  introduces half of the incoming particle renormalizations that are absent from $\psi^{q\to qg}$.  The combination
\begin{equation}
\frac{\sqrt{Z_q(p^+)}}{\sqrt{Z_q(p'^+)Z_g(q^+)}} 
 \psi^{q\to qg}(\pvec,\qvec)
\end{equation}
corresponds to the covariant theory one-particle irreducible vertex multiplied by the wave function  renormalization constants. This is  the quantity appearing in a cross section calculation in covariant theory and the whose UV-divergences can be absorbed into a renormalization of the coupling constant. This will be explicitly 
checked in Sec.~\ref{sec:results}.

\section{Diagram calculations}\label{sec:diags}
\subsection{Leading order gluon emission}

\begin{figure}[t]
\centerline{\includegraphics[width=0.35\textwidth]{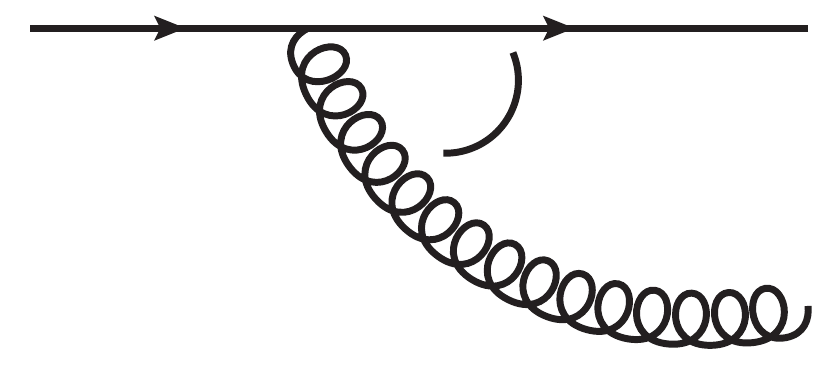}
\begin{tikzpicture}[overlay]
\node[anchor=south east] at (-5cm,2.6cm) {$\pvec,h,i$};
\node[anchor=south west] at (-2cm,2.6cm) {$\ppvec \equiv \pvec-\qvec,h,j$};
\node[anchor=south west] at (-2cm,0.7cm) {$\qvec,\lambda,a; \quad q^+ = z p^+$};
\node[anchor=north west] at (-2.5cm,2cm) {$\nt \equiv \qt - z\pt$};
\draw [dashed] (-4.5,3) -- (-4.5,0.0);
\node[anchor=north] at (-4.5cm,0cm) {0};
\draw [dashed] (-2,3) -- (-2,0.0);
\node[anchor=north] at (-2cm,0cm) {1};
\end{tikzpicture}
}
\caption{Gluon emission at leading order. Momentum  is conserved $\pvec = \ppvec + \qvec$, and the momentum fraction and natural transverse momentum for the emitted gluon are  $z = q^+/p^+$ and $\nt = \qt - z\pt$
.}\label{fig:lovertexqtoqg}
\end{figure}
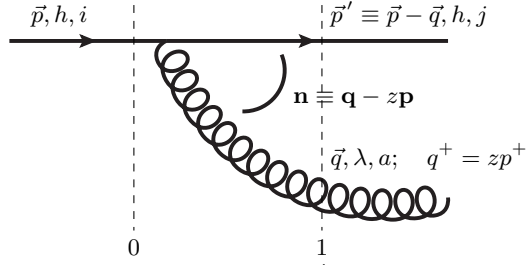

Using the LCPT rules, the leading order (LO) gluon emission wave functionshown in \fig\ref{fig:lovertexqtoqg} (without overall momentum conservation) can be written in the form 
\begin{equation}
\psi^{q\to qg}_{\lo}(\nt,z) = 
\frac{V_{\lambda,h}^{i;j,a}(\nt,z)}{
\Delta^-_{01}
},
\end{equation}
where the vertex function $V_{\lambda,h}^{i;j,a}(\nt,z)$ is defined in \eq\nr{eq:vertexqtoqg}
and the LC energy denominator $\Delta^-_{01} = p^- - p'^--k^-$ simplifies to
\begin{equation}
 \Delta^-_{01} = \frac{-1}{2p^+} \frac{\nt^2}{z(1-z)}.
\end{equation}
Thus the LO gluon emission wave function is given by 
\begin{equation}\label{eq:psilo}
\psi^{q\to qg}_{\lo}(\nt,z) = 
4p^+ gt^a_{ji}\sqrt{1-z}\biggl [\delta_{\lambda,h} + (1-z)\delta_{\lambda,-h}\biggr ]
 \frac{\nt \cdot \epst^{\ast}_{\lambda}}{\nt^2}.
\end{equation}

\subsection{Wave function renormalization and propagator corrections}

\subsubsection{Quark wave function renormalization $Z_q(p^+)$
to order $g^2$}
\label{sec:Zq}

As discussed in Sec.~\ref{sec:fock}, the propagator corrections for initial state particles are not considered a part of the wave function. We will therefore calculate the quark wave function renormalization coefficient using the normalization condition for the one quark state, \eq\nr{eq:defZ}. At the order $g^2$ for $Z_q(p^+)$ we only need to consider the $qg$ state, for which the coefficient has just been calculated in \eq\nr{eq:psilo}. We get
\begin{equation}
\label{eq:renormconst}
 Z_q^{-1}(p^+) = 1 + \frac{1}{\langle q(\pvec)| q(\pvec) \rangle}
\int \dq  \dk 
\frac{\left|\langle q(\qvec) g(\kvec)| H | q(\pvec) \rangle\right|^2}
{(\Delta^-)^2},
\end{equation}
where $\Delta^-= p^--k^--p'^-$ and 
\begin{equation}
\langle q(\pvec)| q(\pvec) \rangle = 2 p^+ (2\pi)^3 \delta^{(3)}(\ovec)
\end{equation}
and 
\begin{equation}
\frac{\langle q(\qvec) g(\kvec)| H | q(\pvec) \rangle}{\Delta^-}
=  (2\pi)^3 \delta^{(3)}(\pvec-\qvec-\kvec)
\psi^{q\to qg}_{\lo}(\kt-z\pt,z=k^+/p^+).
\end{equation}
Thus \eq\nr{eq:renormconst} simplifies to
\begin{equation}
\label{eq: renormconstv2}
 Z_q^{-1}(p^+) = 1 + \frac{1}{2p^+}\int \dq  \dk (2\pi)^3 \delta^{(3)}(\pvec-\qvec-\kvec) \left| \psi^{q\to qg}_{\lo}(\kt-z\pt,z=k^+/p^+)\right|^2,
\end{equation}
where the phase-space measure is 
\begin{equation}
\int \dq  \dk (2\pi)^3 \delta^{(3)}(\pvec-\qvec-\kvec)  = \int_{0}^{p^+} \frac{\ud k^+}{2k^+}\int \frac{\ud^{2} \kt}{(2\pi)^3}\frac{1}{2q^+} = \frac{\mu^{2-d_{\perp}}}{8\pi p^+}\int_{0}^{1} \frac{\ud z}{z(1-z)}\int \frac{\ud^{d_{\perp}}\kt}{(2\pi)^{d_{\perp}}}.
\end{equation}
Regulating the collinear IR-divergence with a mass parameter $\lambda_m > 0$, which is chosen in such a way that the particle carrying $zp^+$ amount of longitudinal momentum is regulated (in this case the gluon with momentum $\kvec$), we get 
\begin{equation}
\Delta^- = \frac{-1}{2p^+z(1-z)}\left[(\kt-z\pt)^2  + (1-z)\lambda_m^2\right],
\end{equation}
and substituting \eq\nr{eq:psilo} into \eq\nr{eq: renormconstv2}  leads to
\begin{eqnarray}
\label{eq:renormconstv3}
 Z_q^{-1}(p^+) &=& 1
+
\frac{g^2}{\pi}t^a_{ji}t^a_{ij} 
\int_{0}^{1} \frac{\ud z}{z}\left[\delta_{\lambda,h} + (1-z)^2\delta_{\lambda,-h}\right]
\mu^{2-d_\perp}\int \frac{\ud^{d_\perp} \kt }{ (2\pi)^{d_\perp}}
\frac{
(\kt-z\pt)\cdot\epst^*_\lambda (\kt-z\pt)\cdot\epst_\lambda
}{[(\kt-z\pt)^2  
+ (1-z)\lambda_m^2]^2}
\\ &=&
1 + \frac{g^2\cf }{\pi}\int_{0}^{1} \frac{\ud z}{z}\left[1 + (1-z)^2\right]\mu^{2-d_\perp}
\int \frac{\ud^{d_\perp} \mt }{ (2\pi)^{d_\perp}} 
\frac{\mt^2/d_\perp}{[\mt^2 + (1-z)\lambda_m^2]^2},
\end{eqnarray}
where we have changed the integration  variable to $\mt = \kt - z\pt$ with $m^{i}m^{j}$ equivalent to  $(\mt^2/d_{\perp})\delta^{ij}$ under the integral,  and summed over the internal gluon helicity $\lambda$ and the colors $j$ and $a$ (but not $i$ since we are calculating the norm of the state with a particle in the fixed color state $i$). Here we should note that while $\varepsilon > 0$ regularizes the UV transverse momentum divergence; the longitudinal $z$ (soft) divergence is an IR one, and would require $\varepsilon < 0$ to regularize it. However, we want to regularize these soft IR divergences by an explicit cutoff instead. This means that all longitudinal momenta should be  larger than $ \alpha p^+$ with $\alpha > 0$. Regulating the soft IR-divergence in $z \rightarrow 0$ by a cutoff with $\alpha < z < 1$ and applying \eq\nr{eq:oltransint4} with $d_{\perp} = 2 - 2\varepsilon$, the quark wave function renormalization constant in \eq\nr{eq:renormconstv3} becomes
\begin{equation}\label{eq:zqv0}
Z^{-1}_q(p^+) = 1 + \frac{g^2\cf }{8\pi^2}\biggl[
\left( \frac{1}{\epsmsbar}
 + \log \frac{\mu^2}{\lambda_m^2} \right)
\left( - \frac{3}{2} - 2\log(\alpha)\right)
  - \frac{5}{4} + \frac{\pi^2}{3}\biggr ]
\end{equation}
i.e.
\begin{equation}\label{eq:zqv1}
Z_q(p^+) = 1 + \frac{ g^2\cf}{8\pi^2}\biggl[
\left( \frac{1}{\epsmsbar}
 + \log \frac{\mu^2}{\lambda_m^2} \right)
\left(  \frac{3}{2} + 2\log(\alpha)\right)
  + \frac{5}{4} - \frac{\pi^2}{3}\biggr ].
\end{equation}
Note that if also the collinear IR divergence was also regulated in the FDH scheme scheme with $\varepsilon < 0$ in stead of the mass $\lambda_m$, we would have $Z_q=1$.


\subsubsection{Final state quark wave function correction}

\begin{figure}[t]
\centerline{
\includegraphics[width=6.4cm]{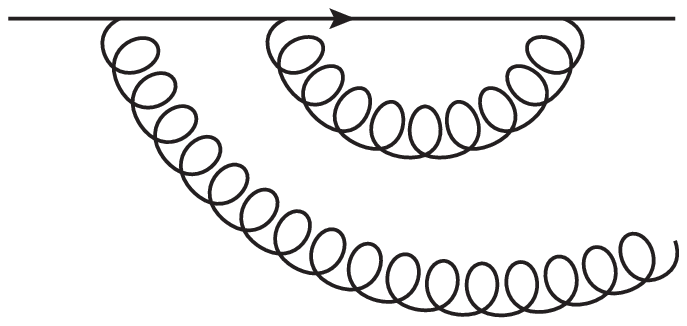}
\begin{tikzpicture}[overlay]
\draw [dashed] (-5.8,3.5) -- (-5.8,0);
\node[anchor=north] at (-5.8cm,-0.2cm) {0};
\draw [dashed] (-4.5,3.5) -- (-4.5,0);
\node[anchor=north] at (-4.5cm,-0.2cm) {1};
\draw [dashed] (-2.5,3.5) -- (-2.5,0);
\node[anchor=north] at (-2.5cm,-0.2cm) {2};
\draw [dashed] (-0.8,3.5) -- (-0.8,0);
\node[anchor=north] at (-0.8cm,-0.2cm) {3};
\node[anchor=east] at (-5.8,3) {$\pvec,h,i$};
\node[anchor=west] at (-0.8,3) {$\ppvec,h,j$};
\node[anchor=center] at (-4.3,3) {$\pppvec,k$};
\node[anchor=center] at (-2,3) {$\kpvec,l$};
\node[anchor=west] at (-0.3,0.8) {$\qvec,\lambda,a$};
\node[anchor=west] at (-2.3,1.4) {$\kvec,\lambda',b$};
\node[anchor=west] at (-3.2,2.5) {$\mt$};
\node[anchor=west] at (-.5,1.8) {$\nt\equiv \qt - z\pt$};
\node[anchor=south west] at (-7cm,0cm) {\namediag{diag:qwavefeafter}};
\end{tikzpicture}
}
\rule{0pt}{1ex}
\caption{Contribution to gluon emission from final state quark propagator correction, diagram \ref{diag:qwavefeafter}, with energy denominator intermediate states and kinematics. Momentum conservation: $\pvec = \ppvec + \qvec$ and 
$\ppvec = \kpvec+\kvec$. The momentum fraction and  natural momentum scale for the emitted gluon are $z = q^+/p^+$ and $\nt = \qt-z\pt$. The momentum fraction for the gluon in the loop is $z' = k^+/p'^+$, i.e. $k^+ = z'(1-z)p^+$  and the natural momentum $\mt = \kt - z'\ptp$.}
\label{fig:qwavefeafter}
 \end{figure}

We now look at quark wave function in a different way, by an explicit calculation of the final state quark propagator correction diagram. The renormalizability of the theory requires that the UV-divergent part of the final state quark propagator correction matches the wave function renormalization~\nr{eq:zqv1}, but as we will see this is not the case for the UV-finite parts.

The contribution to final state quark wave function correction in \fig\ref{fig:qwavefeafter} is given by 

\begin{equation}
\psi^{q\to qg}_{\ref{diag:qwavefeafter}} = \int \dppp \dkp \dk 
(2\pi)^3 \delta^3(\pppvec-\kvec-\kpvec)  
(2\pi)^3 \delta^3(\ppvec -\kvec-\kpvec)  \frac{
V^{i;k,a}_{\lambda,h}(\nt,z)
V^{k;l,b}_{h,\lambda'}(\mt,z')
V^{l,b;j}_{h,\lambda'}(\mt,z')
}{\Delta^-_{01} \Delta^-_{02} \Delta^-_{03}},
\end{equation}
where the LC energy denominators are
\begin{eqnarray}
 \Delta^-_{01} &=& \Delta^-_{03} = \frac{-1}{2p^+} \frac{\nt^ 2}{z(1-z)}
\\
 \Delta^-_{02} &=&\frac{-1}{2p^+}
\left[ \frac{\nt^2}{z(1-z)} + \frac{1}{1-z}\frac{\mt^2}{z'(1-z')}
\right],
\end{eqnarray}
and the phase space simplifies to
\begin{equation}
\begin{split}
\int \dppp \dkp \dk 
(2\pi)^3 \delta^3(\pppvec-\kvec-\kpvec)  
(2\pi)^3 \delta^3(\ppvec -\kvec-\kpvec) & =
\int_{0}^{p^+} \frac{\mathrm{d}k^+}{2k^+}\int \frac{\mathrm{d}^2\kt}{(2\pi)^3} \frac{1}{2p''^+2k'^+}\\
& =\frac{\mu^{2-d_{\perp}}}{16 \pi (p^+)^2}\int_{0}^{1} \frac{\ud z'}{z'(1-z')(1-z)^2}\int\frac{\ud^{d_\perp}\mt}{(2\pi)^{d_\perp}}.
\end{split}
\end{equation}
The product of vertices at the end of the loop (summing over internal polarization $\lambda' $) give
\begin{equation}
\begin{split}
V^{k;l,b}_{h,\lambda'}(\mt,z')V^{l,b;j}_{h,\lambda'}(\mt,z') & = \frac{4g^2t^{b}_{j\ell}t^{b}_{\ell k}}{(z')^2(1-z')}\biggl [\delta_{h,\lambda'} + (1-z')\delta_{h,-\lambda'}\ \biggr ]\biggl [\delta_{h,\lambda'} + (1-z')\delta_{h,-\lambda'}\ \biggr ]\mt \cdot \epst^*_{\lambda'} \mt \cdot \epst_{\lambda'}\\
& = \frac{4g^2\cf\delta_{jk}}{(z')^2(1-z')}\frac{\mt^2}{d_{\perp}}\biggl [1 + (1-z')^2 \biggr ].
\end{split}
\end{equation}
Recalling that our FDH scheme requires us to first compute the momentum integral, we have here in a slight abuse of notation anticipated this by replacing $m^im^j$ with $\mt^2 \delta^{ij}/d_\perp$, which is true under the integration over $\mt$. This leaves the polarization vectors in the structure $\epst^*_{\lambda'}  \cdot \epst_{\lambda'} =1$ (no sum over $\lambda'$). This has already been used on the second line, although strictly speaking the polarization sum is performed only after the $\mt$-integral. 
Putting things together and performing the $\mt$-integral we have 
\begin{equation}
\psi^{q\to qg}_{\ref{diag:qwavefeafter}}=
\psi^{q\to qg}_\lo(\nt,z) \left (\frac{g^2 \cf}{8\pi^2}\right )
\left(-\frac{\Gamma(\varepsilon)}{(1-\varepsilon)(4 \pi)^{-\varepsilon}}\right)
\biggl [z\frac{ \mu^2}{\nt^2}\biggr ]^\varepsilon
\int_0^1\frac{\ud z'}{(z'(1-z'))^\varepsilon}\frac{1}{z'}
\biggl [1 + (1-z')^2\biggr ].
\end{equation}
Regulating the IR-divergence $z'\to 0$ by a cutoff $z'>\alpha/(1-z)$ we get the contribution
\begin{equation}\label{eq:qwavefafterfinalv1}
\begin{split}
\psi^{q\to qg}_{\ref{diag:qwavefeafter}} =  \psi^{q\to qg}_{\lo}(\nt,z)\left (\frac{g^2\cf}{8\pi^2} \right ) \biggl \{\biggl [ \frac{1}{\varepsilon_{\overline{\rm MS}}}  + \log\left (\frac{z\mu^2}{\nt^2} \right )\biggr ]&\left (\frac{3}{2}-2\log\left (\frac{1-z}{\alpha}\right ) \right )\\
& + \frac{27}{6} - \frac{\pi^2}{3} -2\log\left ( \frac{1-z}{\alpha}\right ) - \log^2\left ( \frac{1-z}{\alpha}\right ) \biggr \}.
\end{split}
\end{equation}
Note that the UV-divergent parts are the same as for the quark wave function renormalization constant \nr{eq:zqv1}, but the finite parts are different.

\subsubsection{Gluon propagator corrections}

\begin{figure}[t]
\centerline{
\includegraphics[width=6.4cm]{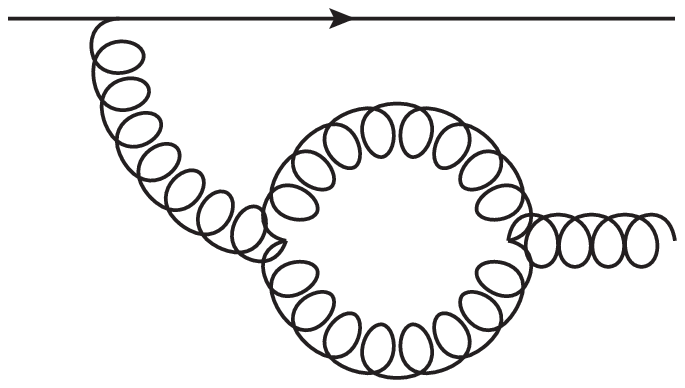}
\begin{tikzpicture}[overlay]
\draw [dashed] (-5.8,3.5) -- (-5.8,0);
\node[anchor=north] at (-5.8cm,-0.2cm) {0};
\draw [dashed] (-4.5,3.5) -- (-4.5,0);
\node[anchor=north] at (-4.5cm,-0.2cm) {1};
\draw [dashed] (-2.8,3.5) -- (-2.8,0);
\node[anchor=north] at (-2.8cm,-0.2cm) {2};
\draw [dashed] (-1,3.5) -- (-1,0);
\node[anchor=north] at (-1cm,-0.2cm) {3};
\node[anchor=east] at (-6,3.5) {$\pvec,h,i$};
\node[anchor=west] at (-1,3.6) {$\ppvec=\pvec-\qvec,h,j$};
\node[anchor=west] at (-0.2,1.4) {$\qvec,\lambda,a$};
\node[anchor=west] at (-3.8,1.4) {$\mt \equiv \kt-z'\qt$};
\node[anchor=west] at (-0.2,2) {$\nt\equiv \qt-z\pt$};
\node[anchor=south west] at (-4,2.5) {$\kvec,\sigma,c \quad \quad k^+ = z' q^+$};
\node[anchor=north west] at (-2.4,0.3) {$\kpvec,\sigma',d$};
\node[anchor=east] at (-4.6,1.2) {$\qpvec,\lambda',b$};
\node[anchor=south west] at (-7cm,0cm) {\namediag{diag:gtoggloop}};
 \end{tikzpicture}
}
\rule{0pt}{1ex}
\caption{Gluon polarization diagram \ref{diag:gtoggloop} with intermediate states for the energy denominators and kinematics. Momentum  conservation dictates: $\pvec = \ppvec + \qvec$ and $\qpvec = \kvec + \kpvec$. The momentum fraction and natural momentum scale for the emitted gluon are $z =q^+/p^+$ and $\nt = \qt - z\pt$. The momentum fraction for the gluon in the loop is $z' = k^+/q^+$, i.e. $k^+ = zz'p^+$, and the natural momentum $\mt = \kt - z'\qt$.}
\label{fig:gtoggloop}
 \end{figure}

\begin{figure}
\centerline{
\includegraphics[width=6.4cm]{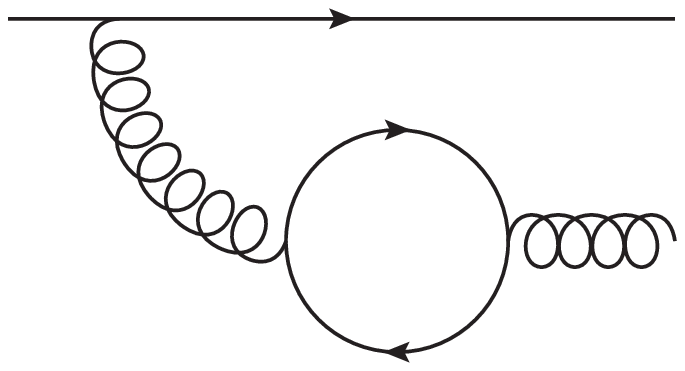}
\begin{tikzpicture}[overlay]
\draw [dashed] (-5.8,3.5) -- (-5.8,0);
\node[anchor=north] at (-5.8cm,-0.2cm) {0};
\draw [dashed] (-4.5,3.5) -- (-4.5,0);
\node[anchor=north] at (-4.5cm,-0.2cm) {1};
\draw [dashed] (-2.8,3.5) -- (-2.8,0);
\node[anchor=north] at (-2.8cm,-0.2cm) {2};
\draw [dashed] (-1,3.5) -- (-1,0);
\node[anchor=north] at (-1cm,-0.2cm) {3};
\node[anchor=east] at (-6,3.4) {$\pvec,h,i$};
\node[anchor=west] at (-1,3.45) {$\ppvec,h,j$};
\node[anchor=west] at (-0.2,1.4) {$\qvec,\lambda,a$};
\node[anchor=west] at (-3.8,1.4) {$\mt \equiv \kt-z'\qt$};
\node[anchor=west] at (-0.5,2) {$\nt\equiv \qt-z\pt$};
\node[anchor=south west] at (-4,2.3) {$\kvec,\sigma,m$};
\node[anchor=north west] at (-2.4,0.3) {$\kpvec,-\sigma,n$};
\node[anchor=east] at (-4.6,1.2) {$\qpvec,\lambda',b$};
\node[anchor=south west] at (-7cm,0cm) {\namediag{diag:gtoqqbloop}};
 \end{tikzpicture}
}
\rule{0pt}{1ex}
\caption{Gluon polarization diagram \ref{diag:gtoqqbloop} with the same energy denominators and kinematics as in \fig\ref{fig:gtoggloop}.}
\label{fig:gtoqqbloop}
 \end{figure}

Next we calculate the contributions to the gluon emission wave function  $\psi^{q\to qg}$ from the final state  gluon propagator correction diagrams, the gluon loop \ref{diag:gtoggloop} shown in \fig\ref{fig:gtoggloop} and the quark loop \ref{diag:gtoqqbloop} in \fig\ref{fig:gtoqqbloop}.

\subsubsection*{Gluon polarization diagram \ref{diag:gtoggloop}}

The contribution from diagram \ref{diag:gtoggloop}, including the symmetry factor $1/2$ for two interchangeable gluon lines, is
\begin{equation}
\label{eq:wfgtoggloop}
 \psi^{q\to qg}_{\ref{diag:gtoggloop}} = \half \int \dk \dqp \dkp 
(2\pi)^3 \delta^3(\qpvec-\kpvec-\kvec)  
(2\pi)^3 \delta^3(\qvec -\kpvec-\kvec) \frac{V^{i;j,b}_{\lambda',h}(\nt,z)}{\Delta_{01}^{-}\Delta_{02}^{-}\Delta_{03}^{-}}\Gamma^{cd;a}_{\lambda';\sigma\sigma'}(\mt,z')\Gamma^{b;cd}_{\sigma\sigma';\lambda}(\mt,z')
\end{equation}
with the vertices given in \eqs\nr{eq:vertexqtoqg}, \nr{eq:vertexgtogg} and~\nr{eq:vertexggtog}. The phase space measure simplifies to
\begin{equation}
\label{eq:PSgtoggloop}
\begin{split}
\int \dk \dqp \dkp 
(2\pi)^3 \delta^3(\qpvec-\kpvec-\kvec)  
(2\pi)^3 \delta^3(\qvec -\kpvec-\kvec)  & = \int_{0}^{q^+} \frac{\mathrm{d}k^+}{2k^+}\int \frac{\mathrm{d}^2\kt}{(2\pi)^3} \frac{1}{2q'^+2k'^+}\\
& = \frac{\mu^{2-d_{\perp}}}{16\pi(p^+)^2}\int_{0}^{1} \frac{\mathrm{d}z'}{z^2z'(1-z')}\int \frac{\mathrm{d}^{d_{\perp}}\mt}{(2\pi)^{d_{\perp}}},
\end{split}
\end{equation}
and the LC energy denominators are given by
\begin{eqnarray}
\label{eq:LCgtoggloop}
 \Delta^-_{01} &=& \Delta^-_{03} = \frac{-1}{2p^+} \frac{\nt^ 2}{z(1-z)}
\\
 \Delta^-_{02} &=&\frac{-1}{2p^+}
\left[ \frac{\nt^2}{z(1-z)} + \frac{1}{z}\frac{\mt^2}{z'(1-z')}
\right].
\end{eqnarray}
The product of two gluon vertices (summed over internal polarizations $\sigma$ and $\sigma'$) in \eq\nr{eq:wfgtoggloop} simplifies to
\begin{equation}
\begin{split}
\Gamma^{cd;a}_{\sigma\sigma';\lambda}(\mt,z') \Gamma^{b;cd}_{\lambda';\sigma\sigma'}(\mt,z') & = 4g^2f^{cda}f^{bcd}  \bigg[ \mt^2 \delta_{\lambda,\lambda'}
\left(\frac{1}{(1-z')^2} + \frac{1}{z'^2} \right) + 2 (\mt\cdot \epst_{\lambda'})( \mt \cdot \epst_{\lambda}^*)
\bigg]\\
& = 4g^2\ca \delta^{ab} \mt^2 \delta_{\lambda,\lambda'} \bigg[ \frac{1}{(1-z')^2} + \frac{1}{z'^2} + \frac{2}{d_{\perp}} \bigg],
\end{split}
\end{equation}
where, in the first line, the factor 2 comes from summing over two polarization states for the gluon in the loop. Putting everything together gives  
\begin{equation}
 \psi^{q\to qg}_{\ref{diag:gtoggloop}}=  \psi^{q\to qg}_{\lo}(\nt,z)\left (\frac{g^2\ca}{2\pi} \right )\left (\frac{1-z}{\nt^2} \right ) \int_{0}^{1}{\rm d}z'\biggl [\frac{1}{(1-z')^2} + \frac{1}{z'^2} + \frac{2}{d_{\perp}}  \biggr ] \mu^{2-d_{\perp}}\int \frac{\rm{d}^{d_{\perp}}\mt}{(2\pi)^{d_{\perp}}}\frac{
\mt^2}{\biggl [\mt^2 + \frac{z'(1-z')}{1-z}\nt^2\biggr ]},
\end{equation}
and performing the $\mt$-integration we obtain
\begin{equation}
 \psi^{q\to qg}_{\ref{diag:gtoggloop}}=  \psi^{q\to qg}_{\lo}(\nt,z)\left (\frac{g^2\ca}{8\pi^2} \right )\left (-\frac{\Gamma(\varepsilon)}{(4\pi)^{-\varepsilon}} \right )\biggl [(1-z)\frac{\mu^2}{\nt^2}\biggr ]^{\varepsilon}   \int_{0}^{1} \frac{{\rm d}z'}{(z'(1-z'))^{\varepsilon}}   \biggl [\frac{z'}{1-z'} + \frac{1-z'}{z'} + \frac{z'(1-z')}{1-\varepsilon}  \biggr ]. 
\end{equation}
Regulating the soft $z'\rightarrow 0$ and $z'\rightarrow 1$ divergences by a cutoff $\alpha/z < z' < 1- \alpha/z$, the above integral becomes
\begin{equation}
\label{eq:gwavefafterfinalv1}
\begin{split}
\psi^{q\to qg}_{\ref{diag:gtoggloop}}=  \psi^{q\to qg}_{\lo}(\nt,z)\left (\frac{g^2\ca}{8\pi^2} \right ) \biggl \{\biggl [\frac{1}{\varepsilon_{\overline{\rm MS}}}  + \log\left(\frac{(1-z)\mu^2}{\nt^2} \right )\biggr  ]\left (\frac{11}{6}-2\log\left (\frac{z}{\alpha}\right ) \right )
+ \frac{32}{9} - \frac{\pi^2}{3} - \log^2\left (\frac{z}{\alpha} \right ) \biggr \}. 
\end{split}
\end{equation}

\subsubsection*{Gluon polarization diagram \ref{diag:gtoqqbloop}}

The contribution from diagram \ref{diag:gtoqqbloop} is
\begin{equation}
\label{eq:wfgtoqqbloop}
 \psi^{q\to qg}_{\ref{diag:gtoqqbloop}} = (-1) \nf \int \dk \dqp \dkp 
(2\pi)^3 \delta^3(\qpvec-\kpvec-\kvec)  
(2\pi)^3 \delta^3(\qvec -\kpvec-\kvec) \frac{V^{i;jb}_{\lambda',h}(\nt,z)}{\Delta_{01}^{-}\Delta_{02}^{-}\Delta_{03}^{-}}A^{b;m,n}_{\lambda',\sigma}(\mt,z')A^{m,n;a}_{\lambda,\sigma}(\mt,z'),
\end{equation}
where $\nf$ is the number of quark flavours, the factor $(-1)$ reflects the presence of a quark loop, and the  gluon splitting $A^{b;m,n}_{\lambda',\sigma}(\mt,z')$ and quark-antiquark annihilation $A^{m,n;a}_{\lambda,\sigma}(\mt,z')$ vertices are given in \eqs\nr{eq:vertexgtoqqb} and \nr{eq:vertexqqbtog}.

The LC energy denominators and phase space are same as in diagram \ref{diag:gtoggloop}. The product of gluon quark-antiquark vertices in \eq\nr{eq:wfgtoqqbloop}, summed over internal polarization $\sigma$, simplifies to 
\begin{equation}
\label{eq:AAproduct}
\begin{split}
A^{b;m,n}_{\lambda',\sigma}(\mt,z')A^{m,n;a}_{\lambda,\sigma}(\mt,z') & = \frac{-4g^2\tr(t^{a}t^{b})}{z'(1-z')}\biggl [z'\delta_{\lambda',\sigma} - (1-z')\delta_{\lambda',-\sigma} \biggr ]\biggl [z'\delta_{\lambda',\sigma} - (1-z')\delta_{\lambda',-\sigma} \biggr ]  \mt\cdot \epst_{\lambda'} \mt \cdot \epst_{\lambda}^*\\
& = \frac{-4g^2\tf\delta^{ab}}{z'(1-z')}\frac{\mt^2}{d_{\perp}}\delta_{\lambda',\lambda}\biggl [2z'(z'-1) + 1\biggr ],
\end{split}
\end{equation}
where $\tf = 1/2$. Substituting \eq\nr{eq:AAproduct} into \eq\nr{eq:wfgtoqqbloop} and following the same steps as in diagram \ref{diag:gtoggloop}, we obtain an IR-safe expression for the diagram \ref{diag:gtoqqbloop}
\begin{equation}
\psi^{q\to qg}_{\lo}(\nt,z)\left (\frac{g^2\tf\nf}{8\pi^2} \right )\left (-\frac{\Gamma(\varepsilon)}{(1-\varepsilon)(4\pi)^{-\varepsilon}} \right )\biggl [(1-z)\frac{\mu^2}{\nt^2}\biggr ]^{\varepsilon}   \int_{0}^{1} \frac{{\rm d}z'}{(z'(1-z'))^{\varepsilon}}   \biggl [2z'(z'-1) + 1  \biggr ],
\end{equation}
which simplifies to 
\begin{equation}
\label{eq:gbwavefafterfinalv1}
\begin{split}
\psi^{q\to qg}_{\ref{diag:gtoqqbloop}}=  \psi^{q\to qg}_{\lo}(\nt,z)\left (-\frac{g^2\tf\nf}{8\pi^2} \right ) \biggl \{\biggl [\frac{1}{\varepsilon_{\overline{\rm MS}}}  + \log\left(\frac{(1-z)\mu^2}{\nt^2} \right )\biggr  ]\frac{2}{3}
+ \frac{19}{9}\biggr \}. 
\end{split}
\end{equation}
Adding the contributions from \eqs\nr{eq:gwavefafterfinalv1} and \nr{eq:gbwavefafterfinalv1} together, the final state gluon propagator corrections in  the FDH scheme take the form:
\begin{multline} \label{eq:gtoggloop}
\psi^{q\to qg}_{\ref{diag:gtoggloop}} + \psi^{q\to qg}_{\ref{diag:gtoqqbloop}} =  \psi^{q\to qg}_{\lo}(\nt,z)
\Bigg\{
1 + \frac{g^2}{8\pi^2}\biggl \{\biggl [\frac{1}{\varepsilon_{\overline{\rm MS}}}  + \log\left(\frac{(1-z)\mu^2}{\nt^2} \right )  \biggr ] \left (\frac{11\ca}{6}  - \frac{2\tf\nf }{3} - 2\ca\log\left ( \frac{z}{\alpha}\right ) \right )\\
 + \ca\left (\frac{32}{9} - \frac{\pi^2}{3} - \log^2\left (\frac{z}{\alpha} \right ) \right ) 
- \frac{19\tf\nf }{9}  \biggr \}
\Bigg\} .
\end{multline}
From this we know that the UV-divergent parts of the gluon wave function renormalization constant are 
\begin{equation}\label{eq:zgpole}
 Z_g^{\textnormal{pole}}(zp^+) = 1 + \frac{g^2}{8\pi^2}\frac{1}{\varepsilon_{\overline{\rm MS}}}
 \left (\frac{11\ca}{6}  - \frac{2\tf\nf }{3} + 2\ca\log\left ( \frac{\alpha}{z}\right ) \right ).
\end{equation}
The full gluon wave function renormalization constant in the FDH scheme can be calculated in a straightforward manner similarly to the calculation for quarks done in Sec.~\ref{sec:Zq}.  We will not repeat the details here but just quote the result:
\begin{multline}
\label{eq:zg}
 Z_g(zp^+) = 1 + \frac{g^2}{8\pi^2}
\Bigg\{
\left[\frac{1}{\varepsilon_{\overline{\rm MS}}} + \ln \frac{\mu^2}{\lambda_m^2}\right]
 \left (\frac{11\ca}{6}  - \frac{2\tf\nf }{3} + 2\ca\log\left ( \frac{\alpha}{z}\right ) \right )
\\
-\left(\frac{5}{36} + \frac{\pi^2}{6} + 2\ln \frac{\alpha}{z}+ \half \ln^2\frac{\alpha}{z}\right)\ca - \frac{13\tf\nf}{18}
\Bigg\}.
\end{multline}
Comparing \eqs\nr{eq:gtoggloop} and \nr{eq:zg} one sees that 
indeed the UV-divergent parts are the same as in the propagator correction diagrams, but the finite parts are different, as was the case for the quarks.

\subsection{One loop vertex corrections to gluon emission}
\label{sec:vertcor}

\begin{figure}[t]
\centerline{
\includegraphics[width=6.4cm]{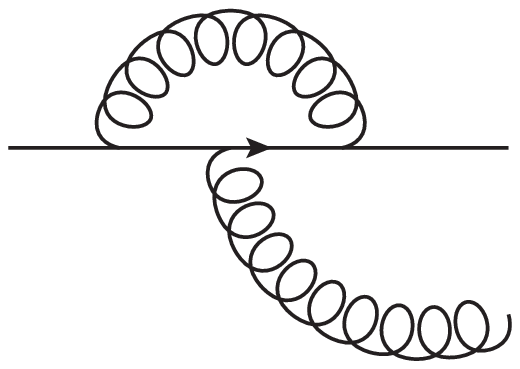}
\begin{tikzpicture}[overlay]
\draw [dashed] (-5.8,4.5) -- (-5.8,0);
\node[anchor=north] at (-5.8cm,-0.2cm) {0};
\draw [dashed] (-4.4,4.5) -- (-4.4,0);
\node[anchor=north] at (-4.4cm,-0.2cm) {1};
\draw [dashed] (-3,4.5) -- (-3,0);
\node[anchor=north] at (-3cm,-0.2cm) {2};
\draw [dashed] (-1,4.5) -- (-1,0);
\node[anchor=north] at (-1,-0.2) {3};
\node[anchor=south] at (-6,2.6) {$\pvec,h,i$};
\node[anchor=north] at (-4.5,2.7) {$\kpvec,h,k$};
\node[anchor=north] at (-2.8,2.7) {$\kppvec,h,l$};
\node[anchor=west] at (-0.5,0.5) {$\qvec,\lambda,a$};
\node[anchor=west] at (-4.5,3.0) {$\mt $};
\node[anchor=east] at (-2.7,3.0) {$\lt$};
\node[anchor=west] at (-0.8,1.5) {$\nt = \qt-z\pt$};
\node[anchor=west] at (-3,1.8) {$\hht$};
\node[anchor=south] at (-1.5,2.6) {$\ppvec,h,j$};
\node[anchor=south] at (-3.6,4.1) {$\kvec,\lambda',b$};
\node[anchor=south west] at (-7cm,0cm) {\namediag{diag:vertexc1}};
 \end{tikzpicture}
}
\rule{0pt}{1ex}
\caption{One loop vertex correction diagram \ref{diag:vertexc1} with LC energy denominators and kinematics. Momentum 
conservation: $\pvec = \kvec + \kpvec$, $\kpvec = \kppvec + \qvec$ and
$\ppvec = \kppvec + \kvec$. The momentum fraction in  first gluon emission is 
$z'$, defined as $k^+ = z'p^+$ and the natural momentum is
$\mt = \kt -z'\pt$. In the absorption of the same gluon the momentum fraction
is $k^+ /p'^+ = z'/(1-z)$ and the natural momentum 
$\lt = \kt - (z'/(1-z))\ptp$. For the emission of the final state gluon the momentum fraction 
is $q^+/k'^+ = z/(1-z')$ and the natural momentum $\hht = \qt - (z/(1-z'))\ktp$.
The natural momentum for the whole diagram is $\nt = \qt-z\pt$.
In order to use $\mt$ as the integration variable we need to know that
$\lt = \mt +(z'/(1-z))\nt$ and $\hht = \nt + (z/(1-z'))\mt$.}
\label{fig:vertexc1kin}
\end{figure}

Next, we calculate the one loop vertex correction diagrams \ref{diag:vertexc1}, \ref{diag:vertexc2}, \ref{diag:vertexcggg1} and \ref{diag:vertexcggg2} shown in \figs\ref{fig:vertexc1kin}-\ref{fig:vertexcgg2kin}.

\subsubsection*{Diagram \ref{diag:vertexc1}}

For diagram  \ref{diag:vertexc1}, with kinematical variables as in \fig\ref{fig:vertexc1kin},  the LC wave function is 
\begin{equation}
\psi^{q\to qg}_{\ref{diag:vertexc1}} = \int \dk \dkp \dkpp
(2\pi)^3 \delta^3(\pvec-\kvec-\kpvec)  
(2\pi)^3 \delta^3(\ppvec -\kppvec-\kvec) \frac{V^{i;k,b}_{\lambda',h}\left(\mt,z'\right)V^{k;l,a}_{\lambda,h}\left(\hht, \frac{z}{1-z'}\right)V^{l,b;j}_{\lambda',h}\left(\lt,\frac{z'}{1-z}\right)
 }{ \Delta_{01}^- \Delta_{02}^- \Delta_{03}^-},
\end{equation}
where the phase space measure is
\begin{equation}
 \int \dk \dkp \dkpp
(2\pi)^3 \delta^3(\pvec-\kvec-\kpvec)  
(2\pi)^3 \delta^3(\ppvec -\kppvec-\kvec)  =   \frac{\mu^{2-d_{\perp}}}{16 \pi (p^+)^2}\int_0^{1-z} \frac{\ud z'}{z'(1-z')(1-z-z')}\int\frac{\ud^{d_\perp}\mt}{(2\pi)^{d_\perp}}
\end{equation}
and the LC energy denominators are
\begin{eqnarray}
 \Delta_{01}^- &=& \frac{-1}{2p^+}\frac{\mt^2}{z'(1-z')}
\\
 \Delta_{02}^- &=&\frac{-1}{2p^+} \left[\frac{\nt^2}{z(1-z)} 
+ \frac{\lt^2}{z'\left(1- \frac{z'}{1-z}\right)}\right]
\\
 \Delta_{03}^- &=& \frac{-1}{2p^+}\frac{\nt^2}{z(1-z)}.
\end{eqnarray}
Putting everything together and using
\begin{equation}
t^{b}_{jl}t^{a}_{lk}t^{b}_{ki}  = \left (\cf - \frac{\ca}{2}\right )t^{a}_{ji}  
\end{equation}
we get
\begin{widetext}
\begin{multline}
\psi^{q\to qg}_{\ref{diag:vertexc1}} = 
\frac{-2g^3t^{a}_{ji}}{\Delta^-_{03}}  \left(\cf -\frac{\ca}{2}\right)
\frac{\mu^{2-d_{\perp}}}{\pi}
\int_0^{1-z} \frac{\ud z'}{z'(1-z')(1-z-z')}
\int\frac{\ud^{d_\perp}\mt}{(2\pi)^{d_\perp}}
\frac{(z')^2(1-z')(1-\frac{z'}{1-z})}{\mt^2\biggl [\lt^2 + \frac{z'(1-z-z')\nt^2}{z(1-z)^2} \biggr ]}\frac{1}{z'\sqrt{1-z'}}
\\
 \frac{(1-z)}{z'\sqrt{1-\frac{z'}{1-z}}} \frac{(1-z')}{z\sqrt{1-\frac{z}{1-z'}}}
\left[\delta_{\lambda',h}+
 \left(1-\frac{z'}{1-z}\right)(1-z') \delta_{\lambda',-h}\right]
\left[\delta_{\lambda,h}+ 
  \left(1-\frac{z}{1-z'}\right) \delta_{\lambda,-h}\right](\mt \cdot \epst^{\ast}_{\lambda'})(\hht  \cdot\epst^*_{\lambda})(\lt\cdot\epst_{\lambda'}) 
\end{multline}
which is simplified to
\begin{multline}
\label{eq:pdiagint}
\psi^{q\to qg}_{\ref{diag:vertexc1}} = 
\frac{-2g^3t^{a}_{ji}}{\Delta^-_{03}}  \left(\cf -\frac{\ca}{2}\right)
\frac{\mu^{2-d_{\perp}}}{\pi}
\int_0^{1-z}\ud z' \frac{\sqrt{1-z}}{z'(1-z-z')}
\int\frac{\ud^{d_\perp}\mt}{(2\pi)^{d_\perp}}
\frac{m^i\left(m^j+\frac{z'}{1-z}n^j\right)
   \left(m^k+\frac{1-z'}{z}n^k\right)}{
 \mt^2\biggl [\left(\mt+\frac{z'}{1-z}\nt\right)^2  + \frac{z'(1-z-z')}{z(1-z)^2 }\nt^2 \biggr ]}
\\
\left[\delta_{\lambda',h}+
 \left(\frac{1-z-z'}{1-z}\right)(1-z') \delta_{\lambda',-h}\right]
\left[\delta_{\lambda,h}+ 
  \frac{1-z -z'}{1-z'} \delta_{\lambda,-h}\right]
\varepsilon_{\lambda'}^{*i} \varepsilon_{\lambda'}^{j}
\varepsilon_{\lambda}^{*k}.
\end{multline}
\end{widetext}
Applying the loop integral \nr{eq:1loopint} from Appendix~\ref{sec:integrals} the UV-divergent part of the $\mt$-integral becomes
\begin{equation}
I^{\rm pole}_{3,{\ref{diag:vertexc1}}}(\pt,\qt,\kt) = \frac{1}{16\pi }\biggl [\frac{1}{\varepsilon_{\overline{\rm MS}}} + \log\left (\frac{\mu^2}{\nt^2} \right ) \biggr ]   \left[
-\frac{z'}{1-z}n^i\delta^{jk}
+\frac{z'}{1-z}n^j\delta^{ik}
+\frac{2(1-z)(1-z')-z'z}{z(1-z)}n^k\delta^{ij}
\right]
\end{equation}
with
\begin{equation}
\label{eq:pdiagvar}
\pt = \kt =-\frac{z'}{1-z}\nt, \quad \qt = -\frac{1-z'}{z}\nt, \quad M^2 = \frac{z'(1-z-z')}{z(1-z)^2}\nt^2,
\end{equation}
and performing the polarization sums (part in $\{\}$) we get
\begin{equation}
\begin{split}
\psi^{q\to qg}_{\ref{diag:vertexc1},pole} = 
 \frac{-2g^3t^{a}_{ji}}{\Delta^-_{03}}  \left(\cf -\frac{\ca}{2}\right)& 
\frac{\nt \cdot \epst^*_\lambda}{8\pi^2}\biggl [\frac{1}{\varepsilon_{\overline{\rm MS}}} + \log\left (\frac{\mu^2}{\nt^2} \right ) \biggr ]   \int_0^{1-z}\ud z' \frac{\sqrt{1-z}}{z'(1-z-z')}\left\{
\frac{(1-z-z')}{z(1-z)^2} 
\right\}
\\
&
\biggl \{
(1-z)\biggl [1+(1-z')^2\biggr ]\delta_{\lambda,h}
+ 
\biggl [1+(1-z')^2-2z(2-z-z')\biggr ]\delta_{\lambda,-h}
\biggr \}.
\end{split}
\end{equation}
The finite part of the $\mt$-integral in \eq\nr{eq:pdiagint} can be performed by using \eq\nr{eq:finiteintpart} which gives the following expression
\begin{equation}
\label{eq:finatep}
\psi^{q\to qg}_{\ref{diag:vertexc1}, \rm finite}= 
\frac{-2g^3t^{a}_{ji}}{\Delta^-_{03}}  \left(\cf -\frac{\ca}{2}\right)
\frac{\nt \cdot \epst^*_\lambda}{\pi}
\int_0^{1-z}\ud z' \frac{\sqrt{1-z}}{z'(1-z-z')} \times \mathcal{H}_{\ref{diag:vertexc1}}(z',z;\lambda,h),
\end{equation}
where the function $\mathcal{H}_{\ref{diag:vertexc1}}$ summed over the internal polarization $\lambda'$ is 
\begin{equation}
\mathcal{H}_{\ref{diag:vertexc1}}(z',z;\lambda,h) = A_{\ref{diag:vertexc1}}(z,z')\delta_{\lambda,h} + B_{\ref{diag:vertexc1}}(z,z')\delta_{\lambda,-h}
\end{equation}
and the coefficients $A_{\ref{diag:vertexc1}}$ and $B_{\ref{diag:vertexc1}}$ corresponding to different polarization states are given by
\begin{equation}
\begin{split}
A_{\ref{diag:vertexc1}}(z,z') & =  \frac{1}{16\pi z(1-z)^2}\biggl [4(1-z)^2  - 4(2-z)(1-z)z' + \biggl \{6-z(10-3z) \biggr \}z'^2-(2-3z)z'^3 \\
& -2(1-z)(1-z-z')\biggl \{(1-z')^2\log\left (\frac{(1-z)(1-z')}{1-z-z'} \right ) - (1+(1-z')^2)\log\left (\frac{z(1-z)^2}{z'(1-z-z')} \right ) \biggr \}\biggr ]
\end{split}
\end{equation}
and
\begin{equation}
\begin{split}
B_{\ref{diag:vertexc1}}(z,z') & =  \frac{(1-z-z')}{16\pi z(1-z)^2(1-z')}\biggl [4(1-z)^2  - 4(2-z)(1-z)z' + \biggl \{6-z(2+z) \biggr \}z'^2-(2+z)z'^3 \\
& -2(1-z')\biggl \{(1-z)^2\log\left (\frac{(1-z)(1-z')}{1-z-z'} \right ) - \left (1+(1-z')^2-2z(2-z-z') \right )\log\left (\frac{z(1-z)^2}{z'(1-z-z')} \right ) \biggr \}\biggr ].
\end{split}
\end{equation}

\subsubsection*{Diagram \ref{diag:vertexc2}}

\begin{figure}[t]
\centerline{
\includegraphics[width=6.4cm]{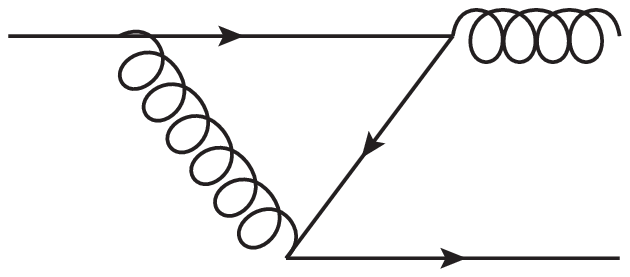}
\begin{tikzpicture}[overlay]
\draw [dashed] (-5.8,2.5) -- (-5.8,0);
\node[anchor=north] at (-5.8cm,-0.2cm) {0};
\draw [dashed] (-4.0,2.5) -- (-4.0,0);
\node[anchor=north] at (-4.0cm,-0.2cm) {1};
\draw [dashed] (-3,2.5) -- (-3,0);
\node[anchor=north] at (-3cm,-0.2cm) {2};
\draw [dashed] (-1,2.5) -- (-1,0);
\node[anchor=north] at (-1,-0.2) {3};
\node[anchor=south] at (-6,2.4) {$\pvec,h,i$};
\node[anchor=north east] at (-4.6,1.5) {$\kvec,\lambda',b$};
\node[anchor=south] at (-3.5,2.4) {$\kpvec,h,k$};
\node[anchor=north] at (-2.0,1.6) {$\kppvec,h,l$};
\node[anchor=west] at (-0.5,0.5) {$\ppvec,h,j$};
\node[anchor=west] at (-4.6,2.0) {$\mt $};
\node[anchor=east] at (-2.5,2.0) {$\hht$};
\node[anchor=west] at (-1.0,1.2) {$\nt = \qt-z\pt$};
\node[anchor=south] at (-3.4,0.7) {$\lt$};
\node[anchor=south] at (-1.2,2.6) {$\qvec,\lambda,a$};
\node[anchor=south west] at (-7cm,0cm) {\namediag{diag:vertexc2}};
 \end{tikzpicture}
}
\rule{0pt}{1ex}
\caption{One loop vertex correction diagram \ref{diag:vertexc2} with LC energy denominators and  kinematics. Momentum 
conservation: $\pvec = \kvec + \kpvec$, $\kpvec + \kppvec = \qvec$ and
$\kvec = \kppvec + \ppvec$. The momentum fraction in  first gluon emission is 
$z'$, defined as $k^+ = z'p^+$ and the natural momentum is
$\mt = \kt -z'\pt$. In the absorption of the same gluon the momentum fraction (of the quark)
is $p'^+ /k^+ = (1-z)/z'$ and the natural momentum 
$\lt = \ptp - ((1-z)/z')\kt$. For the emission of the final state gluon the momentum fraction 
(of the quark) is $k'^+/q^+ = (1-z')/z$ and the natural momentum $\hht = \ktp - ((1-z')/z)\qt$.
 The natural momentum for the whole diagram is $\nt = \qt-z\pt$.
In order to use $\mt$ as the integration variable we need to know that
$\lt = -\nt -((1-z)/z')\mt$ and $\hht = -\mt - ((1-z')/z)\nt$.
}
\label{fig:vertexc2kin}
\end{figure}

Then we turn to diagram \ref{diag:vertexc2}, whose kinematics is presented in \fig\ref{fig:vertexc2kin}.  The LC wave function is 
\begin{equation}
\psi^{q\to qg}_{\ref{diag:vertexc2}} = \int \dk \dkp \dkpp
(2\pi)^3 \delta^3(\pvec-\kvec-\kpvec)  
(2\pi)^3 \delta^3(\qvec -\kpvec-\kppvec) \frac{V^{i;k,b}_{\lambda',h}\left(\mt,z'\right)A^{l,k;a}_{\lambda,h}\left(\hht, \frac{1-z'}{z}\right)A^{b;j,l}_{\lambda',h}\left(\lt,\frac{1-z}{z'}\right)
 }{ \Delta_{01}^- \Delta_{02}^- \Delta_{03}^-},
\end{equation}
where the phase space measure simplifies to
\begin{equation}
\int \dk \dkp \dkpp(2\pi)^3 \delta^3(\pvec-\kvec-\kpvec)  
(2\pi)^3 \delta^3(\qvec -\kpvec-\kppvec) =  \frac{\mu^{2-d_{\perp}}}{16 \pi (p^+)^2}\int_{1-z}^1\frac{\ud z' }{z'(1-z')(z'+z-1)}\int\frac{\ud^{d_\perp}\mt}{(2\pi)^{d_\perp}}
\end{equation}
and the LC energy denominators are 
\begin{eqnarray}
 \Delta_{01}^- &=& \frac{-1}{2p^+}\frac{\mt^2}{z'(1-z')}
\\
 \Delta_{02}^- &=&\frac{-1}{2p^+} \biggl [\frac{\nt^2}{z(1-z)} 
+ \frac{\hht^2}{(1-z')\left(1- \frac{1-z'}{z}\right)}\biggr ]
\\
 \Delta_{03}^- &=& \frac{-1}{2p^+}\frac{\nt^2}{z(1-z)}.
\end{eqnarray}
Following the same steps as in diagram \ref{diag:vertexc1} we find
\begin{widetext}
\begin{multline}
\label{eq:diagqint}
\psi^{q\to qg}_{\ref{diag:vertexc2}} = 
\frac{-2g^3t^{a}_{ji}}{\Delta^-_{03}}  \left(\cf -\frac{\ca}{2}\right)
\frac{\mu^{2-d_{\perp}}}{\pi}
\int_{1-z}^{1}\ud z' 
\frac{\sqrt{1-z}}{z'(z'+z-1)} 
\int\frac{\ud^{d_\perp}\mt}{(2\pi)^{d_\perp}}
\frac{m^i \left (m^j + \frac{z'}{1-z}n^j\right ) \left (m^k + \frac{1-z'}{z}n^k\right )}{\mt^2
\biggl [(\mt + \frac{1-z'}{z}\nt)^2 + \frac{(1-z')(z+z'-1) }{z^2(1-z)} \nt^2\biggr ]}
\\
\left[\frac{1-z}{z'}  \delta_{\lambda',h}
- \left (\frac{z'+z-1}{z'}\right )(1-z')\delta_{\lambda',-h}
   \right]
\left[-  \frac{1-z'}{z}  \delta_{\lambda,h}+
    \frac{z'+z-1}{z}\delta_{\lambda,-h}
 \right]
\varepsilon_{\lambda'}^{*i} \varepsilon_{\lambda'}^{j}
\varepsilon_{\lambda}^{*k}. 
\end{multline}
\end{widetext}
The UV-divergent part of the $\mt$-integral above is given by
\begin{equation}
I^{\rm pole}_{3,\ref{diag:vertexc2}}(\pt,\qt,\kt) = \frac{1}{16\pi}\biggl [\frac{1}{\varepsilon_{\overline{\rm MS}}} + \log\left (\frac{\mu^2}{\nt^2} \right ) \biggr ] \left[
-\frac{1-z'}{z} n^i\delta^{jk}
+  \frac{z'z+z+z'-1}{z(1-z)}n^j\delta^{ik}
+ \frac{1-z'}{z} n^k\delta^{ij}
\right]
\end{equation}
with 
\begin{equation}
\label{eq:qdiagvar}
\pt = \frac{z'}{1-z}\nt, \quad \qt=\kt = \frac{1-z'}{z}\nt, \quad M^2 = \frac{(1-z')(z+z'-1)}{z^2(1-z)}\nt^2.
\end{equation}
Performing the internal polarization sums we get
\begin{equation}
\begin{split}
\psi^{q\to qg}_{\ref{diag:vertexc2},pole} = 
\frac{-2g^3t^{a}_{ji}}{\Delta^-_{03}}  \left(\cf -\frac{\ca}{2}\right)
\frac{\nt \cdot \epst^*_\lambda}{8 \pi^2}\biggl [\frac{1}{\varepsilon_{\overline{\rm MS}}} + \log\left (\frac{\mu^2}{\nt^2} \right ) \biggr ]  &  \int_{1-z}^1\ud z' \frac{\sqrt{1-z}}{z'(z'+z-1)} \\
&\left\{
\frac{(z+z'-1)}{z(1-z)}
\right\}
\left\{
\left(
(1-z')^2
\right)
\delta_{\lambda,h}
+
\left( 
1-z
\right)
\delta_{\lambda,-h}
\right\}.
\end{split}
\end{equation}
Again, the finite $\mt$-integral in \eq\nr{eq:diagqint} can be performed by using \eq\nr{eq:finiteintpart}, and thus we obtain 
\begin{equation}
\label{eq:finateq}
\psi^{q\to qg}_{\ref{diag:vertexc2}, \rm finite}= 
\frac{-2g^3t^{a}_{ji}}{\Delta^-_{03}}  \left(\cf -\frac{\ca}{2}\right)
\frac{\nt \cdot \epst^*_\lambda}{\pi}
\int_{1-z}^{1}\ud z' \frac{\sqrt{1-z}}{z'(z'+z-1)} \times \mathcal{H}_{\ref{diag:vertexc2}}(z',z;\lambda,h)
\end{equation}
with
\begin{equation}
\mathcal{H}_{\ref{diag:vertexc2}}(z',z;\lambda,h) = A_{\ref{diag:vertexc2}}(z,z')\delta_{\lambda,h} + B_{\ref{diag:vertexc2}}(z,z')\delta_{\lambda,-h}
\end{equation}
and
\begin{equation}
\begin{split}
A_{\ref{diag:vertexc2}} & =  \frac{-1}{16\pi z^2(1-z)z'}\biggl [(1-z')^2\left (1 + (1-z')^2\right ) + z^2(1-z')\left (2-3(1-z')z' \right ) - z(1-z')\biggl \{4-z'\left (7-(8-3z')z'\right ) \biggr \}\\
& +2zz'(z'+z-1)\biggl \{  \left ( 1+(1-z')^2 \right )  \log\left (\frac{zz'}{z'+z-1} \right ) - (1-z')^2\log\left (\frac{z^2(1-z)}{(1-z')(z'+z-1)} \right ) \biggr \}\biggr ]
\end{split}
\end{equation}
and
\begin{equation}
\begin{split}
B_{\ref{diag:vertexc2}} & =  \frac{(z'+z-1)}{16\pi z^2(1-z)^2z'}\biggl [(1-z)(1-z')\left (1 + (1-z')^2\right ) + z^2(1-z)\left (2-z'(3+z')\right ) -z(1-z)\biggl \{4-\left ( 7-z'^2\right )z' \biggr \}\\
& -2zz'\biggl \{\left (1+(1-z')^2-2z(2-z-z') \right )\log\left (\frac{zz'}{z'+z-1} \right ) - (1-z)^2\log\left (\frac{(1-z)z^2}{(1-z')(z'+z-1)} \right )  \biggr \}\biggr ].
\end{split}
\end{equation}

\subsubsection*{Sum of diagrams \ref{diag:vertexc1} and \ref{diag:vertexc2} }

The diagrams  \ref{diag:vertexc1} and \ref{diag:vertexc2} have the same color structure, in fact they are represented by the same diagram in covariant perturbation theory. 
Adding \ref{diag:vertexc1} and \ref{diag:vertexc2} and regulating the IR-divergence $z'\rightarrow 0$ in \ref{diag:vertexc1} by $\alpha < z' < 1-z$,  we obtain for the UV divergent part
\begin{equation} 
\label{eq:vertexno3gfinalv1}
\psi^{q\to qg}_{\ref{diag:vertexc1},\rm pole}+
\psi^{q\to qg}_{\ref{diag:vertexc2},\rm pole}
= \psi^{q\to qg}_{\lo}(\nt, z)
\frac{g^2}{8 \pi^2}\biggl [\frac{1}{ \varepsilon_{\overline{\rm MS}}}+\log \left (\frac{\mu^2}{\nt^2}\right )\biggr ]
\left(\cf -\frac{\ca}{2}\right)
\left[
-\frac{3}{2}+ \ln(1-z)-2\ln \alpha
\right].
\end{equation}
Similarly, the remaining finite $z'$-integrals in \eqref{eq:finatep} and \eqref{eq:finateq} simplifies to
\begin{equation}
\label{eq:finaled}
\begin{split}
 \psi^{q\to qg}_{\ref{diag:vertexc1},\rm finite} & +  \psi^{q\to qg}_{\ref{diag:vertexc2},\rm finite} = \psi^{q\to qg}_{\lo}(\nt, z)\frac{g^2}{48\pi^2} \left(\cf-\frac{\ca}{2}\right) 
\mathcal{H}_{\ref{diag:vertexc1}\ref{diag:vertexc2}}(z,\alpha)\\
 & + \frac{-2g^3t^{a}_{ji}}{\Delta^-_{03}}  \left(\cf-\frac{\ca}{2}\right)
\frac{-3(\nt \cdot \epst^*_\lambda)}{48\pi^2z\sqrt{1-z}}\biggl [\left (z + (2+z)\log(1-z) \right )\delta_{\lambda, h}  + z\log(z)\delta_{\lambda, -h}\biggr ]
\end{split}
\end{equation}
with
\begin{equation}
\begin{split}
\mathcal{H}_{\ref{diag:vertexc1}\ref{diag:vertexc2}}(z,\alpha)
= \pi^2-15 & + 3\log^2(1-z) -9\log(z) + \frac{12}{z}\log(1-z) - 6\log(1-z) + 12\log(1-z)\log\left ( \frac{z}{\alpha}\right ) \\
&-12\log(z)\log(\alpha) -12\log(\alpha)  + 6\log^2(\alpha) + 6\mathrm{Li}_2\left (1- z\right ).
\end{split}
\end{equation}
Here $\mathrm{Li}_2(z)$ is the standard dilogarithm function, defined as 
\begin{equation}
\mathrm{Li}_2\left (z \right) \equiv -\int_0^z \mathrm{d}y \frac{\log(1-y)}{y}, \quad z \in \Re \backslash [1,\infty),
\end{equation}
and in particular, for $0 < z < 1$ we have 
\begin{equation}
\mathrm{Li}_2\left (z \right)  + \mathrm{Li}_2\left (1-z \right) = \frac{\pi^2}{6} - \log(z)\log(1-z). 
\end{equation}
The UV-finite parts of the expression  \eqref{eq:finaled} depend on the renormalization scheme. In the FDH scheme the NLO part of the wave function has a part that is not proportional to the 
$\delta_{\lambda,h} + (1-z)\delta_{\lambda,-h}$ helicity structure of the leading order part.

\subsubsection*{Diagram \ref{diag:vertexcggg1}}

\begin{figure}[t]
\centerline{
\includegraphics[width=6.4cm]{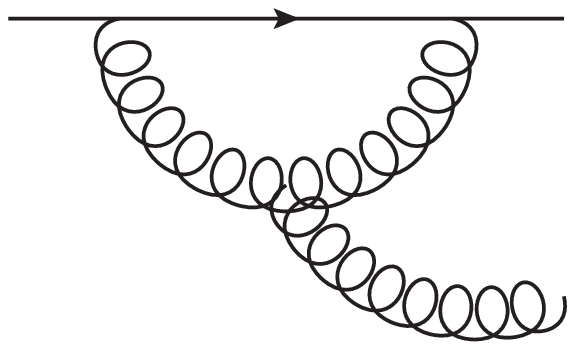}
\begin{tikzpicture}[overlay]
\draw [dashed] (-5.8,3.5) -- (-5.8,0);
\node[anchor=north] at (-5.8cm,-0.2cm) {0};
\draw [dashed] (-4.0,3.5) -- (-4.0,0);
\node[anchor=north] at (-4.0cm,-0.2cm) {1};
\draw [dashed] (-2.2,3.5) -- (-2.2,0);
\node[anchor=north] at (-2.2cm,-0.2cm) {2};
\draw [dashed] (-0.8,3.5) -- (-0.8,0);
\node[anchor=north] at (-0.8,-0.2) {3};
\node[anchor=east] at (-5.6,3.8) {$\pvec,h,i$};
\node[anchor=west] at (-1,3.8) {$\ppvec,h,j$};
\node[anchor=center] at (-3,3.8) {$\pppvec,h,k$};
\node[anchor=west] at (-0.3,0.4) {$\qvec,\lambda,a$};
\node[anchor=south west] at (-4.8,3.0) {$\mt $};
\node[anchor=south east] at (-1.9,3.0) {$\lt$};
\node[anchor=west] at (-0.8,2) {$\nt = \qt-z\pt$};
\node[anchor=west] at (-2.6,1.4) {$\hht$};
\node[anchor=north west] at (-1.5,3.0) {$\kpvec,\lambda_2,c$};
\node[anchor=east] at (-5.1,2.5) {$\kvec,\lambda_1,b$};
\node[anchor=south west] at (-7cm,0cm) {\namediag{diag:vertexcggg1}};
 \end{tikzpicture}
}
\rule{0pt}{1ex}
\caption{One loop vertex correction diagram \ref{diag:vertexcggg1} with LC energy denominators and kinematics. Momentum conservation: $\ppvec=\pvec-\qvec$, $\pppvec=\pvec-\kvec$ and 
$\kpvec=\kvec-\qvec$.  Momentum fractions are defined by
$ k^+ = z' p^+$, $k'^+ = \xi p'^+$ and $q^+ = z p^+$.
The momentum fraction of the 3 gluon splitting is $q^+/k^+ = z/z'$.
The natural momenta for the three vertices are
$\mt \equiv \kt-z'\pt$, $\lt \equiv \ktp-\xi \ptp$ and
 $\hht\equiv \qt-(z/z')\kt$, and the natural momentum 
scale for the whole diagram is 
$\nt\equiv \qt-z\pt$.
Note that $z'=z+\xi(1-z)$ so that $p''^+ = (1-\xi)(1-z)p^+$.
To choose $\mt$ as the integration variable we need
$\lt = \mt - (1-\xi)\nt$ and $\hht = \nt - (z/z')\mt$.}
\label{fig:vertexcgg1kin}
\end{figure}

Diagram \ref{diag:vertexcggg1}, shown  in \fig\ref{fig:vertexcgg1kin}, is
\begin{equation}
\label{eq:vertexcggg1}
\psi^{q\to qg}_{\ref{diag:vertexcggg1}} = 
\int \dk \dkp  \dppp
(2\pi)^3 \delta^3(\qpvec-\kpvec-\kvec)  
(2\pi)^3 \delta^3(\qvec -\kpvec-\kvec)\\ 
\frac{V^{i;k,b}_{\lambda_1,h}(\mt,z')V^{k,c;j}_{\lambda_2,h}(\lt,\xi) }{\Delta^-_{01}\Delta^-_{02}\Delta^-_{03}}
\Gamma^{b;c,a}_{\lambda_1;\lambda_2,\lambda}(\hht,z/z')
\end{equation}
where the phase space measure is 
\begin{equation}
\begin{split}
\int \dk \dkp  \dppp
(2\pi)^3 \delta^3(\qpvec-\kpvec-\kvec)  
(2\pi)^3 \delta^3(\qvec -\kpvec-\kvec)
& =
\frac{\mu^{2-d_{\perp}}}{16 \pi (p^+)^2}\int_z^1 \frac{\ud z' }{z'(1-z)^2\xi(1-\xi)} \int \frac{\ud^{d_\perp}\kt}{(2\pi)^{d_{\perp}}}\\
& =
\frac{\mu^{2-d_{\perp}}}{16 \pi (p^+)^2}\int_z^1 \frac{\ud z' }{z'(z'-z)(1-z')}\int \frac{\ud^{d_\perp}\mt}{(2\pi)^{d_{\perp}}},
\end{split}
\end{equation}
the LC energy denominators are
\begin{equation}
\begin{split}
 \Delta^-_{01} &=  \frac{-1}{2p^+} \frac{\mt^ 2}{z'(1-z')}\\
 \Delta^-_{02} &= \frac{-1}{2p^+}\left[ \frac{\nt^2}{z(1-z)} + \frac{1}{1-z}\frac{\lt^2}{\xi(1-\xi)}
\right]\\ 
 \Delta^-_{03} &= \frac{-1}{2p^+} \frac{\nt^ 2}{z(1-z)},
\end{split}
\end{equation}
and the variable $\xi$ is defined in \fig\ref{fig:vertexcgg1kin}. Putting everything together and using
\begin{equation}
it^{c}_{jk}t^{b}_{ki}f^{bca} = \frac{\ca}{2}t^{a}_{ji}
\end{equation}
the expression in  \eq\nr{eq:vertexcggg1} simplifies to 
\begin{multline}
 \psi^{q\to qg}_{\ref{diag:vertexcggg1}} = 
 \frac{+2g^3t^{a}_{ji}}{\Delta_{03}^{-}} \left (\frac{\ca}{2}\right )\frac{\mu^{2-d_\perp}}{\pi} 
\int_z^1 \ud z' \frac{z\sqrt{1-z}}{(z')^2(z'-z)}
\int \frac{\ud^{d_\perp}\mt}{(2\pi)^{d_\perp}}
\frac{m^i \left (m^j-(1-\xi)n^j\right ) \left (m^k-(z'/z)n^k\right )}{\mt^2\biggl [(\mt-(1-\xi)\nt)^2 + \frac{\xi(1-\xi)}{z}\nt^2\biggr ]}
\\ \
\biggl [\delta_{\lambda_1,h} + (1-z')\delta_{\lambda_1,-h}\biggr ]
\biggl [\delta_{\lambda_2,h} + (1-\xi)\delta_{\lambda_2,-h}\biggr ]
\left[
\frac{\varepsilon^{*k}_{\lambda_2} \delta_{\lambda_1,\lambda}}{1-z/z'}
+ \frac{\varepsilon^{*k}_{\lambda}\delta_{\lambda_1,\lambda_2}}{z/z'}
- \varepsilon^k_{\lambda_1}\delta_{\lambda_2,-\lambda}
\right]\varepsilon_{\lambda_1}^{*i}
\varepsilon_{\lambda_2}^j.
\end{multline}
The UV divergent part of the $\mt$-integral becomes
\begin{equation}
 I_{3,\ref{diag:vertexcggg1}}^{\rm pole}(\pt,\qt,\kt) =  \frac{1}{16\pi}\biggl [\frac{1}{\varepsilon_{\overline{\rm MS}}} + \log\left (\frac{\mu^2}{\nt^2} \right )\biggr ]\biggl [ (1-\xi)n^i\delta^{jk} -(1-\xi)n^j\delta^{ik}
+(1-\xi-2z'/z)n^k\delta^{ij}\biggr ],
\end{equation}
where  
\begin{equation}
\pt = \kt = (1-\xi)\nt, \quad \qt = \frac{z'}{z}\nt, \quad M^2 = \frac{\xi(1-\xi)}{z}\nt^2.
\end{equation}
Performing the helicity sums and index contractions then gives for the pole part
\begin{multline}\label{eq:vcgg1}
\psi^{q\to qg}_{\ref{diag:vertexcggg1}, \rm pole} = 
\frac{+2g^3t^{a}_{ji}}{\Delta_{03}^{-}} \left (\frac{\ca}{2}\right )\frac{\nt \cdot \epst^{\ast}_{\lambda}}{8\pi^2}\biggl [\frac{1}{\varepsilon_{\overline{\rm MS}}} + \log\left (\frac{\mu^2}{\nt^2} \right )\biggr ]
\int_z^1 \ud z' \frac{z\sqrt{1-z}}{(z')^2(z'-z)}
\bigg\{
 z'\frac{  z(1-z')^2 - z' (1+(1-z')^2)  }{(1-z)z^2}
\delta_{\lambda,h}
\\ +
z' \frac{ z (1-z)^2 + z z'(2-z) - z' (1+(1-z')^2)  }{(1-z)^2 z^2}
\delta_{\lambda,-h}
\bigg\}.
\end{multline}
For the finite $\mt$-integral we obtain
\begin{equation}
\psi^{q\to qg}_{\ref{diag:vertexcggg1}, \rm finite}= 
\frac{+2g^3t^{a}_{ji}}{\Delta^-_{03}}  \left(\frac{\ca}{2}\right)
\frac{\nt \cdot \epst^*_\lambda}{\pi}
\int_z^{1}\ud z' \frac{z\sqrt{1-z}}{(z')^2(z'-z)} \times \mathcal{H}_{\ref{diag:vertexcggg1}}(z',z;\lambda,h),
\end{equation}
where 
\begin{equation}
\mathcal{H}_{\ref{diag:vertexcggg1}}(z',z;\lambda,h) = A_{\ref{diag:vertexcggg1}}(z,z')\delta_{\lambda,h} + B_{\ref{diag:vertexcggg1}}(z,z')\delta_{\lambda,-h}
\end{equation}
with the coefficients 
\begin{equation}
\begin{split}
A_{\ref{diag:vertexcggg1}}(z,z') & =  \frac{1}{16\pi (1-z)^2 z^2 (z-z')}\biggl [2(z')^3\left (1+(1-z')^2 \right ) + z^3\left (2-z' (5-3(2-z')z' ) \right )-z(z')^2\left (6-z'(2+3(1-z')z') \right )\\
& -z^2z'\left (2-z'(11-z'(13-6z')) \right )  + 2(1-z)(z-z')z'\biggl \{ \left (z'(1-z')^2-z(2-z'(2-z')\right )\log\left (\frac{(1-z)z'}{z'-z} \right )\\
& + \left (z(1-z')^2 -z'(1 + (1-z')^2) \right )\log\left (\frac{z(1-z)^2}{(1-z')(z'-z)} \right ) \biggr \}\biggr ]
\end{split}
\end{equation}
and
\begin{equation}
\begin{split}
B_{\ref{diag:vertexcggg1}}(z,z') & =  \frac{1}{16\pi (1-z)^2z^2(z-z')}\biggl [2(1-z)z^3-(1-z)z^2(2+3z)z'-z(z')^2\left (6-z(11-(9-z)z) \right )\\
& +(z')^3\left (4+z(2+3z) \right ) - (z')^4(4+5z) + (z')^5(2+z)  -2z'\biggl \{(1-z)^2\left (2z^2-3zz'+(z')^2 \right )\log\left (\frac{(1-z)z'}{z'-z} \right )\\
& - (z-z')\left (z(1-z)^2+ zz'(2-z)- z'(1+(1-z')^2) \right )\log\left (\frac{z(1-z)^2}{(1-z')(z'-z)} \right ) \biggr \}\biggr ].
\end{split}
\end{equation}

\subsubsection*{Diagram \ref{diag:vertexcggg2}}

\begin{figure}[t]
\centerline{
\includegraphics[width=6.4cm]{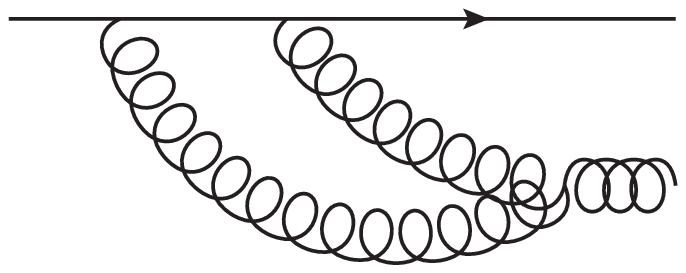}
\begin{tikzpicture}[overlay]
\draw [dashed] (-5.8,2.8) -- (-5.8,0);
\node[anchor=north] at (-5.8cm,-0.2cm) {0};
\draw [dashed] (-4.5,2.8) -- (-4.5,0);
\node[anchor=north] at (-4.5cm,-0.2cm) {1};
\draw [dashed] (-2.6,2.8) -- (-2.6,0);
\node[anchor=north] at (-2.6cm,-0.2cm) {2};
\draw [dashed] (-0.8,3.5) -- (-0.8,0);
\node[anchor=north] at (-0.8,-0.2) {3};
\node[anchor=east] at (-6,2.5) {$\pvec,h,i$};
\node[anchor=west] at (-1.6,2.5) {$\ppvec,h,j$};
\node[anchor=center] at (-4.5,2.5) {$\pppvec,h,k$};
\node[anchor=west] at (-0.2,0.8) {$\qvec,\lambda,a$};
\node[anchor=west] at (-5.6,2.5) {$\mt $};
\node[anchor=east] at (-3.6,2.5) {$\lt$};
\node[anchor=west] at (-0.8,1.5) {$\nt = \qt-z\pt$};
\node[anchor=west] at (-1.5,1.2) {$\hht$};
\node[anchor=south west] at (-3,1.5) {$\kpvec,\lambda_2,c$};
\node[anchor=east] at (-5.1,1.2) {$\kvec,\lambda_1,b$};
\node[anchor=south west] at (-7cm,0cm) {\namediag{diag:vertexcggg2}};
 \end{tikzpicture}
}
\rule{0pt}{1ex}
\caption{One loop vertex correction diagram \ref{diag:vertexcggg2} with energy denominators and kinematics. Momentum conservation: $\ppvec=\pvec-\qvec$, $\pppvec=\pvec-\kvec$ and
$\kpvec=\pppvec-\ppvec= \qvec-\kvec$.   Momentum fractions are defined by
$ k^+ = z' p^+$, $k'^+ = \xi p''^+$ and $q^+ = z p^+$.
The momentum fraction of the 3 gluon merging is $k^+/q^+ = z'/z$.
The natural momenta for the three vertices are
$\mt \equiv \kt-z'\pt$, $\lt \equiv \ktp-\xi \ptpp$ and
 $\hht\equiv \kt-(z'/z)\qt$, and the natural momentum 
scale for the whole diagram is 
$\nt\equiv \qt-z\pt$.
Note that $z=z'+\xi(1-z')$ so that $p''^+ = (1-z')p^+$.
To choose $\mt$ as the integration variable we need
$\lt = \nt - (1-\xi)\mt$ and $\hht = \mt - (z'/z)\nt$.}
\label{fig:vertexcgg2kin}
 \end{figure}

The LC wave function  for the diagram \ref{diag:vertexcggg2} with the kinematics shown in \fig\ref{fig:vertexcgg2kin} is 
\begin{equation}
\begin{split}
\psi^{q\to qg}_{\ref{diag:vertexcggg2}} = \int \dk \dkp  \dppp(2\pi)^3 \delta^3(\kpvec-\pppvec + \ppvec)  
(2\pi)^3 \delta^3(\pppvec -\pvec + \kvec)\frac{V^{i;k,b}_{\lambda_1,h}(\mt,z')V^{k;j,c}_{\lambda_2,h}(\lt,\xi) }{\Delta^-_{01}\Delta^-_{02}\Delta^-_{03}}
\Gamma^{cb;a}_{\lambda_2,\lambda_1;\lambda}(\hht,z'/z),
\end{split}
\end{equation}
where the phase space measure is 
\begin{equation}
\begin{split}
\int \dk \dkp  \dppp(2\pi)^3 \delta^3(\kpvec-\pppvec + \ppvec)  
(2\pi)^3 \delta^3(\pppvec -\pvec + \kvec)  =
\frac{\mu^{2-d_{\perp}}}{16 \pi (p^+)^2}\int_0^z \frac{\ud z' }{z'(z-z')(1-z')}\int \frac{\ud^{d_\perp}\mt}{(2\pi)^{d_{\perp}}}
\end{split}
\end{equation}
and the LC energy denominators are
\begin{eqnarray}
 \Delta^-_{01} &=&  \frac{-1}{2p^+} \frac{\mt^ 2}{z'(1-z')}
\\
 \Delta^-_{02} &=&
\frac{-1}{2p^+}
\left[ \frac{\nt^2}{z(1-z)} + \frac{1}{z}\frac{\hht^2}{(z'/z)(1-(z'/z))}
\right]
\\
 \Delta^-_{03} &=&  \frac{-1}{2p^+} \frac{\nt^ 2}{z(1-z)}.
\end{eqnarray}
Adding everything together yields 
\begin{multline}
   \psi^{q\to qg}_{\ref{diag:vertexcggg2}} = 
\frac{+2g^3t^{a}_{ji}}{\Delta^-_{03}}
\left (\frac{\ca}{2}\right )
\frac{\mu^{2-d_{\perp}}}{\pi}
\int_0^z \ud z' \frac{\sqrt{1-z}}{z(z-z')}
\int \frac{\ud^{d_\perp}\mt}{(2\pi)^{d_\perp}}
\frac{m^i \left(m^j-\left (\frac{1-z'}{1-z}\right )n^j\right) \left(m^k-\frac{z'}{z}n^k \right)  }
{\mt^2\biggl [\left (\mt - \frac{z'}{z}\nt\right)^2 + \frac{z'(z-z')}{z^2(1-z)}\nt^2\biggr ] }
\\ 
\biggl [\delta_{\lambda_1,h} + (1-z')\delta_{\lambda_1,-h}\biggr ]
\biggl [\delta_{\lambda_2,h} + \left (\frac{1-z}{1-z'} \right )\delta_{\lambda_2,-h}\biggr ]
\left[
\frac{\varepsilon^k_{\lambda_2} \delta_{\lambda_1,\lambda}}{1-z'/z}
+ \frac{\varepsilon^k_{\lambda_1}\delta_{\lambda,\lambda_2}}{z'/z}
-  \varepsilon^{*k}_{\lambda}\delta_{\lambda_1,-\lambda_2}
\right]\varepsilon^{*i}_{\lambda_1}\varepsilon^{*j}_{\lambda_2}.
\end{multline}
The UV divergent part of the transverse integral is
\begin{equation}
I_{3,\ref{diag:vertexcggg2}}^{\rm pole}(\pt,\qt,\kt)  = \frac{1}{16\pi}\biggl [\frac{1}{\varepsilon_{\overline{\rm MS}}} + \log\left (\frac{\mu^2}{\nt^2} \right )\biggr  ]\biggl [ \frac{z'}{z}n^i\delta^{jk} -\frac{z-z' +z(1-z')}{z(1-z)}n^j\delta^{ik}
-\frac{z'}{z}n^k\delta^{ij}\biggr ],
\end{equation}
where 
\begin{equation}
\pt = \left ( \frac{1-z'}{1-z} \right )\nt, \quad \qt =\kt= \left (\frac{z'}{z}\right )\nt, \quad M^2 = \frac{z'(z-z')}{z^2(1-z)}\nt^2.
\end{equation}
Using mathematica for the polarization sums the pole part of the diagram becomes
\begin{equation}
\begin{split}
\psi^{q\to qg}_{\ref{diag:vertexcggg2}}  = 
\frac{+2g^3t^{a}_{ji}}{\Delta^-_{03}}
\left (\frac{\ca}{2}\right )\frac{\nt \cdot \epst^*_\lambda }{8 \pi^2} \biggl [\frac{1}{\varepsilon_{\overline{\rm MS}}} + \log\left (\frac{\mu^2}{\nt^2} \right )\biggr ]
\int_0^z \ud z' \frac{\sqrt{1-z}}{z(z-z')}&
\bigg\{
 \frac{z'(1-z')^2-z(1+(1-z')^2)}{z'(1-z)}\delta_{\lambda,h} \\
 & + 
 \frac{z'-2z}{z'}\delta_{\lambda,-h}
\bigg\}.
\end{split}
\end{equation}
Like before the remaining finite part can be cast in the following form
\begin{equation}
\psi^{q\to qg}_{\ref{diag:vertexcggg2}, \rm finite}= 
\frac{+2g^3t^{a}_{ji}}{\Delta^-_{03}}  \left(\frac{\ca}{2}\right)
\frac{\nt \cdot \epst^*_\lambda}{\pi}
\int_0^{z}\ud z' \frac{\sqrt{1-z}}{z(z-z')} \times \mathcal{H}_{\ref{diag:vertexcggg2}}(z',z;\lambda,h),
\end{equation}
where
\begin{equation}
\mathcal{H}_{\ref{diag:vertexcggg2}}(z',z;\lambda,h) = A_{\ref{diag:vertexcggg2}}(z,z')\delta_{\lambda,h} + B_{\ref{diag:vertexcggg2}}(z,z')\delta_{\lambda,-h}
\end{equation}
and the coefficients are given by
\begin{equation}
\begin{split}
A_{\ref{diag:vertexcggg2}}(z,z') & =   \frac{1}{16\pi (1-z)(1-z')zz'(z-z')}\biggl [-(z')^3\left (1+(1-z')^2 \right ) + z(z')^2\left (2+3(1-z')^2z' \right ) -z^3(1-z')\left (4-z'(4-3z') \right ) \\
& +z^2z'\left (6-z'(17-2(8-3z')z') \right )  + 2z(1-z')(z-z')\biggl \{ \left (z(1-z')^2-z'(1+(1-z')^2) \right )\log\left (\frac{z(1-z')}{z-z'} \right ) \\	
& + \left (z'(1-z')^2-z(1+(1-z')^2) \right )\log \left ( \frac{(1-z)z^2}{z'(z-z')} \right ) \biggr \} \biggr ]
\end{split}
\end{equation}
and
\begin{equation}
\begin{split}
B_{\ref{diag:vertexcggg2}}(z,z') & =  \frac{1}{16\pi (1-z)^2(1-z')zz'(z'-z)}\biggl [(1-z)(z')^3(1+(1-z')^2) + 2(1-z)z^3(2+z'-5(z')^2)\\
& -z(1-z)(z')^2(2-z'+4(z')^2-(z')^3) -3(1-z)z^2z'(2-z'(3+z'))-(1-z)z^4(4-z'(4+z'))\\
& +2z(1-z')\biggl \{(1-z)^2(2z^2-3zz'+(z')^2)\log \left (\frac{(1-z)z^2}{z'(z-z')} \right )\\
& -(z-z')\left (z(1-z)^2 + zz'(2-z) -z'(1+(1-z')^2) \right )\log \left ( \frac{z(1-z')}{z-z'}\right ) \biggr \}\biggr ].
\end{split}
\end{equation}

\subsubsection*{Sum of diagrams \ref{diag:vertexcggg1} and \ref{diag:vertexcggg2} }
Again, diagrams  \ref{diag:vertexcggg1} and  \ref{diag:vertexcggg2} have the same color structure and correspond to the same covariant theory diagram. 
The sum of  \ref{diag:vertexcggg1} and  \ref{diag:vertexcggg2}, with the soft divergence regulated by $z+\alpha<z'<1$ in \ref{diag:vertexcggg1} and  $\alpha < z'<z-\alpha$ in \ref{diag:vertexcggg2} has the UV-divergent part
\begin{equation} \label{eq:vertexw3gfinalv1}
 \psi^{q\to qg}_{\ref{diag:vertexcggg1},\rm pole}
+  \psi^{q\to qg}_{\ref{diag:vertexcggg2}, \rm pole}
=  \psi^{q\to qg}_\lo(\nt,z) \frac{g^2}{8\pi^2}\left (\frac{\ca}{2}\right )\biggl [\frac{1}{ \varepsilon_{\overline{\rm MS}}}+\log \left (\frac{\mu^2}{\nt^2}\right )\biggr ]
 \left[-\frac{3}{2}- 4 \ln \alpha + \ln(1-z) +2\ln z \right].
\end{equation}
And the finite ones 
\begin{equation}
\label{eq:finalfg}
\begin{split}
 \psi^{q\to qg}_{\ref{diag:vertexcggg1},\rm finite} & +  \psi^{q\to qg}_{\ref{diag:vertexcggg2},\rm finite}  =  
-\psi^{q\to qg}_\lo(\nt,z) \frac{g^2}{48\pi^2} \left(\frac{\ca}{2}\right)
\mathcal{H}_{\ref{diag:vertexcggg1}\ref{diag:vertexcggg2}}(z,\alpha)\\
& +  \frac{2g^3t^{a}_{ji}}{\Delta^-_{03}}  \left(\frac{\ca}{2}\right)
\frac{3(\nt \cdot \epst^*_\lambda)}{48\pi^2z\sqrt{1-z}} \biggl [ \left ((1+2z)\log(1-z)  - 1 \right )\delta_{\lambda, h} +  2z(2-z)\log(z)\delta_{\lambda, -h} \biggr ], 
\end{split}
\end{equation}
where 
\begin{equation}
\begin{split}
\mathcal{H}_{\ref{diag:vertexcggg1}\ref{diag:vertexcggg2}}(z,\alpha) = & 12 - 4\pi^2 + \frac{12(1-z)z}{\alpha} + 3\left ( \frac{1-3z^2}{z}\right )\log(1-z) -3\left (\frac{1+z+z^2}{1-z} \right )\log(z) + 24\log(\alpha) - 18\log^2(\alpha)\\
& - 36\log\left ( \frac{z}{1-z}\right )\log(\alpha) -30\log(1-z)\log(z) - 9\log^2(1-z) -12\log^2(z) - 6\mathrm{Li}_2\left (z\right ).
\end{split}
\end{equation}

\begin{figure}[t]
\includegraphics[height=2cm]{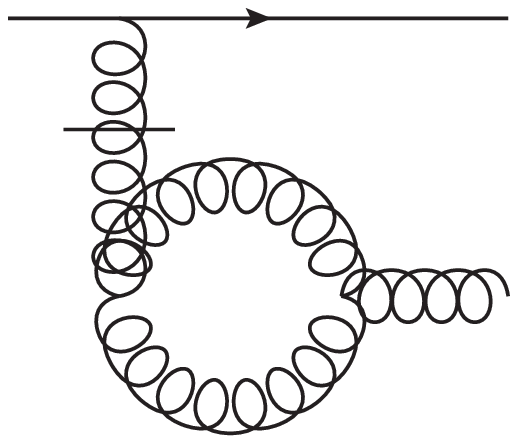}
\begin{tikzpicture}[overlay]
\node[anchor=south east] at (-0cm,0cm) {\namediag{diag:instggloop}};
\end{tikzpicture}
\includegraphics[height=2cm]{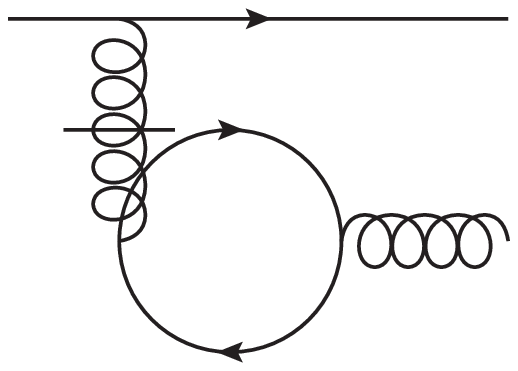}
\begin{tikzpicture}[overlay]
\node[anchor=south east] at (-0cm,0cm) {\namediag{diag:instqqbarloop}};
\end{tikzpicture}
\includegraphics[height=2cm]{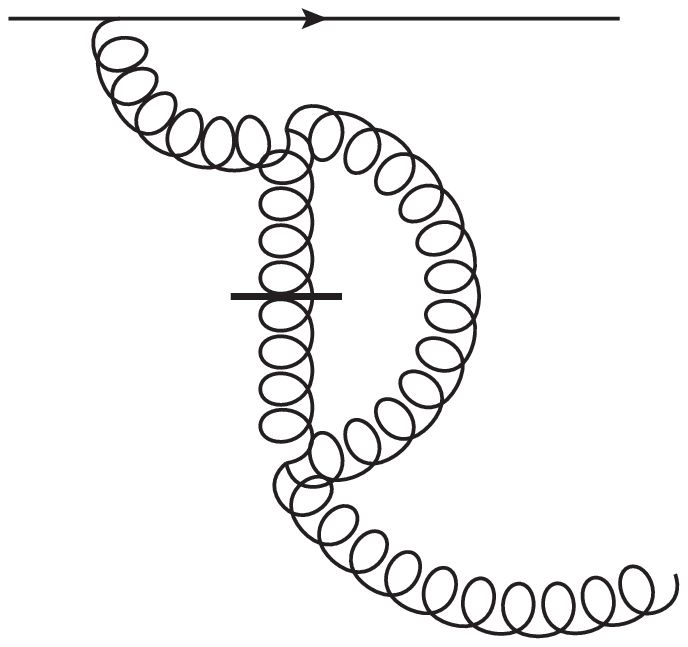}
\begin{tikzpicture}[overlay]
\node[anchor=south west] at (-2.5cm,0cm) {\namediag{diag:instinggloop}};
\end{tikzpicture}
\includegraphics[height=2cm]{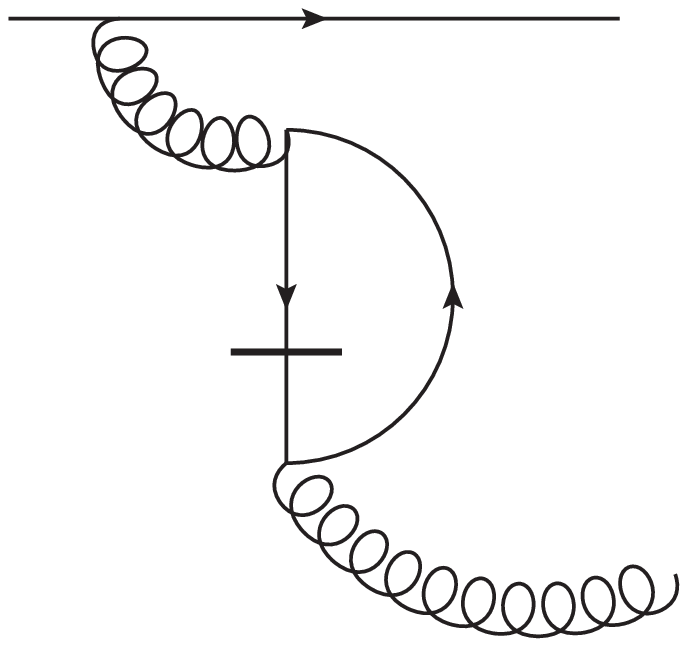}
\begin{tikzpicture}[overlay]
\node[anchor=south west] at (-2.5cm,0cm) {\namediag{diag:instinqqloop}};
\end{tikzpicture}
\includegraphics[height=2cm]{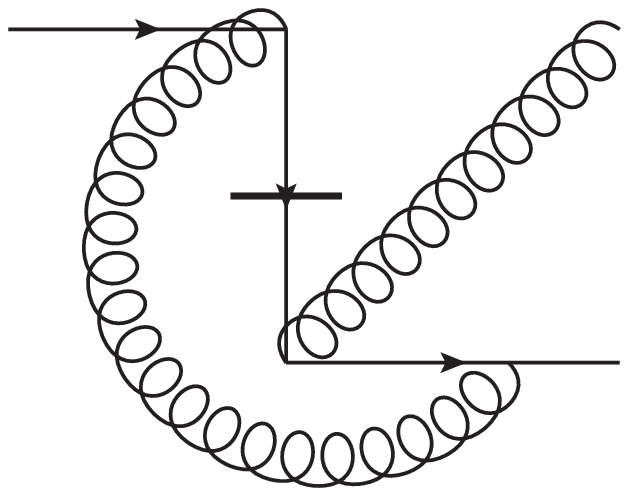}
\begin{tikzpicture}[overlay]
\node[anchor=south west] at (-0.5cm,1cm) {\namediag{diag:instqinvertexc1}};
\end{tikzpicture}
\includegraphics[height=2cm]{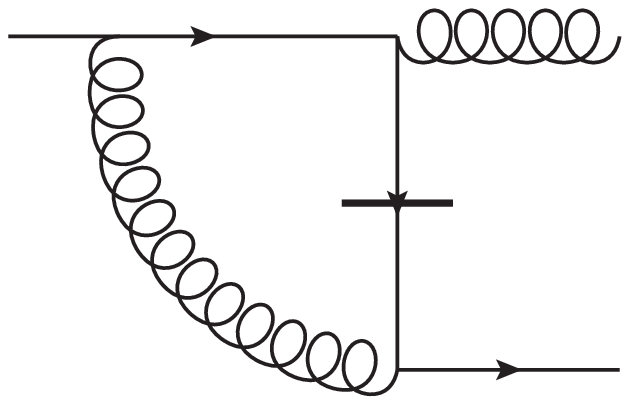}
\begin{tikzpicture}[overlay]
\node[anchor=south west] at (-0.5cm,0.8cm) {\namediag{diag:instqinvertexc2}};
\end{tikzpicture}
\includegraphics[height=2cm]{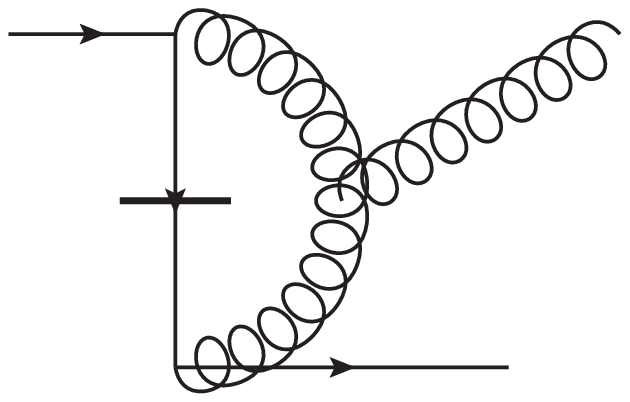}
\begin{tikzpicture}[overlay]
\node[anchor=south west] at (-0.5cm,0.5cm) {\namediag{diag:instqinvertexc3}};
\end{tikzpicture}
\includegraphics[height=2cm]{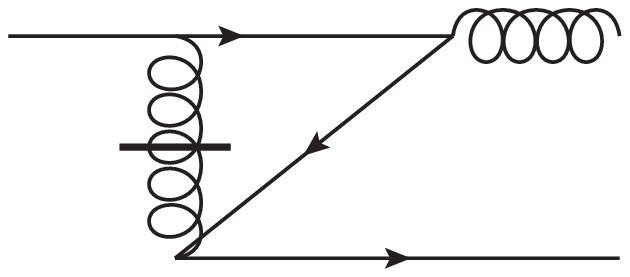}
\begin{tikzpicture}[overlay]
\node[anchor=south west] at (-1cm,0.2cm) {\namediag{diag:instginvertex1}};
\end{tikzpicture}
\includegraphics[height=2cm]{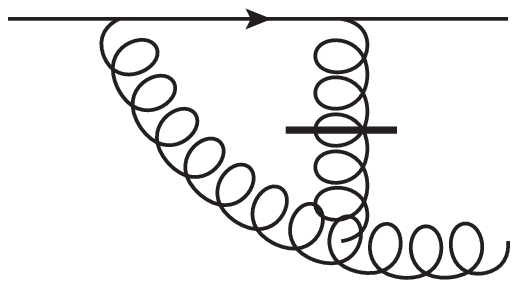}
\begin{tikzpicture}[overlay]
\node[anchor=south west] at (-1cm,0.5cm) {\namediag{diag:instginvertex2}};
\end{tikzpicture}
\includegraphics[height=2cm]{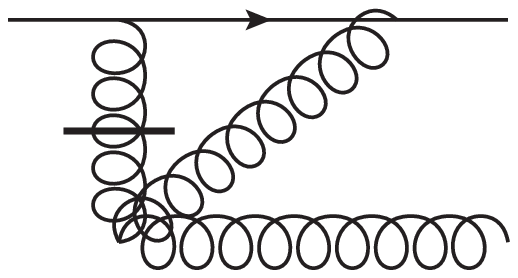}
\begin{tikzpicture}[overlay]
\node[anchor=south west] at (-1cm,0.6cm) {\namediag{diag:instginvertex3}};
\end{tikzpicture}
\includegraphics[height=2cm]{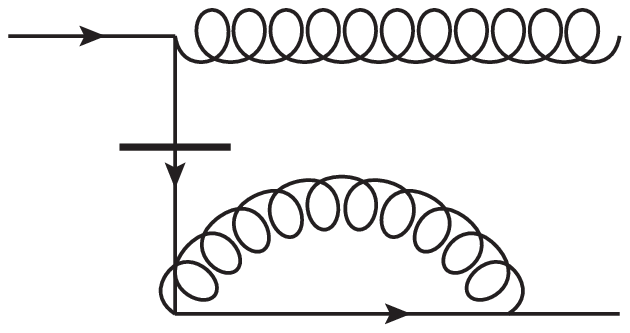}
\begin{tikzpicture}[overlay]
\node[anchor=south west] at (-1cm,0.2cm) {\namediag{diag:instqwavefafter}};
\end{tikzpicture}
\caption{
Loop diagrams with an instantaneous interaction, vanishing in dimensional regularization.
}\label{fig:instloop}
\end{figure}

\subsection{One loop vertex corrections to gluon emission with an  instantaneous diagram}

In addition to the one loop vertex corrections computed in section \ref{sec:vertcor} there are several one loop diagrams containing either quark or gluon instantaneous vertex, see \fig\ref{fig:instloop}. It is, however, straightforward to show that at one loop all of these corrections are only linearly proportional to the transverse momentum in the loop, and hence the possible  $d_{\perp}$ dimensional transverse integral gives zero in dimensional regularization framework.

\subsection{Three-particle final states}
\label{sec:3partfinal}

\begin{figure}[t]
\centerline{
\includegraphics[width=6.4cm]{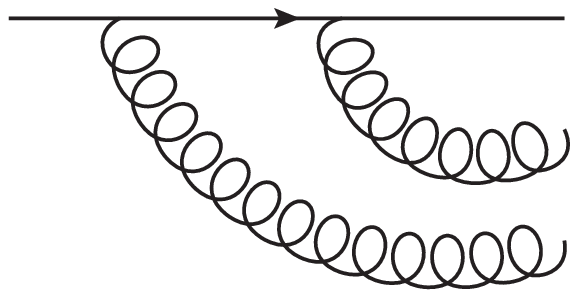}
\begin{tikzpicture}[overlay]
\draw [dashed] (-5.8,3.5) -- (-5.8,0);
\node[anchor=north] at (-5.8cm,-0.2cm) {0};
\draw [dashed] (-3.6,3.5) -- (-3.6,0);
\node[anchor=north] at (-3.6cm,-0.2cm) {1};
\draw [dashed] (-1.5,3.5) -- (-1.5,0);
\node[anchor=north] at (-1.5cm,-0.2cm) {2};
\node[anchor=south] at (-6.2,3) {$\pvec,h,i$};
\node[anchor=south] at (-0.5,3) {$\ppvec,h,j$};
\node[anchor=south] at (-4,3) {$\pppvec,h,k$};
\node[anchor=west] at (-0.4,1.6) {$\kvec,\lambda,a$};
\node[anchor=west] at (-0.4,0.6) {$\kpvec,\lambda',b$};
\node[anchor=south] at (-5,3) {$\hht$};
\node[anchor=south] at (-2.8,3) {$\lt$};
\node[anchor=south west] at (-7cm,0cm) {\namediag{diag:doubleg}};
 \end{tikzpicture}
}
\rule{0pt}{1ex}
\caption{
Two-gluon emission diagram \ref{diag:doubleg} with LC energy denominators and kinematics. 
Energy conservation: $\pvec = \pppvec  + \kpvec$ 
and $\pppvec = \ppvec + \kvec$. The momentum fraction and natural 
momentum scale for the emitted gluon $\kpvec$ are $z(1-z') = k'^+/p^+$ and
 $\hht = \ktp - z(1-z')\pt$, respectively. 
Similarly, for the emitted gluon $\kvec$ we have $zz'/(1-z(1-z'))
 = k^+/{p''}^{+}$ 
and $\lt = \kt - zz'/(1-z(1-z'))\ptpp$. 
}
\label{fig:doublegkin}
\end{figure}

\begin{figure}[t]
\centerline{
\includegraphics[width=6.3cm]{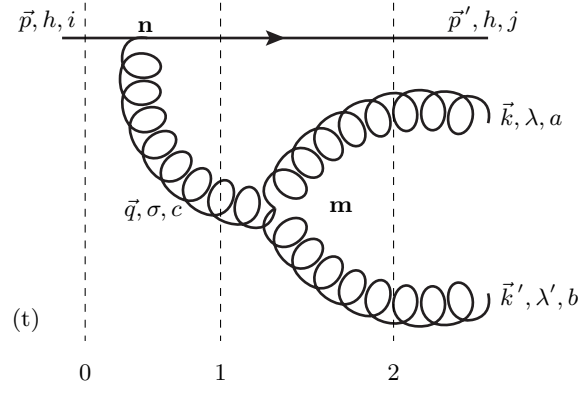}
\begin{tikzpicture}[overlay]
\draw [dashed] (-5.8,4.5) -- (-5.8,0);
\node[anchor=north] at (-5.8cm,-0.2cm) {0};
\draw [dashed] (-4,4.5) -- (-4,0);
\node[anchor=north] at (-4cm,-0.2cm) {1};
\draw [dashed] (-1.7,4.5) -- (-1.7,0);
\node[anchor=north] at (-1.7cm,-0.2cm) {2};
\node[anchor=south] at (-6.3,4) {$\pvec,h,i$};
\node[anchor=south] at (-0.5,4) {$\ppvec, h, j$};
\node[anchor=north east] at (-4.4,2) {$\qvec,\sigma,c$};
\node[anchor=west] at (-0.4,3) {$\kvec,\lambda,a$};
\node[anchor=west] at (-0.4,0.6) {$\kpvec,\lambda',b$};
\node[anchor=south] at (-5,4) {$\nt$};
\node[anchor=south] at (-2.4,1.6) {$\mt$};
\node[anchor=south west] at (-7cm,0cm) {\namediag{diag:qtoqgg3g}};
 \end{tikzpicture}
}
\rule{0pt}{1ex}
\caption{
Two-gluon splitting diagram \ref{diag:qtoqgg3g} with intermediate states for LC energy denominators and kinematics. Energy conservation: $\pvec = \ppvec + \qvec$ and 
$\qvec = \kpvec + \kvec$. 
The momentum fraction and natural momentum scale for the emitted gluon $\qvec$ 
are $z = q^+/p^+$ and $\mathbf{n} = \mathbf{q} - z\mathbf{p}$, respectively.  
The momentum fraction of the 3 gluon splitting is $k^+/q^+ = z'$ and the natural momentum  $\mt = \kt - z'\qt$.  
}\label{fig:3gkin}
\end{figure}

Finally, we write down for reference the wave functions for possible three-particle final states shown in \figs\ref{fig:doublegkin}, \ref{fig:3gkin} and \ref{fig:qtoqqbqkin}.
The wave function for the two-gluon emission diagram  \ref{diag:doubleg}, with kinematical variables shown in  \fig\ref{fig:doublegkin}, is
\begin{equation}
\psi^{q\rightarrow qgg}_{\ref{diag:doubleg}} =  
\int \dppp (2\pi)^3\delta^{(3)}(\pvec - \pppvec - \kpvec)
\frac{
V^{i;\ell,b}_{\lambda',h}\left(\hht,z(1-z')\right)
V^{\ell;j,a}_{\lambda,h}\left(\lt,\frac{zz'}{1-z(1-z')}\right)
}{\Delta_{01}^{-} \Delta_{02}^{-} },
\end{equation}
where the LC energy denominators are given by 
\begin{eqnarray}
\Delta_{01}^{-} &= &\frac{-1}{2p^+}\frac{\hht^2}{z(1-z')(1-z(1-z'))}
\\
\label{eq:totaldelta3part1}
\Delta_{02}^{-}  & = & 
\frac{-1}{2p^+}
\frac{(1-z(1-z'))}{zz'(1-z)}
\biggl [
\lt^2 + \frac{z'(1-z)}{(1-z')(1-z(1-z'))^2}\hht^2 \biggl],
\end{eqnarray}
and the phase space measure simplifies to 
\begin{equation}
\int \dppp(2\pi)^3\delta^{(3)}(\pvec - \pppvec - \kpvec) = \frac{1}{2p^{+}(1-z(1-z'))}.  
\end{equation}
Putting everything together we get 
\begin{multline}
\label{eq:doubleg}
\psi^{q\rightarrow qgg}_{\ref{diag:doubleg}} = 
8p^{+}   g^2 t^{a}_{jk} t^{b}_{ki}
 \sqrt{1-z}
\biggl [\delta_{\lambda', h} + (1-z(1-z'))\delta_{\lambda', -h} \biggr ]
\biggl [\delta_{\lambda, h} + \left (\frac{1-z}{1-z(1-z')} \right )\delta_{\lambda, -h} \biggr ]
\\ 
\times \frac{(\lt\cdot \epst_{\lambda}^{\ast})(\hht\cdot \epst_{\lambda'}^{\ast} )}{
\hht^2 
\biggl [
\lt^2 + \frac{z'(1-z)}{(1-z')(1-z(1-z'))^2}\hht^2 \biggl]}.
\end{multline}
Note that in the full Fock state decomposition, the wave function $\Psi^{q\to qgg}$ also needs to include a second copy of diagram \ref{diag:doubleg} with the final state gluon momentum labels interchanged.

For diagram \ref{diag:qtoqgg3g}, with kinematical variables as in \fig\ref{fig:3gkin}, the LC wave function is given by 
\begin{equation}
\psi^{q\rightarrow qgg}_{\ref{diag:qtoqgg3g}} =  
\int \dq (2\pi)^3\delta^{(3)}(\pvec- \ppvec - \qvec)
\frac{
V^{i;j,c}_{\sigma,h}(\nt,z)
\Gamma^{c;a,b}_{\sigma, \lambda, \lambda'}(\mt,z') 
}{\Delta_{01}^{-} \Delta_{02}^{-} },
\end{equation}
and the LC energy denominators are
\begin{eqnarray}
\label{eq:totaldelta3partinterm}
\Delta_{01}^{-} & = & \frac{-1}{2p^+}\frac{\nt^2}{z(1-z)} \\
\label{eq:totaldelta3part2}
\Delta_{02}^{-} & =& \frac{-1}{2p^+}\frac{1}{z(1-z)}\biggl [\nt^2+\frac{(1-z)}{z'(1-z')}\mt^2\biggr ],
\end{eqnarray}
where the second LC energy difference \nr{eq:totaldelta3part2} is naturally the same as for the other emission diagram to the same final state in \nr{eq:totaldelta3part1}, although expressed in terms of different variables.
The phase space integration gives 
\begin{equation}
\label{eq:PSqgg}
\int \dq (2\pi)^3\delta^{(3)}(\pvec- \ppvec - \qvec)
 = \frac{1}{2zp^{+}} . 
\end{equation}
Putting everything together we obtain
\begin{multline}
\label{eq:dia3g}
\psi^{q\rightarrow qgg}_{\ref{diag:qtoqgg3g}} 
= 8ip^+g^2t^c_{ji}f^{cab}(1-z)^{3/2}
\biggl [\delta_{\sigma, h} + (1-z)\delta_{\sigma, -h} \biggr ]\biggl [
\frac{\mt\cdot \epst^{\ast}_{\lambda}}{1-z'} \delta_{\sigma, \lambda'} 
+ \frac{\mt\cdot \epst^{\ast}_{\lambda'}}{z'} \delta_{\sigma, \lambda} 
- \mt\cdot \epst_{\sigma} \delta_{\lambda, -\lambda'}\biggr ]
\\
\times
\frac{(\mathbf{n}\cdot \epst_{\lambda}^{\ast})
}{
\nt^2 \biggl [\nt^2 + \frac{(1-z)}{z'(1-z')}\mt^2 \biggr ].
}
\end{multline}
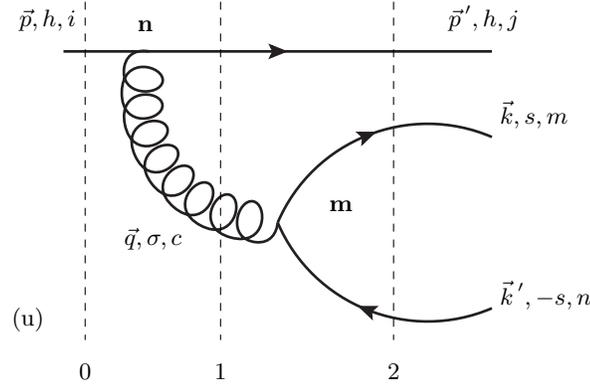
\begin{figure}[t]
\centerline{
\includegraphics[width=6.3cm]{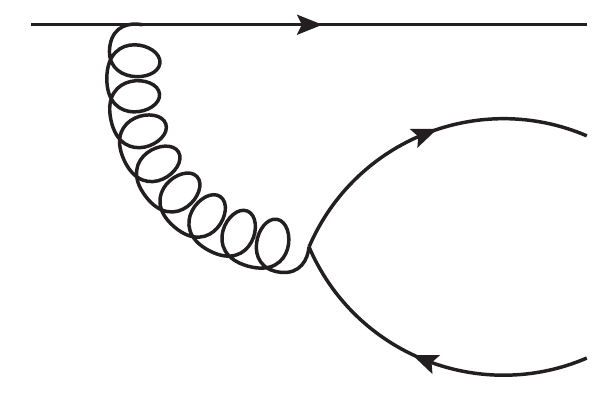}
\begin{tikzpicture}[overlay]
\draw [dashed] (-5.8,4.5) -- (-5.8,0);
\node[anchor=north] at (-5.8cm,-0.2cm) {0};
\draw [dashed] (-4,4.5) -- (-4,0);
\node[anchor=north] at (-4cm,-0.2cm) {1};
\draw [dashed] (-1.7,4.5) -- (-1.7,0);
\node[anchor=north] at (-1.7cm,-0.2cm) {2};
\node[anchor=south] at (-6.3,4) {$\pvec,h,i$};
\node[anchor=south] at (-0.5,4) {$\ppvec, h, j$};
\node[anchor=north east] at (-4.4,1.6) {$\qvec,\sigma,c$};
\node[anchor=west] at (-0.4,3) {$\kvec,s,m$};
\node[anchor=west] at (-0.4,0.6) {$\kpvec,-s,n$};
\node[anchor=south] at (-5,4) {$\nt$};
\node[anchor=south] at (-2.4,1.6) {$\mt$};
\node[anchor=south west] at (-7cm,0cm) {\namediag{diag:qtoqqbarq}};
 \end{tikzpicture}
}
\rule{0pt}{1ex}
\caption{
Quark-antiquark splitting diagram \ref{diag:qtoqqbarq} with intermediate states and kinematics. Momentum conservation dictates $\pvec = \ppvec + \qvec$ and $\qvec = \kpvec + \kvec$. The momentum fraction and natural momentum scale for the emitted gluon $\qvec$ 
are $z = q^+/p^+$ and $\mathbf{n} = \mathbf{q} - z\mathbf{p}$, respectively.  
The momentum fraction of the gluon to quark pair splitting is $k^+/q^+ = z'$ and the natural momentum  $\mt = \kt - z'\qt$.  
}\label{fig:qtoqqbqkin}
\end{figure}

The wave function for quark-antiquark splitting diagram \ref{diag:qtoqqbarq} shown in \fig\ref{fig:qtoqqbqkin} can be written as 
 
\begin{equation}
\psi^{q\rightarrow qq\bar{q}}_{\ref{diag:qtoqqbarq}} =  
\int \dq (2\pi)^3\delta^{(3)}(\pvec- \ppvec - \qvec)
\frac{
V^{i;j,c}_{\sigma,h}(\nt,z)A^{c;n,m}_{\sigma,s}(\mt,z') 
}{\Delta_{01}^{-} \Delta_{02}^{-} },
\end{equation}
where the integration over the phase space simplifies to \eq\nr{eq:PSqgg}, and the LC energy denominators are the same as for the three gluon final state, \eqs\nr{eq:totaldelta3partinterm} and \nr{eq:totaldelta3part2}. Adding everything together, the wave function for diagram  \ref{diag:qtoqqbarq}  can be cast in the following form
\begin{multline}
\label{eq:diag2q}
\psi^{q\rightarrow qq\bar{q}}_{\ref{diag:qtoqgg3g}} 
= 8p^+g^2t^c_{ji}t^c_{nm}\frac{(1-z)^{3/2}}{\sqrt{z'(1-z')}}
\biggl [\delta_{\sigma, h} + (1-z')\delta_{\sigma, -h} \biggr ]
\biggl [z'\delta_{\sigma, s} - (1-z')\delta_{\sigma, -s} \biggr ]
\frac{(\mathbf{n}\cdot \epst_{\sigma}^{\ast})(\mathbf{m}\cdot \epst_{\sigma})
}{
\nt^2 \biggl [\nt^2 + \frac{(1-z)}{z'(1-z')}\mt^2 \biggr ]
}.
\end{multline}
In addition to these, there are three contributions from the instantaneous diagrams in \figs\ref{fig:qtoqgginstq}, \ref{fig:qtoqgginst} and \ref{fig:qtoqqbarqinst}. Since these diagrams do not have an intermediate state, there is no phase space integral. The  contribution to the wave function is given directly by the instantaneous vertex (\eqs\nr{eq:qtoqgginstq}, \nr{eq:qtoqgginst} and \nr{eq:qtoqqbinst}) divided by the light cone energy denominator between the initial and final states (which is equal to $\Delta^-_{02}$  in \nr{eq:totaldelta3part1} or \nr{eq:totaldelta3part2}).

\section{Results} \label{sec:results}

Let us first collect the UV-divergent pieces of our results and show that they can be absorbed into a renormalization of the QCD coupling constant. The pole part of the quark wave function renormalization constant \nr{eq:zqv1} is
\begin{equation}
Z_q^{\textnormal{pole}}(k^+=\xi p^+) = 1 + \frac{g^2\cf}{8\pi^2}\left(\frac{1}{\epsmsbar} + \ln \mu^2 \right) 
\left[
\left( \frac{3}{2}+2\log(\alpha/\xi) \right) 
\right]
+\dots
\end{equation}
Note the presence of the  longitudinal soft cutoff $\alpha$ from the integration over the momentum fractions in the UV-divergent term. This mixing of divergences is a well known annoying aspect of light cone perturbation theory.
The UV-pole part of the gluon wave function renormalization is \nr{eq:zgpole}
\begin{equation}
Z_g^{\textnormal{pole}}(k^+=\xi p^+) = 1 + \frac{g^2}{8\pi^2}\left(\frac{1}{\epsmsbar} + \ln \mu^2 \right) 
\left[
\left( \frac{11}{6} +2\log(\alpha/\xi) \right) \ca
- \frac{2\tf\nf}{3}
\right] + \mathcal{O}(g^4)
\end{equation}
The UV-divergent part of the splitting wave function, with UV-divergent parts from the final state propagator correction diagrams \ref{diag:qwavefeafter}, \ref{diag:gtoggloop}, \ref{diag:gtoqqbloop}
(\eqs\nr{eq:qwavefafterfinalv1} and \nr{eq:gtoggloop}) and  the vertex correction diagrams \ref{diag:vertexc2}, \ref{diag:vertexc1}, \ref{diag:vertexcggg1} 
 and  \ref{diag:vertexcggg2} 
(\eqs\nr{eq:vertexno3gfinalv1}, \nr{eq:vertexw3gfinalv1}) 
is
\begin{multline}
\psi^{q\to qg}_{\textnormal{pole}}
=
\psi^{q\to qg}_{\lo}  
 \bigg[
1+
\frac{g^2}{8 \pi^2 }\left(\frac{1}{\epsmsbar} + \ln \mu^2 \right) \bigg\{
\left(\cf -\frac{\ca}{2}\right)
\left[
-\frac{3}{2}+ \ln(1-z)-2\ln \alpha
\right]
\\
+
\frac{\ca}{2}
 \left[-\frac{3}{2}- 4 \ln \alpha + \ln(1-z) +2\ln z \right]
\bigg\}
+ \left[Z^{\textnormal{pole}}_g(z)-1\right] + \left[Z^{\textnormal{pole}}_q(1-z)-1\right]
\bigg].
\end{multline}
From these the combination 
\begin{equation}
\frac{\sqrt{Z_q(1)}}{\sqrt{Z_q(1-z)}\sqrt{Z_g(z)}} \psi^{q\to qg}
= \psi^{q\to qg}_{\lo}\left[
1+
\frac{g^2}{16 \pi^2 } \left( \frac{11}{6} \ca - \frac{2\tf\nf}{3} \right) \left(\frac{1}{\epsmsbar} + \ln \mu^2 \right) 
+ \mathcal{O}(g^4)\right]
\end{equation}
is independent of the soft regulator $\alpha$ and can be absorbed into a replacement of the coupling $g$ in $ \psi^{q\to qg}_{\lo}$ by a renormalized running coupling:
\begin{equation}
 g_R(Q^2) = g\left[
1+
\frac{g^2}{16 \pi^2 } \left( \frac{11}{6} \ca - \frac{2\tf\nf}{3} \right) \left(\frac{1}{\epsmsbar} 
+ \ln \frac{\mu^2}{Q^2} \right) \right].
\end{equation}

We will not repeat the expressions for the UV finite parts here. The finite parts of the quark and gluon wavefunction renormalization constants can be found in \eqs\nr{eq:zqv1} and~\nr{eq:zg}. The light cone wave function is obtained by summing the contributions in \eqs\nr{eq:qwavefafterfinalv1}, \nr{eq:gtoggloop}, \nr{eq:finaled} and \nr{eq:finalfg}. The finite terms contain single logarithmic, double logarithmic, and power like soft divergences in terms of the longitudinal cut-off parameter $\alpha$. In the context of low-$x$ QCD, the single logarithmic divergences can be dealt by an appropriate low-$x$ evolution equation at the cross-section level (see e.g. \cite{Balitsky:1995ub,Kovchegov:1999yj,Kovchegov:1999ua,Balitsky:2008zza,Chirilli:2011km}). As could have been expected, the double logarithmic terms cancel each other in Eqs. (54), (67), (99) and (129). What remains, however, is a power law infrared divergence $\sim 1/\alpha$. The appearence of such divergences is a well known troublesome feature of calculations in light cone gauge. Without performing a full cross section calculation it is not fully possible to say whether this term would cancel in the final result. It would also be interesting to study further whether the appearence of such a term is related to our implementation of the FDH scheme for regulating the UV divergences. This could be done e.g. by calculating the same loop diagram in the conventional dimensional regularization scheme.

\section{Discussion} \label{sec:disc}

Let us now briefly return to the question of applications for the results obtained here. As stated earlier, the primary motivation for the light cone wave function formulation is for calculations off an (potentially nonperturbatively strong) classical field target in the high energy limit. We have here obtained the $qg$ final state wave function to order $g^3$, and the $ggg$, $gq\bar{q}$ one to order $g^2$. These are sufficient to calculate to order $g^4$, i.e. to NLO, cross sections for gluon emission from a quark probe. A simple example would be the gluon brehmsstrahlung process $q\gamma^* \to q g$, e.g. the conventional DIS process. Another example, where the only increase in complexity would be on the target side, would be the process $qg^* \to qg$, i.e. a dihadron production in forward proton.-nucleus collisions~\cite{Marquet:2007vb,Lappi:2012nh} to NLO.

There are other calculations where the results obtained here would be useful, but that would require also the quark and gluon wave function renormalization to order $g^4$. This would be the case e.g. for the NLO DGLAP splitting functions~\cite{Curci:1980uw} (or the quark anomalous dimension to 2 loops). A rederivation of the NLO BK equation would also require  a similar expression for a quark-antiquark dipole, but only in the soft gluon limit $z\to 0$. The wave function renormalization constant can be obtained from the state normalization constraint, i.e. from the square of the 3-particle final state wave functions obtained in this paper, and an interference term between the LO and NLO 1-gluon emission wave functions also written down here. There is no shortage of interesting things to calculate.

An important  goal of this paper has been to formulate the LCPT rules in helicity space, independently of spinor representation, and demonstrate their use. We have found that this is often not done in the literature, but presents a major  technological improvement for  loop calculations, replacing the Dirac algebra by simple helicity sums. We believe that this will make many calculations such as the one recently performed in \cite{Beuf:2016wdz} technically simpler.
As a part of our work we rederived the QCD beta function in LCPT~\cite{Perry:1992sw}. More importantly we also extracted the finite parts of the one loop wave function correction diagrams which, to our knowledge, have not appeared in the previous literature.

\begin{acknowledgments}
  We are grateful to H. M\"antysaari and R. Venugopalan for discussions.
  This work has been  supported by the Academy of Finland, projects 267321,  273464 and  303756 and  by the European Research Council, grants
 ERC-2011-StG-279579 and  ERC-2015-CoG-681707 and by the Ministerio de Ciencia e
 Innovacion of Spain under project FPA2014-58293-C2-1-P and the Xunta de Galicia
 - Strategic Unit AGRUP2015/11.
\end{acknowledgments}

\appendix

\section{Loop integrals}
\label{sec:integrals}

We work in the four dimensional helicity (FDH) scheme, where all polarization vectors, the momenta of the observed particles, as well as the helicity sums (or Dirac matrix algebra in the numerator) are kept in four dimensions. Although we keep all the spins in 4 spacetime dimensions, we regularize UV divergences by integrating over the momenta of all particles ub $d$ dimensions. Consequently the integral measure appearing in \eq\nr{eq:phasespace} is replaced by
\begin{equation}
\int \dk \rightarrow  \frac{\mu^{2-d_{\perp}}}{2\pi}\int\frac{\ud k^+}{2k^+} \int \frac{\ud^{d_{\perp}} \kt}{(2\pi)^{d_{\perp}}} \quad \text{with}\quad k^{i}k^{j}\rightarrow \frac{\delta^{ij}}{d_{\perp}}\kt^2,
\end{equation}
where the transverse integral is dimensionaly regularized near two dimensions $d_{\perp} = 2 - 2\varepsilon$, and an arbitrary scale $\mu$ is introduced so that the transverse integrals preserve their natural dimensions.

At one loop order, we need the following set of transverse integrals in $d_{\perp}$ dimensions:
\begin{eqnarray}
\label{eq:oltransint1}
 \mu^{2 \varepsilon} \int \frac{\ud^{2-2\varepsilon} \kt}{(2\pi)^{2-2\varepsilon}}
\frac{1}{\kt^2 + M^2} 
&=& \frac{\Gamma(\varepsilon)}{(4\pi)^{1-\varepsilon}} \left(\frac{\mu^2}{M^2}\right)^\varepsilon
\\
\label{eq:oltransint2}
 \mu^{2 \varepsilon} \int \frac{\ud^{2-2\varepsilon} \kt}{(2\pi)^{2-2\varepsilon}}
\frac{1}{(\kt^2 + M^2)^2} 
&=& \frac{1}{M^2}\frac{\varepsilon\Gamma(\varepsilon)}{(4\pi)^{1-\varepsilon}} \left(\frac{\mu^2}{M^2}\right)^\varepsilon
\\
\label{eq:oltransint3}
 \mu^{2 \varepsilon} \int \frac{\ud^{2-2\varepsilon} \kt}{(2\pi)^{2-2\varepsilon}}
\frac{\kt^2}{\kt^2 + M^2} 
&=& - M^2 \frac{\Gamma(\varepsilon)}{(4\pi)^{1-\varepsilon}} \left(\frac{\mu^2}{M^2}\right)^\varepsilon
\\
\label{eq:oltransint4}
 \mu^{2 \varepsilon} \int \frac{\ud^{2-2\varepsilon} \kt}{(2\pi)^{2-2\varepsilon}}
\frac{\kt^2}{(\kt^2 + M^2)^2} 
&=&  \frac{\Gamma(\varepsilon)(1-\varepsilon)}{(4\pi)^{1-\varepsilon}} 
\left(\frac{\mu^2}{M^2}\right)^\varepsilon
\\
\label{eq:oltransint5}
 \mu^{2 \varepsilon} \int \frac{\ud^{2-2\varepsilon} \kt}{(2\pi)^{2-2\varepsilon}}
\frac{k^i k^j}{(\kt^2 + M^2)^2} 
&=&  \frac{\Gamma(\varepsilon)\delta^{ij}}{2(4\pi)^{1-\varepsilon}} 
\left(\frac{\mu^2}{M^2}\right)^\varepsilon.
\end{eqnarray}
For the computation of one loop vertex corrections with 3 external legs we need the generic integral
\begin{equation}\label{eq:1loopint}
I_3(\pt,\qt,\kt,M) =  \mu^{2\varepsilon}\int \frac{\ud^{2-2\varepsilon}\mt}{(2\pi)^{2-2\varepsilon}}
\frac{m^i (m-p)^j (m-q)^k}{\mt^2\biggl [(\mt-\kt)^2 + M^2\biggr ]}.
\end{equation}
This can be Feynman parametrized as
\begin{equation}
I_3 (\pt,\qt,\kt,M)
=\int_0^1 \ud x \int \frac{\ud^{2-2\varepsilon}\mt}{(2\pi)^{2-2\varepsilon}} 
\frac{\mu^{2\varepsilon} \quad m^i (m-p)^j (m-q)^k}{\left[(1-x)\mt^2 + x\left((\mt-\kt)^2 + M^2\right)\right]^2}
\end{equation}
and with a variable change $\rt\equiv \mt-x\kt$, the integral $I_3$ becomes
\begin{equation}
I_3 (\pt,\qt,\kt,M)
=\int_0^1 \ud x \mu^{2\varepsilon}\int \frac{\ud^{2-2\varepsilon}\rt}{(2\pi)^{2-2\varepsilon}} 
\frac{(r+xk)^i (r+xk-p)^j (r+xk-q)^k}{\left[\rt^2 + x(1-x)\kt^2 + x  M^2 \right]^2}.
\end{equation}
At this point what survives is a UV finite part
\begin{equation}
\label{eq:finitepart}
I_3^{\rm finite}
=
\frac{\varepsilon\Gamma(\varepsilon)\mu^{2\varepsilon}}{(4\pi)^{1-\varepsilon}}
\int_0^1 \ud x  
\frac{xk^i (xk-p)^j (xk-q)^k}{\biggl [ x(1-x)\kt^2 + x M^2\biggr ]^{1+\varepsilon}}
\end{equation}
and a UV divergent one
\begin{equation}
\label{eq:polepart}
I_3^{pole}=\frac{\Gamma(\varepsilon)}{2(4\pi)^{1-\varepsilon}}
\int_0^1 \ud x
\left( xk^i\delta^{jk} + (xk-p)^j \delta^{ik} + (xk-q)^k\delta^{ij}\right)
\left(\frac{\mu^2}{ x(1-x)\kt^2 + x M^2}\right)^\varepsilon.
\end{equation}
Expanding the integrands in \eqref{eq:finitepart} and \eqref{eq:polepart} in powers of $\varepsilon$, and performing the final elementary integrals we obtain
\begin{equation}
 I_3(\pt,\qt,\kt,M) =  I^{\rm UV}_3(\pt,\qt,\kt) +  I^{\rm finite}_3(\pt,\qt,\kt,M),
\end{equation}
where the UV pole part is 
\begin{equation}
I^{\rm UV}_3(\pt,\qt,\kt) = \frac{1}{16\pi}\biggl [\frac{1}{\varepsilon_{\overline{\rm MS}}} + \log\left ( \frac{\mu^2}{\nt^2}\right ) \biggr ]
\biggl [ k^i\delta^{jk} + (k - 2p)^j \delta^{ik} + (k - 2q)^k\delta^{ij}\biggr ],
\end{equation}
and we have defined, $1/\varepsilon_{\overline{\rm MS}} = 1/\varepsilon - \gamma_{\rm E} + \log(4\pi)$, which correspond to the standard modified minimal subtraction ($\overline{\rm MS}$) scheme choice.
Here $\nt^2$ is an arbitrary scale at which one divides the logarithm between the  pole and finite terms.
 The finite part simplifies to
\begin{equation}
\label{eq:finiteintpart}
\begin{split}
I^{\rm finite}_3(\pt,\qt,\kt,M) =  I_3^{a}  + I_3^{b} ,
\end{split}
\end{equation}
where 
\begin{equation}
\begin{split}
I_3^{a} =    \frac{1}{16\pi} \log\left (\frac{\nt^2}{M^2}\right )\biggl [k^i\delta^{jk} + (k - 2p)^j \delta^{ik} + (k - 2q)^k\delta^{ij}\biggr ] &+ \frac{\log(1+\xi)}{4\pi M^2\xi^3}\biggl [ (1+\xi)^2\Delta_1^{a} + \xi(1+\xi)\Delta_2^{a} + \xi^2\Delta_3^{a} \biggr ]\\
& -\frac{(1+\xi)\log(1+\xi)}{8\pi \xi}\biggl [ \frac{(1+\xi)}{2\xi}\Delta_1^{b} + \Delta_2^{b}\biggr ] + \mathcal{O}(\varepsilon)
\end{split}
\end{equation}
and 
\begin{equation}
\begin{split}
I_3^{b} =  
 -\frac{1}{4\pi M^2\xi}\biggl [\frac{(2+3\xi)}{2\xi}\Delta_1^{a} + \Delta_2^{a} \biggr ] + \frac{1}{8\pi}\biggl [\frac{(1+2\xi)}{2\xi}\Delta_1^{b}  + 2\Delta_2^{b} \biggr ] + \mathcal{O}(\varepsilon)
\end{split}
\end{equation}
with $\xi  = \kt^2/M^2$ and 
\begin{equation}
\begin{split}
\Delta_1^{a} & =k^{i}k^{j}k^{k} \\
\Delta_2^{a} & =-k^{i}(k^{j}q^{k}+k^{k}p^{j}) \\
\Delta_3^{a} & = k^{i}p^{j}q^{k}\\
\Delta_1^{b} & = k^{i}\delta^{j k} + k^{j}\delta^{ik} + k^{k}\delta^{ij}\\
\Delta_2^{b} & = -\left (p^{j}\delta^{ik} + q^{k}\delta^{ij}\right ).
\end{split}
\end{equation}

\bibliography{spires}

\providecommand{\href}[2]{#2}\begingroup\raggedright\begin{thebibliography}{10}

\bibitem{Kogut:1969xa}
J.~B. Kogut and D.~E. Soper, {\it Quantum electrodynamics in the infinite
  momentum frame},  \href{http://dx.doi.org/10.1103/PhysRevD.1.2901}{{\em Phys.
  Rev.} {\bf D1} (1970) 2901}.

\bibitem{Bjorken:1970ah}
J.~Bjorken, J.~B. Kogut and D.~E. Soper, {\it Quantum electrodynamics at
  infinite momentum: Scattering from an external field},
  \href{http://dx.doi.org/10.1103/PhysRevD.3.1382}{{\em Phys.Rev.} {\bf D3}
  (1971) 1382}.

\bibitem{Lepage:1980fj}
G.~P. Lepage and S.~J. Brodsky, {\it Exclusive processes in perturbative
  quantum chromodynamics},
  \href{http://dx.doi.org/10.1103/PhysRevD.22.2157}{{\em Phys. Rev.} {\bf D22}
  (1980) 2157}.

\bibitem{Brodsky:1997de}
S.~J. Brodsky, H.-C. Pauli and S.~S. Pinsky, {\it Quantum chromodynamics and
  other field theories on the light cone},
  \href{http://dx.doi.org/10.1016/S0370-1573(97)00089-6}{{\em Phys. Rept.} {\bf
  301} (1998) 299} [\href{http://arXiv.org/abs/hep-ph/9705477}{{\tt
  arXiv:hep-ph/9705477 [hep-ph]}}].

\bibitem{Iancu:2003xm}
E.~Iancu and R.~Venugopalan in {\em Quark gluon plasma} (R.~Hwa and X.~N. Wang,
  eds.).
\newblock World Scientific, 2003.
\newblock \href{http://arXiv.org/abs/hep-ph/0303204}{{\tt
  arXiv:hep-ph/0303204}}.

\bibitem{Weigert:2005us}
H.~Weigert, {\it Evolution at small {$x_{\textnormal{bj}}$}: The color glass
  condensate},  \href{http://dx.doi.org/10.1016/j.ppnp.2005.01.029}{{\em Prog.
  Part. Nucl. Phys.} {\bf 55} (2005) 461}
  [\href{http://arXiv.org/abs/hep-ph/0501087}{{\tt arXiv:hep-ph/0501087}}].

\bibitem{Gelis:2010nm}
F.~Gelis, E.~Iancu, J.~Jalilian-Marian and R.~Venugopalan, {\it The color glass
  condensate},
  \href{http://dx.doi.org/10.1146/annurev.nucl.010909.083629}{{\em Ann. Rev.
  Nucl. Part. Sci.} {\bf 60} (2010) 463}
  [\href{http://arXiv.org/abs/1002.0333}{{\tt arXiv:1002.0333 [hep-ph]}}].

\bibitem{Albacete:2014fwa}
J.~L. Albacete and C.~Marquet, {\it Gluon saturation and initial conditions for
  relativistic heavy ion collisions},
  \href{http://dx.doi.org/10.1016/j.ppnp.2014.01.004}{{\em Prog. Part. Nucl.
  Phys.} {\bf 76} (2014) 1} [\href{http://arXiv.org/abs/1401.4866}{{\tt
  arXiv:1401.4866 [hep-ph]}}].

\bibitem{Thorn:1979gv}
C.~B. Thorn, {\it Asymptotic freedom in the infinite momentum frame},
  \href{http://dx.doi.org/10.1103/PhysRevD.20.1934}{{\em Phys. Rev.} {\bf D20}
  (1979) 1934}.

\bibitem{Mustaki:1990im}
D.~Mustaki, S.~Pinsky, J.~Shigemitsu and K.~Wilson, {\it Perturbative
  renormalization of null plane {QED}},
  \href{http://dx.doi.org/10.1103/PhysRevD.43.3411}{{\em Phys. Rev.} {\bf D43}
  (1991) 3411}.

\bibitem{Perry:1992sw}
R.~J. Perry, A.~Harindranath and W.-M. Zhang, {\it Asymptotic freedom in
  {Hamiltonian} light front quantum chromodynamics},
  \href{http://dx.doi.org/10.1016/0370-2693(93)90739-5}{{\em Phys. Lett.} {\bf
  B300} (1993) 8}.

\bibitem{Zhang:1993is}
W.-M. Zhang and A.~Harindranath, {\it Role of longitudinal boundary integrals
  in light front {QCD}},
  \href{http://dx.doi.org/10.1103/PhysRevD.48.4868}{{\em Phys. Rev.} {\bf D48}
  (1993) 4868} [\href{http://arXiv.org/abs/hep-th/9302119}{{\tt
  arXiv:hep-th/9302119 [hep-th]}}].

\bibitem{Zhang:1993dd}
W.-M. Zhang and A.~Harindranath, {\it Light front {QCD}. 2: Two component
  theory},  \href{http://dx.doi.org/10.1103/PhysRevD.48.4881}{{\em Phys. Rev.}
  {\bf D48} (1993) 4881}.

\bibitem{Harindranath:1993de}
A.~Harindranath and W.-M. Zhang, {\it Light front {QCD}. 3: Coupling constant
  renormalization},  \href{http://dx.doi.org/10.1103/PhysRevD.48.4903}{{\em
  Phys. Rev.} {\bf D48} (1993) 4903}.

\bibitem{Balitsky:2008zza}
I.~Balitsky and G.~A. Chirilli, {\it Next-to-leading order evolution of color
  dipoles},  \href{http://dx.doi.org/10.1103/PhysRevD.77.014019}{{\em Phys.
  Rev.} {\bf D77} (2008) 014019} [\href{http://arXiv.org/abs/0710.4330}{{\tt
  arXiv:0710.4330 [hep-ph]}}].

\bibitem{Balitsky:2013fea}
I.~Balitsky and G.~A. Chirilli, {\it Rapidity evolution of {Wilson} lines at
  the next-to-leading order},
  \href{http://dx.doi.org/10.1103/PhysRevD.88.111501}{{\em Phys. Rev.} {\bf
  D88} (2013) 111501} [\href{http://arXiv.org/abs/1309.7644}{{\tt
  arXiv:1309.7644 [hep-ph]}}].

\bibitem{Kovner:2013ona}
A.~Kovner, M.~Lublinsky and Y.~Mulian, {\it {Jalilian-Marian}, {Iancu},
  {McLerran}, {Weigert}, {Leonidov}, {Kovner} evolution at next to leading
  order},  \href{http://dx.doi.org/10.1103/PhysRevD.89.061704}{{\em Phys. Rev.}
  {\bf D89} (2014) 061704} [\href{http://arXiv.org/abs/1310.0378}{{\tt
  arXiv:1310.0378 [hep-ph]}}].

\bibitem{Balitsky:2014mca}
I.~Balitsky and A.~V. Grabovsky, {\it {NLO} evolution of 3-quark {Wilson} loop
  operator},  \href{http://dx.doi.org/10.1007/JHEP01(2015)009}{{\em JHEP} {\bf
  01} (2015) 009} [\href{http://arXiv.org/abs/1405.0443}{{\tt arXiv:1405.0443
  [hep-ph]}}].

\bibitem{Beuf:2014uia}
G.~Beuf, {\it Improving the kinematics for low-x {QCD} evolution equations in
  coordinate space},  \href{http://dx.doi.org/10.1103/PhysRevD.89.074039}{{\em
  Phys. Rev.} {\bf D89} (2014) 074039}
  [\href{http://arXiv.org/abs/1401.0313}{{\tt arXiv:1401.0313 [hep-ph]}}].

\bibitem{Lappi:2015fma}
T.~Lappi and H.~M{\"a}ntysaari, {\it Direct numerical solution of the
  coordinate space {Balitsky-Kovchegov} equation at next to leading order},
  \href{http://dx.doi.org/10.1103/PhysRevD.91.074016}{{\em Phys. Rev.} {\bf
  D91} (2015) 074016} [\href{http://arXiv.org/abs/1502.02400}{{\tt
  arXiv:1502.02400 [hep-ph]}}].

\bibitem{Iancu:2015vea}
E.~Iancu, J.~D. Madrigal, A.~H. Mueller, G.~Soyez and D.~N. Triantafyllopoulos,
  {\it Resumming double logarithms in the {QCD} evolution of color dipoles},
  \href{http://dx.doi.org/10.1016/j.physletb.2015.03.068}{{\em Phys. Lett.}
  {\bf B744} (2015) 293} [\href{http://arXiv.org/abs/1502.05642}{{\tt
  arXiv:1502.05642 [hep-ph]}}].

\bibitem{Iancu:2015joa}
E.~Iancu, J.~D. Madrigal, A.~H. Mueller, G.~Soyez and D.~N. Triantafyllopoulos,
  {\it Collinearly-improved {BK} evolution meets the {HERA} data},
  \href{http://dx.doi.org/10.1016/j.physletb.2015.09.071}{{\em Phys. Lett.}
  {\bf B750} (2015) 643} [\href{http://arXiv.org/abs/1507.03651}{{\tt
  arXiv:1507.03651 [hep-ph]}}].

\bibitem{Lappi:2016fmu}
T.~Lappi and H.~M{\"a}ntysaari, {\it Next-to-leading order {Balitsky-Kovchegov}
  equation with resummation},
  \href{http://dx.doi.org/10.1103/PhysRevD.93.094004}{{\em Phys. Rev.} {\bf
  D93} (2016) 094004} [\href{http://arXiv.org/abs/1601.06598}{{\tt
  arXiv:1601.06598 [hep-ph]}}].

\bibitem{Lublinsky:2016meo}
M.~Lublinsky and Y.~Mulian, {\it High energy {QCD} at {NLO}: from light-cone
  wave function to {JIMWLK} evolution},
  \href{http://arXiv.org/abs/1610.03453}{{\tt arXiv:1610.03453 [hep-ph]}}.

\bibitem{Beuf:2011xd}
G.~Beuf, {\it {NLO} corrections for the dipole factorization of dis structure
  functions at low x},
  \href{http://dx.doi.org/10.1103/PhysRevD.85.034039}{{\em Phys. Rev.} {\bf
  D85} (2012) 034039} [\href{http://arXiv.org/abs/1112.4501}{{\tt
  arXiv:1112.4501 [hep-ph]}}].

\bibitem{Balitsky:2012bs}
I.~Balitsky and G.~A. Chirilli, {\it Photon impact factor and
  $k_t$-factorization for dis in the next-to-leading order},
  \href{http://dx.doi.org/10.1103/PhysRevD.87.014013}{{\em Phys. Rev.} {\bf
  D87} (2013) 014013} [\href{http://arXiv.org/abs/1207.3844}{{\tt
  arXiv:1207.3844 [hep-ph]}}].

\bibitem{Beuf:2016wdz}
G.~Beuf, {\it Dipole factorization for {DIS} at {NLO} {I}: Loop correction to
  the photon to quark-antiquark light-front wave-functions},
  \href{http://dx.doi.org/10.1103/PhysRevD.94.054016}{{\em Phys. Rev.} {\bf
  D94} (2016)~no.~5 054016} [\href{http://arXiv.org/abs/1606.00777}{{\tt
  arXiv:1606.00777 [hep-ph]}}].

\bibitem{Altinoluk:2011qy}
T.~Altinoluk and A.~Kovner, {\it Particle production at high energy and large
  transverse momentum - 'the hybrid formalism' revisited},
  \href{http://dx.doi.org/10.1103/PhysRevD.83.105004}{{\em Phys. Rev.} {\bf
  D83} (2011) 105004} [\href{http://arXiv.org/abs/1102.5327}{{\tt
  arXiv:1102.5327 [hep-ph]}}].

\bibitem{JalilianMarian:2011dt}
J.~Jalilian-Marian and A.~H. Rezaeian, {\it Hadron production in {pA}
  collisions at the {LHC} from the {Color Glass Condensate}},
  \href{http://dx.doi.org/10.1103/PhysRevD.85.014017}{{\em Phys. Rev.} {\bf
  D85} (2012) 014017} [\href{http://arXiv.org/abs/1110.2810}{{\tt
  arXiv:1110.2810 [hep-ph]}}].

\bibitem{Chirilli:2012jd}
G.~A. Chirilli, B.-W. Xiao and F.~Yuan, {\it Inclusive hadron productions in
  {pA} collisions},  \href{http://dx.doi.org/10.1103/PhysRevD.86.054005}{{\em
  Phys. Rev.} {\bf D86} (2012) 054005}
  [\href{http://arXiv.org/abs/1203.6139}{{\tt arXiv:1203.6139 [hep-ph]}}].

\bibitem{Stasto:2013cha}
A.~M. Stasto, B.-W. Xiao and D.~Zaslavsky, {\it Towards the test of saturation
  physics beyond leading logarithm},
  \href{http://dx.doi.org/10.1103/PhysRevLett.112.012302}{{\em Phys. Rev.
  Lett.} {\bf 112} (2014) 012302} [\href{http://arXiv.org/abs/1307.4057}{{\tt
  arXiv:1307.4057 [hep-ph]}}].

\bibitem{Kang:2014lha}
Z.-B. Kang, I.~Vitev and H.~Xing, {\it Next-to-leading order forward hadron
  production in the small-$x$ regime: rapidity factorization},
  \href{http://dx.doi.org/10.1103/PhysRevLett.113.062002}{{\em Phys. Rev.
  Lett.} {\bf 113} (2014) 062002} [\href{http://arXiv.org/abs/1403.5221}{{\tt
  arXiv:1403.5221 [hep-ph]}}].

\bibitem{Altinoluk:2014eka}
T.~Altinoluk, N.~Armesto, G.~Beuf, A.~Kovner and M.~Lublinsky, {\it
  Single-inclusive particle production in proton-nucleus collisions at
  next-to-leading order in the hybrid formalism},
  \href{http://dx.doi.org/10.1103/PhysRevD.91.094016}{{\em Phys. Rev.} {\bf
  D91} (2015) 094016} [\href{http://arXiv.org/abs/1411.2869}{{\tt
  arXiv:1411.2869 [hep-ph]}}].

\bibitem{Ayala:2016lhd}
A.~Ayala, M.~Hentschinski, J.~Jalilian-Marian and M.~E. Tejeda-Yeomans, {\it
  Polarized 3 parton production in inclusive {DIS} at small x},
  \href{http://dx.doi.org/10.1016/j.physletb.2016.08.035}{{\em Phys. Lett.}
  {\bf B761} (2016) 229} [\href{http://arXiv.org/abs/1604.08526}{{\tt
  arXiv:1604.08526 [hep-ph]}}].

\bibitem{Boussarie:2014lxa}
R.~Boussarie, A.~V. Grabovsky, L.~Szymanowski and S.~Wallon, {\it Impact factor
  for high-energy two and three jets diffractive production},
  \href{http://dx.doi.org/10.1007/JHEP09(2014)026}{{\em JHEP} {\bf 09} (2014)
  026} [\href{http://arXiv.org/abs/1405.7676}{{\tt arXiv:1405.7676 [hep-ph]}}].

\bibitem{Boussarie:2016ogo}
R.~Boussarie, A.~V. Grabovsky, L.~Szymanowski and S.~Wallon, {\it On the one
  loop {$ {\gamma}^{\left(\ast \right)}\to q\overline{q} $} impact factor and
  the exclusive diffractive cross sections for the production of two or three
  jets},  \href{http://dx.doi.org/10.1007/JHEP11(2016)149}{{\em JHEP} {\bf 11}
  (2016) 149} [\href{http://arXiv.org/abs/1606.00419}{{\tt arXiv:1606.00419
  [hep-ph]}}].

\bibitem{Pauli:2000gw}
H.~C. Pauli, {\it A compendium of light cone quantization},
  \href{http://dx.doi.org/10.1016/S0920-5632(00)00904-X}{{\em Nucl. Phys. Proc.
  Suppl.} {\bf 90} (2000) 259} [\href{http://arXiv.org/abs/hep-ph/0103106}{{\tt
  arXiv:hep-ph/0103106 [hep-ph]}}].
\newblock [,259(2000)].

\bibitem{Kovchegov:2012mbw}
Y.~V. Kovchegov and E.~Levin, {\em {Quantum chromodynamics at high energy}},
  vol.~33 of {\em Cambridge monographs on particle physics, nuclear physics and
  cosmology}.
\newblock Cambridge University Press, 2012.

\bibitem{Bern:1991aq}
Z.~Bern and D.~A. Kosower, {\it The computation of loop amplitudes in gauge
  theories},  \href{http://dx.doi.org/10.1016/0550-3213(92)90134-W}{{\em Nucl.
  Phys.} {\bf B379} (1992) 451}.

\bibitem{Bern:2002zk}
Z.~Bern, A.~De~Freitas, L.~J. Dixon and H.~L. Wong, {\it Supersymmetric
  regularization, two loop {QCD} amplitudes and coupling shifts},
  \href{http://dx.doi.org/10.1103/PhysRevD.66.085002}{{\em Phys. Rev.} {\bf
  D66} (2002) 085002} [\href{http://arXiv.org/abs/hep-ph/0202271}{{\tt
  arXiv:hep-ph/0202271 [hep-ph]}}].

\bibitem{Chirilli:2011km}
G.~A. Chirilli, B.-W. Xiao and F.~Yuan, {\it One-loop factorization for
  inclusive hadron production in {pA} collisions in the saturation formalism},
  \href{http://dx.doi.org/10.1103/PhysRevLett.108.122301}{{\em Phys. Rev.
  Lett.} {\bf 108} (2012) 122301} [\href{http://arXiv.org/abs/1112.1061}{{\tt
  arXiv:1112.1061 [hep-ph]}}].

\bibitem{Balitsky:1995ub}
I.~Balitsky, {\it Operator expansion for high-energy scattering},
  \href{http://dx.doi.org/10.1016/0550-3213(95)00638-9}{{\em Nucl. Phys.} {\bf
  B463} (1996) 99} [\href{http://arXiv.org/abs/hep-ph/9509348}{{\tt
  arXiv:hep-ph/9509348}}].

\bibitem{Kovchegov:1999yj}
Y.~V. Kovchegov, {\it Small-x {F2} structure function of a nucleus including
  multiple pomeron exchanges},
  \href{http://dx.doi.org/10.1103/PhysRevD.60.034008}{{\em Phys. Rev.} {\bf
  D60} (1999) 034008} [\href{http://arXiv.org/abs/hep-ph/9901281}{{\tt
  arXiv:hep-ph/9901281}}].

\bibitem{Kilgore:2011ta}
W.~B. Kilgore, {\it Regularization schemes and higher order corrections},
  \href{http://dx.doi.org/10.1103/PhysRevD.83.114005}{{\em Phys. Rev.} {\bf
  D83} (2011) 114005} [\href{http://arXiv.org/abs/1102.5353}{{\tt
  arXiv:1102.5353 [hep-ph]}}].

\bibitem{Boughezal:2011br}
R.~Boughezal, K.~Melnikov and F.~Petriello, {\it The four-dimensional helicity
  scheme and dimensional reconstruction},
  \href{http://dx.doi.org/10.1103/PhysRevD.84.034044}{{\em Phys. Rev.} {\bf
  D84} (2011) 034044} [\href{http://arXiv.org/abs/1106.5520}{{\tt
  arXiv:1106.5520 [hep-ph]}}].

\bibitem{Kovchegov:1999ua}
Y.~V. Kovchegov, {\it Unitarization of the {BFKL} pomeron on a nucleus},
  \href{http://dx.doi.org/10.1103/PhysRevD.61.074018}{{\em Phys. Rev.} {\bf
  D61} (2000) 074018} [\href{http://arXiv.org/abs/hep-ph/9905214}{{\tt
  arXiv:hep-ph/9905214}}].

\bibitem{Marquet:2007vb}
C.~Marquet, {\it Forward inclusive dijet production and azimuthal correlations
  in pa collisions},
  \href{http://dx.doi.org/10.1016/j.nuclphysa.2007.09.001}{{\em Nucl. Phys.}
  {\bf A796} (2007) 41} [\href{http://arXiv.org/abs/0708.0231}{{\tt
  arXiv:0708.0231 [hep-ph]}}].

\bibitem{Lappi:2012nh}
T.~Lappi and H.~M{\"a}ntysaari, {\it Forward dihadron correlations in
  deuteron-gold collisions with the {Gaussian} approximation of {JIMWLK}},
  \href{http://dx.doi.org/10.1016/j.nuclphysa.2013.03.017}{{\em Nucl. Phys.}
  {\bf A908} (2013) 51} [\href{http://arXiv.org/abs/1209.2853}{{\tt
  arXiv:1209.2853 [hep-ph]}}].

\bibitem{Curci:1980uw}
G.~Curci, W.~Furmanski and R.~Petronzio, {\it Evolution of parton densities
  beyond leading order: The nonsinglet case},
  \href{http://dx.doi.org/10.1016/0550-3213(80)90003-6}{{\em Nucl. Phys.} {\bf
  B175} (1980) 27}.

\end{thebibliography}\endgroup
\bibliographystyle{JHEP-2modlong}

\end{document}